\documentclass[a4paper,11pt]{article}
\pdfoutput=1 

\usepackage{jheppub} 

\usepackage[T1]{fontenc} 

\usepackage{amsmath,bm}
\usepackage{amsfonts}
\usepackage{amssymb}
\usepackage{graphicx}
\usepackage{float}
\usepackage{wrapfig}
\usepackage{latexsym}
\usepackage{hyperref}
\usepackage{feynmf}
\usepackage{exscale}
\usepackage{relsize}
\usepackage{listings}
\usepackage{upquote}
\usepackage{overpic}
\usepackage{setspace}
\usepackage{caption}
\usepackage{subcaption}
\usepackage{mathtools}
\usepackage[utf8]{inputenc} 
\usepackage{tikz}
\usepackage{url}

\def\K{\mathcal{K}}
\def\be{\begin{equation}}
\def\ee{\end{equation}}
\def\ba{\begin{eqnarray}}
\def\ea{\end{eqnarray}}
\def\nl{\nonumber\\}

\def\a{\alpha}

\def\Res{\mbox{Res}}

\def\CP1{\mathbb{CP}^1}
\def\SL2C{\mathrm{SL}(2,\mathbb{C})}

\def\Z2{\mathbb{Z}_2}

\def\oM{\overline{\mathcal{M}}}

\newcommand{\A}{\mathcal{A}}
\newcommand{\B}{\mathcal{B}}
\newcommand{\M}{\mathcal{M}}
\newcommand{\Proj}{\mathbb{P}}
\newcommand{\R}{\mathbb{R}}
\newcommand{\GL}{\text{GL}}
\newcommand{\sign}{\text{sign}}
\newcommand{\aOmega}{\underline{\Omega}}

\title{\boldmath Scattering Forms and the Positive Geometry of Kinematics, Color and the Worldsheet}

\author[a]{Nima Arkani-Hamed
,}
\author[b]{Yuntao Bai,}
\author[c,d]{Song He,}
\author[e,c]{Gongwang Yan}

\affiliation[a]{School of Natural Sciences, Institute for Advanced Study, Princeton, NJ, 08540, USA}
\affiliation[b]{Department of Physics, Princeton University, Princeton, NJ, 08544, USA}
\affiliation[c]{CAS Key Laboratory of Theoretical Physics, Institute of Theoretical Physics, Chinese Academy
of Sciences, Beijing, 100190, China}
\affiliation[d]{
University of Chinese Academy of Sciences, No.19A Yuquan Road, Beijing 100049, China}
\affiliation[e]{Institute for Advanced Study, Tsinghua University, Beijing, 100084, China}

\emailAdd{arkani@ias.edu}
\emailAdd{ytbai@princeton.edu}
\emailAdd{songhe@itp.ac.cn}
\emailAdd{ygw17@mails.tsinghua.edu.cn}

\abstract{The search for a theory of the S-Matrix over the past five decades has revealed surprising geometric structures underlying scattering amplitudes ranging from the string worldsheet to the amplituhedron, but these are all geometries in auxiliary  spaces as opposed to the kinematical space where amplitudes actually live. Motivated by recent advances providing a reformulation of the amplituhedron and planar $\mathcal{N}=4$ SYM amplitudes directly in kinematic space, we propose a novel geometric understanding of amplitudes in more general theories. The key idea is to think of amplitudes not as functions, but rather as differential forms on kinematic space.  We explore the resulting picture for a wide range of massless theories in general spacetime dimensions. For the bi-adjoint $\phi^3$ scalar theory, we establish a direct connection between its ``scattering form'' and a classic polytope---the associahedron---known to mathematicians since the 1960's. We find an associahedron living naturally in kinematic space, and the tree level amplitude is simply the ``canonical form'' associated with this ``positive geometry''. Fundamental physical properties such as locality and unitarity, as well as novel ``soft'' limits, are fully determined by the combinatorial geometry of this polytope. Furthermore, the moduli space for the open string worldsheet has also long been recognized as an associahedron.  We show that the scattering equations act as a diffeomorphism between the interior of this  old ``worldsheet associahedron'' and the new ``kinematic associahedron'', providing a geometric interpretation and simple conceptual derivation of the bi-adjoint CHY formula.  We also find ``scattering forms'' on kinematic space for Yang-Mills theory and the Non-linear Sigma Model, which are dual to the fully color-dressed amplitudes despite having no explicit color factors. This is possible due to a remarkable fact---``Color is Kinematics''--- whereby kinematic wedge products in the scattering forms satisfy the same Jacobi relations as color factors. Finally, all our scattering forms are well-defined on the projectivized kinematic space, a property which can be seen to provide a geometric origin for color-kinematics duality.}

\begin{document} 
\maketitle

\section{Introduction}

Scattering amplitudes are arguably the most basic observables in fundamental physics. 
Apart from their prominent role in the experimental exploration of the high energy frontier, scattering amplitudes also have a privileged theoretical status as the only known observable of quantum gravity in asymptotically flat space-time. As such it is natural to ask the ``holographic'' questions we have become accustomed to asking (and beautifully answering) in AdS spaces for two decades: given that the observables are anchored to the boundaries at infinity, is there also a  ``theory at infinity'' that directly computes the S-Matrix without invoking a local picture of evolution in the interior of the spacetime? 

Of course this question is famously harder in flat space than it is in AdS space. The (exceedingly well-known) reason for this is the fundamental difference in the nature of the boundaries of the two spaces. The boundary of AdS is an ordinary flat space with completely standard notions of ``time'' and ``locality'', thus we have perfectly natural candidates for what a ``theory on the boundary'' could be---just a local quantum field theory. We do not have these luxuries in asymptotically flat space. We can certainly think of the ``asymptotics'' concretely in any of a myriad of ways by specifying the asymptotic on-shell particle momenta in the scattering process. But whether this is done with Mandelstam invariants, or spinor-helicity variables, or twistors, or using the celestial sphere at infinity, in no case is there an obvious notion of ``locality'' and/or ``time'' in these spaces, and we are left with the fundamental mystery of what principles a putative ``theory of the S-Matrix'' should be based on. 

Indeed, the absence of a good answer to this question was the fundamental flaw that doomed the 1960's S-Matrix program. Many S-Matrix theorists hoped to find some sort of first-principle ``derivation'' of  fundamental analyticity properties encoding unitarity and causality in the S-Matrix, and in this way to find the principles for a theory of the S-Matrix. But to this day we do not know precisely what these ``analyticity properties encoding causality'' should be,  even in perturbation theory, and so it is not surprising that this ``systematic'' approach to the subject hit a dead end not long after it began. 

Keenly wary of this history, and despite the same focus on the S-Matrix as a fundamental observable, much of the modern explosion in our understanding of scattering amplitudes has adopted a fundamentally different and more intellectually adventurous philosophy towards the subject. Instead of hoping to slavishly {\it derive} the needed properties of the S-Matrix {\it from} the principles of unitarity and causality, there is now a different strategy: to look for fundamentally new principles and new laws, very likely associated with new mathematical structures, that produce the S-Matrix as the answer to entirely different kinds of natural questions, and to only later discover space-time and quantum mechanics, embodied in unitarity and (Lorentz-invariant) causality, as derived consequences rather than foundational principles.

The past fifty years have seen the emergence of a few fascinating geometric structures underlying scattering amplitudes in unexpected ways, encouraging this point of view. The first and still in many ways most remarkable example is perturbative string theory \cite{GSW,Pol}, which computes scattering amplitudes by an auxiliary computation of correlation functions in the worldsheet CFT. At the most fundamental level there is a basic geometric object---the moduli space of marked points on Riemann surfaces \cite{DM}---which has a ``factorizing'' boundary structure. This is the primitive origin of the {\it factorization} of scattering amplitudes, which is needed for unitarity and locality in perturbation theory.  More recently, we have seen a new interpretation of the same worldsheet structure first in the context of ``twistor string theory'' \cite{Witten:2003nn}, and much more generally in the program of ``scattering equations''~\cite{Cachazo:2013gna, Cachazo:2013hca}, which directly computes the amplitudes for massless particles using a worldsheet but with no stringy excitations~\cite{Berkovits:2013xba,Mason:2013sva}. 

Over the past five years, we have also seen an apparently quite different set of mathematical ideas~\cite{ArkaniHamed:2009dn,Hodges:2009hk,ArkaniHamed:2012nw} underlying scattering amplitudes in planar maximally supersymmetric gauge theory---the amplituhedron~\cite{Arkani-Hamed:2013jha}. This structure is more alien and unfamiliar than the worldsheet, but its core mathematical ideas are even simpler, of a fundamentally combinatorial nature involving nothing more than grade-school algebra in its construction. Moreover, the amplituhedron as a {\it positive geometry}~\cite{Arkani-Hamed:2017tmz} again produces a ``factorizing'' boundary structure that gives rise to locality and unitarity in a geometric way and makes manifest the hidden infinite-dimensional Yangian symmetry of the theory. 

While the existence of these magical structures is strong encouragement for the existence of a master theory for the S-Matrix, all these ideas have a disquieting feature in common. In all cases, the new geometric structures are {\it not seen directly in the space where the scattering amplitudes naturally live, but in some auxiliary spaces}.  These auxiliary spaces are where all the action is, be it the worldsheet or the generalized Grassmannian spaces of the amplituhedron. We are therefore {\it still} left to wonder: what sort of questions do we have to ask, directly in the space of ``scattering kinematics'', to generate local, unitary dynamics? Clearly we should not be writing down Lagrangians and computing path integrals, but what should we do instead? What mathematical structures breathe scattering-physics-life into the ``on-shell kinematic space''? And is there any avatar of the geometric structures of the worldsheet, or amplituhedra, in this kinematic space? 

\begin{figure}
\centering
\begin{overpic}[width=4cm]
{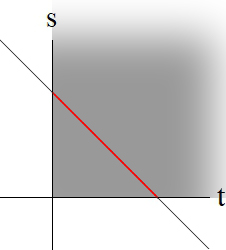}
\put(25,10){$s{+}t{=}c{>}0$}
\end{overpic}
\caption{The one-dimensional associahedron (red line segment) as the intersection of the positive region and the subspace $s+t=c$ where $c>0$ is a constant.}
\label{fig:4pt}
\end{figure}

Recent advances in giving a more intrinsic definition of the amplituhedron \cite{Arkani-Hamed:2017vfh} suggest the beginning of an answer to this question. A key observation is that, instead of thinking about scattering amplitudes merely as {\it functions} on kinematic space, they are to be thought of more fundamentally as {\it differential forms} on kinematic space. In the context of the amplituhedron and planar ${\cal N}=4$ SYM, kinematic space is simply the space of momentum twistors $Z_i$ for the particles $i=1,\ldots,n$~\cite{Hodges:2009hk}. And on this space the differential form has a natural purpose in life---it literally ``bosonizes'' the super-amplitude by treating the on-shell Grassmann variables $\eta_i$ for the $i^\text{th}$ particle as the momentum twistor differential $\eta_i \to dZ_i$. This seemingly innocuous move has dramatic geometric consequences: given a differential form, we can compute residues around singularities, and by now this is well known to reveal the underlying {\it positive geometry}. Indeed,~\cite{Arkani-Hamed:2017vfh} provides a novel description of the amplituhedron {\it purely in the standard momentum twistor kinematic space}, whereby the geometry arises as the intersection of a top-dimensional ``positive region'' in the kinematic space with a certain family of lower-dimensional subspaces with further ``positivity'' properties. The scattering form is defined everywhere in kinematic space, and is completely specified by its behavior when ``pulled back'' to the subspace on which the amplituhedron is revealed, whereby it becomes the {\it canonical form}~\cite{Arkani-Hamed:2017tmz} with logarithmic singularities on the boundaries of this positive geometry. 

In this paper, we will see a virtually identical structure emerge remarkably in a setting very far removed from special theories with maximal supersymmetry in the planar limit. We will consider a wide variety of theories of massless particles in a general number of dimensions, beginning with one of the simplest possible scalar field theories---a theory of {\it bi-adjoint scalars} with cubic interactions~\cite{Cachazo:2013iea}. The words connecting amplitudes to positive geometry are identical, but the cast of characters---the kinematic space, the precise definitions of the top-dimensional ``positive region'' and the ``family of subspaces''---differ in important ways. Happily all the objects involved are simpler and more familiar---the kinematic space is simply the space of Mandelstam invariants, the positive region is imposed by inequalities that demand positivity of physical poles, and the subspaces are cut out by linear equations in kinematic space---so that the resulting positive geometries are ordinary polytopes (as opposed to the generalization of polytopes into the Grassmannian seen in the amplituhedron). When the dust settles, what emerges is the famous and beautiful {\it associahedron} polytope~\cite{Stasheff_1,Stasheff_2}. In fact, the ``kinematic associahedron'' we have discovered is in a precise sense the ``amplituhedron'' for the bi-adjoint $\phi^3$ theory.

\begin{figure}
\begin{subfigure}{.5\textwidth}
\centering
\begin{overpic}[width=5cm]
{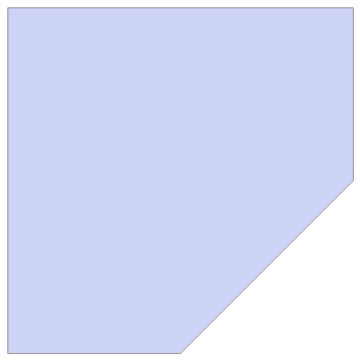}
\put(20,-7){$s_{45}$}
\put(-15,60){$s_{12}$}
\put(30,102){$s_{34}$}
\put(100,74){$s_{15}$}
\put(82,20){$s_{23}$}
\end{overpic}
\end{subfigure}
\begin{subfigure}{0.5\textwidth}
\centering
\begin{overpic}[width=8cm]
{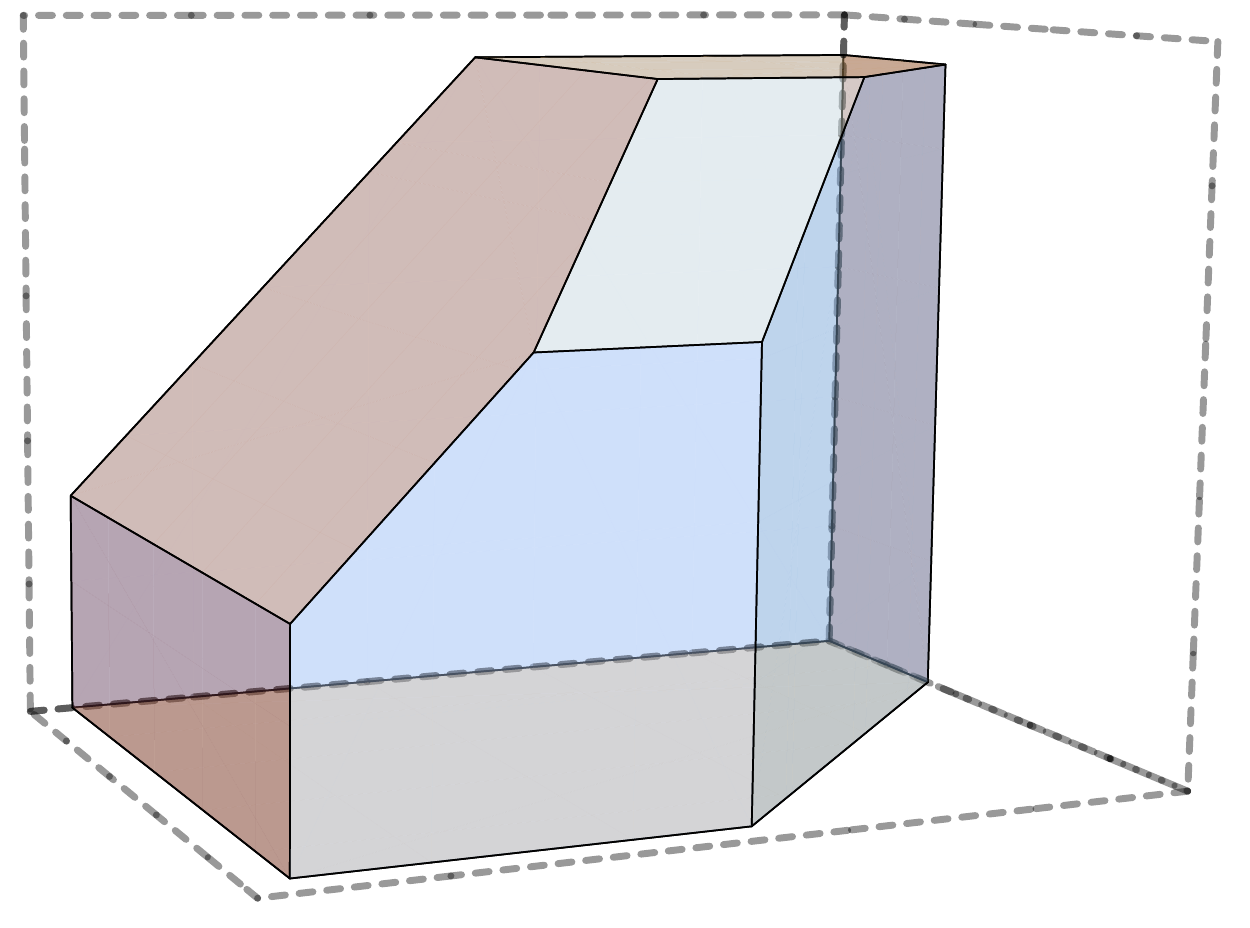}
\put(5,66){$s_{12}$}
\put(76,50){$s_{123}$}
\put(68,11){$s_{45}$}
\put(46,76){$s_{1234}$}
\put(25,46){$s_{34}$}
\put(52,56){$s_{234}$}
\put(40,26){$s_{2345}$}
\put(10,20){$s_{345}$}
\put(63,38){$s_{23}$}
\end{overpic}
\end{subfigure}
\newline
\caption{Pictures for $n{=}5$ (left) and $n{=}6$ (right) associahedra, where we have labeled every facet by the corresponding vanishing planar variable.}\label{fig:assoc_intro}
\end{figure}
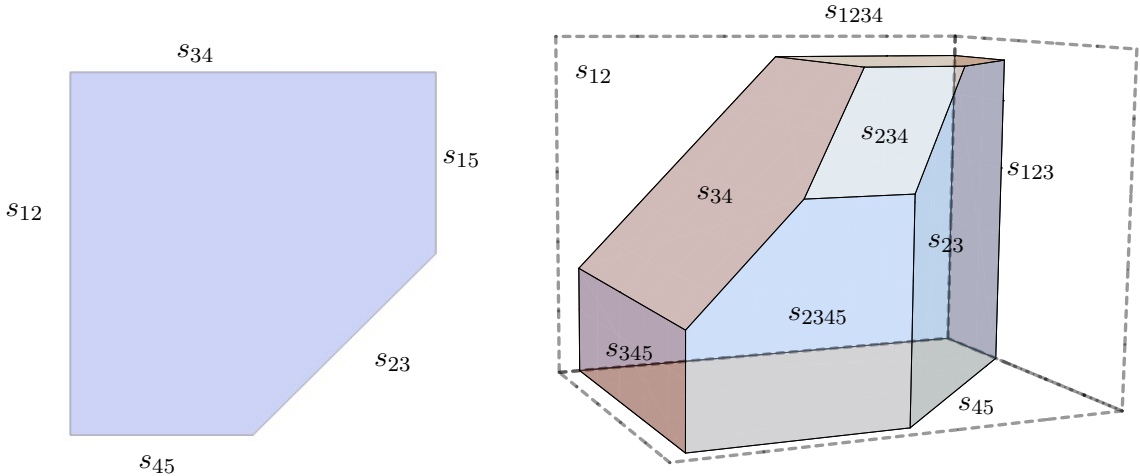
By way of a broad-brush invitation to the rest of the paper, let us illustrate the key ideas in some simple examples. Consider an amplitude for massless scalar particles whose Feynman diagram expansion is simply given by the sum over planar cubic tree graphs. For $n{=}4$ particles, the amplitude would simply be $\frac{1}{s} + \frac{1}{t}$. However, we consider instead a one-form $\Omega^{(1)}_{n{=}4}$ given by 
\begin{equation}
\Omega_{n{=}4}^{(1)}=\frac{ds}{s} - \frac{dt}{t} 
\end{equation}
The structure of the form is of course very natural; we are simply replacing ``1/propagator'' with $d\log$ of the propagator. The relative minus sign is more intriguing and is demanded by an interesting requirement---the differential form must be well-defined, not only on the two-dimensional $(s,t)$ space, but also on the projectivized version of the space; in other words, the form must be invariant under {\it local} ${\rm GL}(1)$ transformations $(s,t) \to \Lambda(s,t) (s,t)$; or said another way, it must only depend on the ratio $(s/t)$. Indeed, the minus sign allows us to rewrite the form as $d\log(s/t)$ which is manifestly projective. At $n$ points, we have an $(n{-}3)$-form obtained by wedging together the $d\log$ of propagators for every planar cubic graph, and summing over all graphs with relative signs fixed by projectivity. 

Returning to four points, we have a one-form defined on the two-dimensional $(s,t)$ space. But how can we extract the ``actual amplitude'' $\frac{1}{s} + \frac{1}{t}$ from this form, and how is it related to any sort of positive geometry? Both questions are answered at once by identifying some natural regions in kinematic space. First, if the poles of the amplitude are to correspond to boundaries of a geometry, it is clear that we should impose a positivity constraint on all the planar poles, which at four points are simply the conditions that $s,t\geq 0$. This brings us to the upper quadrant of the $(s,t)$ plane. But this alone can not correspond to the positive geometry we are seeking---for one thing, the space is two-dimensional while our scattering form is a one-form! This suggests that in addition to imposing these positivity constraints, we should {\it also} identify a one-dimensional subspace on which to pull back our form. Again it is trivial to identify a natural subspace in our four-particle example: we simply impose that $s + t = c$, where $c>0$ is a positive constant. Note that the intersection of this line with the positive region $s,t>0$ is a line segment with two boundaries at $s=0$ and $t=0$,  which is a one-dimensional positive geometry (See Figure~\ref{fig:4pt}).  Furthermore, quite beautifully, pulling back our scattering one-form to this one-dimensional subspace accomplishes two things: (1) this pulled-back form is also the canonical form of the positive geometry of the interval; (2) given that $-u = s+t = c$, we have $ds + dt = 0$ on the line, and so the pullback of the form can be written as {\it e.g.} $ds/s - dt/t = ds (1/s + 1/t)$, whereby factoring out the top form $ds$ on the line segment leaves us with the amplitude!

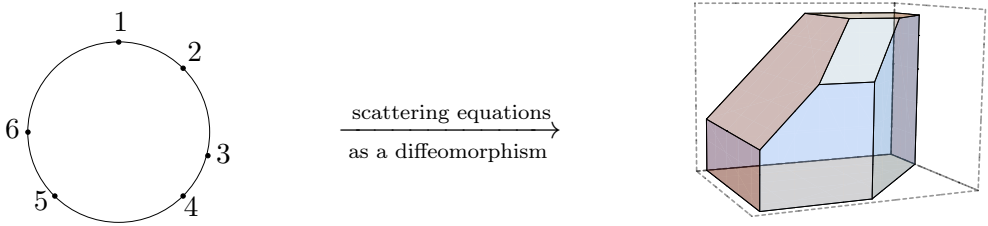
\begin{figure}
\centering
\qquad
\begin{overpic}[width=2.5cm]
{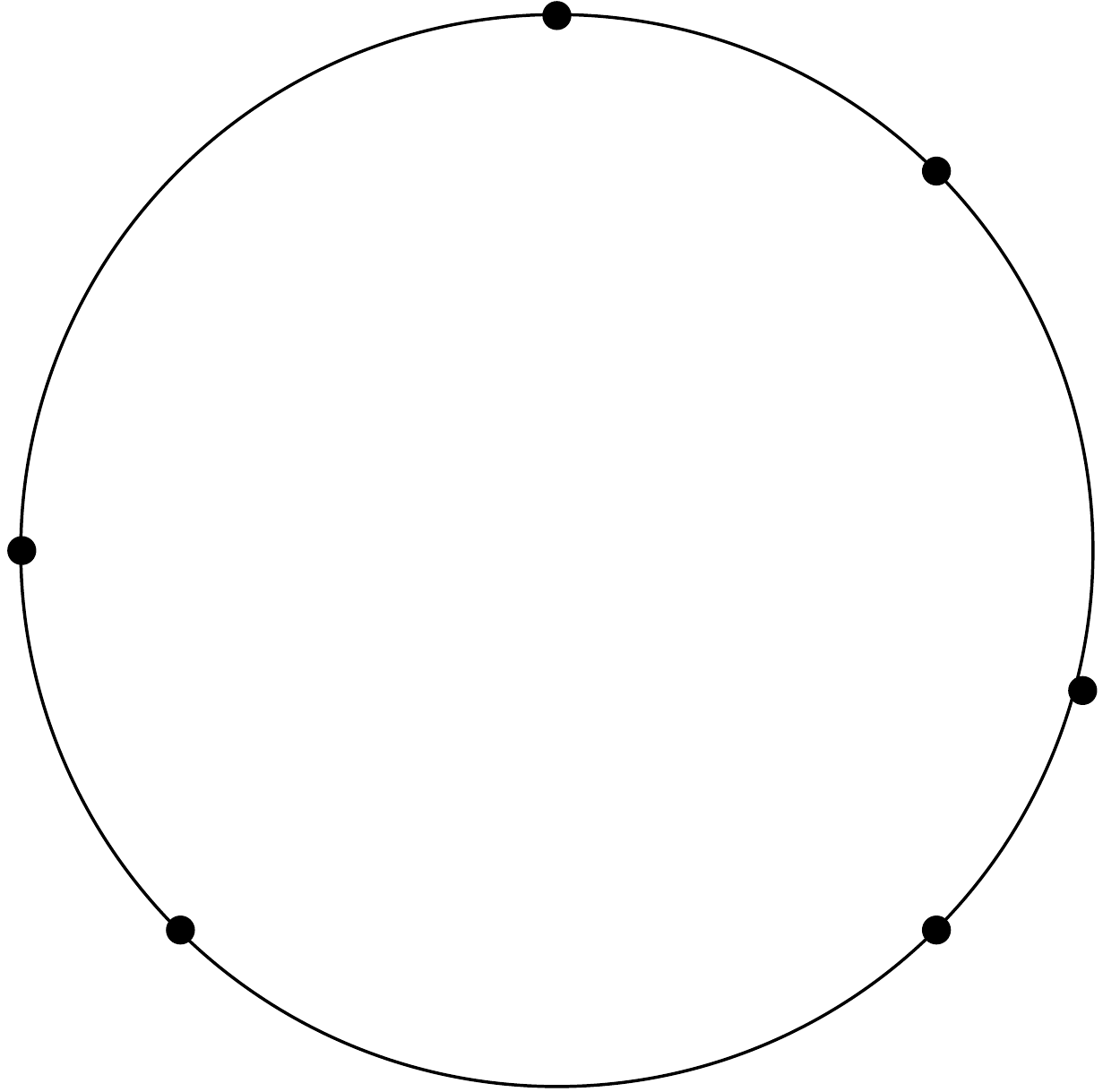}
\put(47,102){$1$}
\put(86,86){$2$}
\put(101,32){$3$}
\put(84,4){$4$}
\put(5,7){$5$}
\put(-10,45){$6$}
\put(165,45){\large $\xrightarrow[\text{as a diffeomorphism}]{\text{ scattering equations}}$}
\end{overpic}
\qquad\qquad\qquad\qquad\qquad\qquad\qquad\qquad
\begin{overpic}[width=4cm]
{6ptreg.pdf}
\end{overpic}
\newline
\caption{The scattering equations provide a diffeomorphism from the worldsheet associahedron to the kinematic associahedron.}
\label{fig:cartoon}
\end{figure}

This geometry generalizes to all $n$ in a simple way. The full kinematic space of Mandelstam invariants is $n(n-3)/2$-dimensional. A nice basis for this space is provided by all planar propagators $s_{a,a{+}1, \ldots,b{-}1}$, and there is a natural ``positive region'' in which all these variables are forced to be positive. There is also a natural $(n{-}3)$-dimensional subspace that is cut out by the equations $-s_{i j} = c_{ij}$ for all non-adjacent $i,j$ excluding the index $n$,  where the $c_{ij}>0$ are positive constants. These equalities pick out an $(n-3)$-dimensional hyperplane in kinematic space whose intersection with the positive region is the associahedron polytope. A picture of $n{=}5,6$  associahedra can be seen in Figure~\ref{fig:assoc_intro}. As we saw for four points, when the scattering form is pulled back to this subspace, it is revealed to be the canonical form with logarithmic singularities on all the boundaries of this associahedron! 

The computation of scattering amplitudes then reduces to triangulating the associahedron. Quite nicely one natural choice of triangulation directly reproduces the Feynman diagram expansion, but other triangulations are of course also possible. As a concrete example, for $n{=}5$ the Feynman diagrams express the amplitude as the sum over 5 cyclically rotated terms: 
\begin{equation}
\frac{1}{s_{12} s_{123}} +\frac{1}{s_{23} s_{234}} +\frac{1}{s_{34} s_{345}} +\frac{1}{s_{45} s_{451}} +\frac{1}{s_{51} s_{512}}
\end{equation}
But there is another triangulation of the $n{=}5$ associahedron that yields a surprising 3-term expression: 
\begin{equation}
\frac{s_{12}+ s_{234}}
{s_{12} s_{34} s_{234}}
+\frac{s_{12}+s_{234}}
{s_{12} s_{234} s_{23}}
+\frac{s_{12}-s_{123}+s_{23}}
{s_{12} s_{23} s_{123}}
\end{equation}
which can not be obtained by any recombination of the Feynman diagram terms. Indeed, we will see that the form enjoys a symmetry that is destroyed by individual terms in the Feynman diagram triangulation and restored only in the full sum. In contrast, this new representation comes from a simple triangulation that keeps this symmetry manifest, much as ``BCFW triangulations'' of the amplituhedron~\cite{Hodges:2009hk,ArkaniHamed:2012nw} make manifest the dual conformal/Yangian symmetries of planar ${\cal N}=4$ SYM that are not seen in the usual Feynman diagram expansion. 

Beyond these parallels to the story of the amplituhedron, the picture of scattering forms on kinematic space appears to have a fundamental role to play in the physics of scattering amplitudes in  more general settings.  For instance, string theorists have long known of an important associahedron, associated with the open string worldsheet; this raises a natural question: Is there a natural diffeomorphism from the (old) worldsheet associahedron to the (new) kinematic space associahedron? The answer is yes, and the map is precisely provided by the scattering equations!  This correspondence gives a one-line conceptual proof of the CHY formulas for bi-adjoint amplitudes~\cite{Cachazo:2013iea} as a ``pushforward'' from the worldsheet ``Parke-Taylor form'' to the kinematic space scattering form.

The scattering forms also give a strikingly simple and direct connection between kinematics and color! This is seen at two levels. First, we can define very general scattering forms as a sum over all possible cubic graphs $g$ in a ``big kinematic space'', with each graph given by the wedge of the $d\log$ of all its propagator factors  weighted with ``kinematic coefficients'' $N(g)$. The first important observation is that the projectivity of the form on this big kinematic space forces the kinematic coefficients $N(g)$ to satisfy the same Jacobi relations as color factors; in other words, projectivity of the scattering form provides a deep geometric origin for and interpretation of the BCJ  {\it color-kinematics duality}~\cite{Bern:2008qj,Bern:2010ue} 

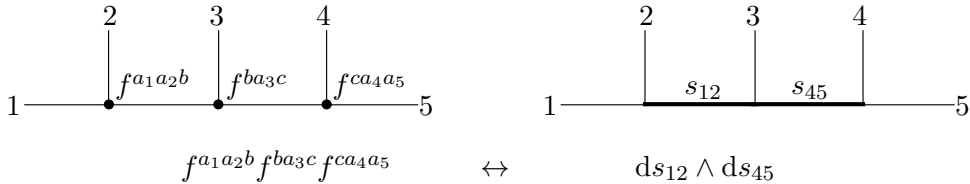
\begin{figure}
\centering
\begin{overpic}[width=5.4cm]
{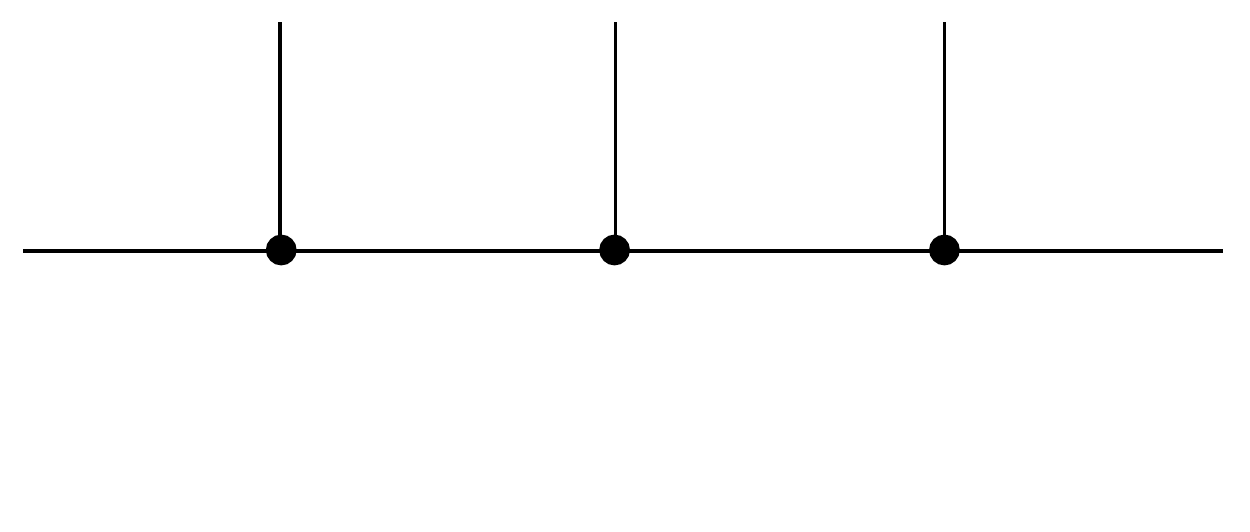}
\put(-2.5,20){$1$}
\put(21,42){$2$}
\put(47,42){$3$}
\put(73,42){$4$}
\put(98,20){$5$}
\put(24,25){$f^{a_1a_2b}$}
\put(51,25){$f^{b a_3 c}$}
\put(77,25){$f^{c a_4 a_5}$}
\put(40,5){$f^{a_1 a_2 b}f^{b a_3 c}f^{c a_4 a_5}$}
\put(113,5){$\leftrightarrow$}
\end{overpic}
\qquad\qquad
\begin{overpic}[width=5.4cm]
{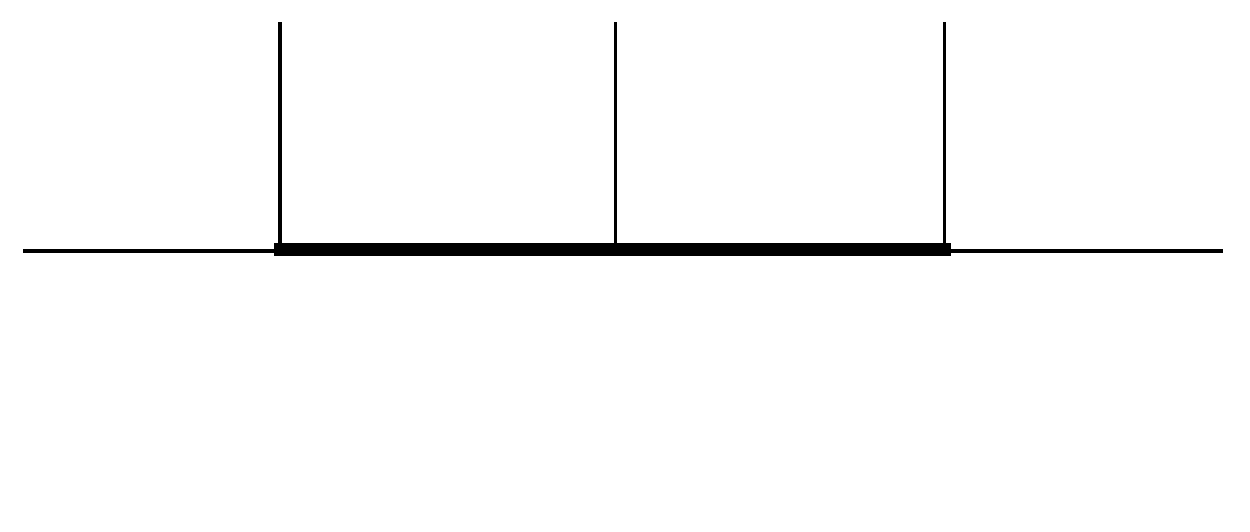}
\put(-2.5,20){$1$}
\put(21,42){$2$}
\put(47,42){$3$}
\put(73,42){$4$}
\put(98,20){$5$}
\put(32,25){$s_{12}$}
\put(58,25){$s_{45}$}
\put(20,5){$\mathrm{d}s_{12}\wedge\mathrm{d}s_{45}$}
\end{overpic}
\caption{An example of the duality between color factors and differential forms}
\label{fig:5ptdual}
\end{figure}

But there is a second, more startling connection to color made apparent by the scattering forms---``Color is Kinematics''. More precisely, as a simple consequence of momentum conservation and on-shell conditions, the wedge product of the $d$(propagator) factors associated with any cubic graph satisfies exactly the same algebraic identities as the color factors associated with the same graph, as indicated in Figure~\ref{fig:5ptdual} for a $n=5$ example.
This ``Color is Kinematics'' connection allows us to speak of the scattering forms for Yang-Mills theory and the Non-linear Sigma Model in a fascinating new way. Instead of thinking about partial amplitudes, or of objects dressed with color factors, we deal with fully permutation invariant differential forms on kinematic space with no color factors in sight! The usual colored amplitudes can be obtained from these forms by replacing the wedges of the $d$ of propagators with color factors in a completely unambiguous way. These forms are furthermore rigid, god-given objects, entirely fixed (at least at tree level) simply by standard dimensional power-counting, gauge-invariance (for YM) or the Adler zero (for the NLSM)~\cite{Arkani-Hamed:2016rak}, and the requirement of projectivity. And of course, these forms are again obtained as the ``pushforward'' via the scattering equations from the familiar differential forms on the worldsheet~\cite{Cachazo:2013iea,Cachazo:2014xea}, in parallel to the bi-adjoint theory.

We now proceed to describe all the ideas sketched above in much more detail before concluding with remarks on avenues for further work in this direction. 

\section{The Planar Scattering Form on Kinematic Space}
We introduce the {\it planar scattering form}, which is a {\it differential form} on the space of kinematic variables that encodes information about on-shell tree-level scattering amplitudes of the bi-adjoint scalar. We emphasize the importance of ``upgrading'' amplitudes to forms, which reveals deep and unexpected connections between physics and geometry that are not seen in the Feynman diagram expansion, leading amongst other things to novel (and in some cases more compact) representations of the amplitudes.  We also find connections to scattering equations and color-kinematics duality as discussed in Sections~\ref{sec:worldsheet} and~\ref{sec:color}, respectively. We generalize to Yang-Mills and Non-linear Sigma Model in Section~\ref{sec:YMNLSM}.

\subsection{Kinematic Space}

We begin by defining the kinematic space $\K_n$ for $n$ massless momenta $p_i$ for $i=1,\ldots, n$ as the space spanned by linearly independent Mandelstam variables in spacetime dimension $D\geq n{-}1$:
\begin{equation}
s_{ij}:=(p_i+p_j)^2=2p_i\cdot p_j
\end{equation}
For $D<n{-}1$ there are further constraints on Mandelstam variables---Gram determinant conditions---so the number of independent variables is lower.
Due to the massless on-shell conditions and momentum conservation, we have $n$ linearly independent constraints
\be
\sum_{j=1; \;j\neq i}^n s_{ij}=0\;\;\;\text{for $i=1,2,\ldots,n$}
\ee
The dimensionality of kinematic space is therefore
\begin{equation}
\dim\K_n=\binom{n}{2}-n=\frac{n(n{-}3)}{2}
\end{equation}
More generally, for any set of particle labels $I\subset\{1,\ldots, n\}$, we define the Mandelstam variable
\begin{equation}
s_I:=\left(\sum_{i\in I} p_i\right)^2=\sum_{i,j\in I;\; i<j}s_{ij}
\end{equation}
It follows from momentum conservation that $s_I=s_{\bar{I}}$, where $\bar{I}$ is the complement of $I$. For mutually disjoint index sets $I_1,\ldots, I_d$, we define $s_{I_1\cdots I_d}:=s_{I_1\cup\cdots \cup I_d}$. We also define, for any pair of index sets $I,J$:
\be
s_{I|J}:=2\left(\sum_{i\in I}p_i\right)\cdot\left(\sum_{j\in J}p_j\right)=\sum_{i\in I, j\in J} s_{ij}
\ee

\subsection{Planar Kinematic Variables}
\label{sec:planar_var}

\begin{figure}
\begin{center}
\begin{overpic}[width=\linewidth]{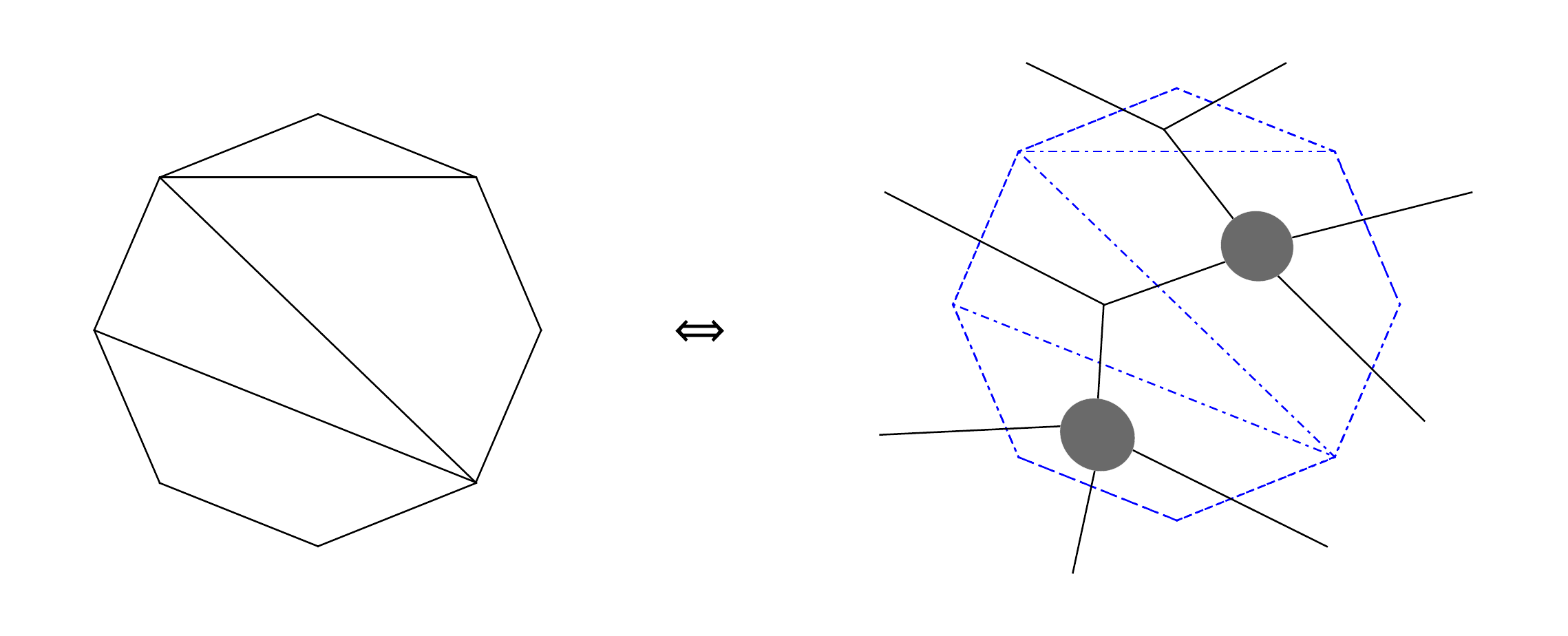}
\put(20,34){$1$}\put(31,29){$2$}\put(35,19){$3$}\put(30,7){$4$}
\put(20,3){$5$}\put(9,7){$6$}\put(4,19){$7$}\put(8,29){$8$}
\put(68,29.5){$s_{234567}$}
\put(72,24.5){$s_{4567}$}
\put(71,17.5){$s_{456}$}
\put(18,26.8){$X_{2,8}$}
\put(20,21){$X_{4,8}$}
\put(16,16){$X_{4,7}$}
\put(82,36){$1$}\put(94,28){$2$}\put(91,13){$3$}\put(85,4){$4$}\put(69,2){$5$}\put(54,12){$6$}\put(54,28){$7$}\put(63,36){$8$}
\end{overpic}
\end{center}
\caption{Correspondence between a 3-diagonal partial triangulation and a triple cut. Note that the vertices are numbered on the left while the edges/particles are numbered on the right.}
\label{fig:triang_cuts}
\end{figure}
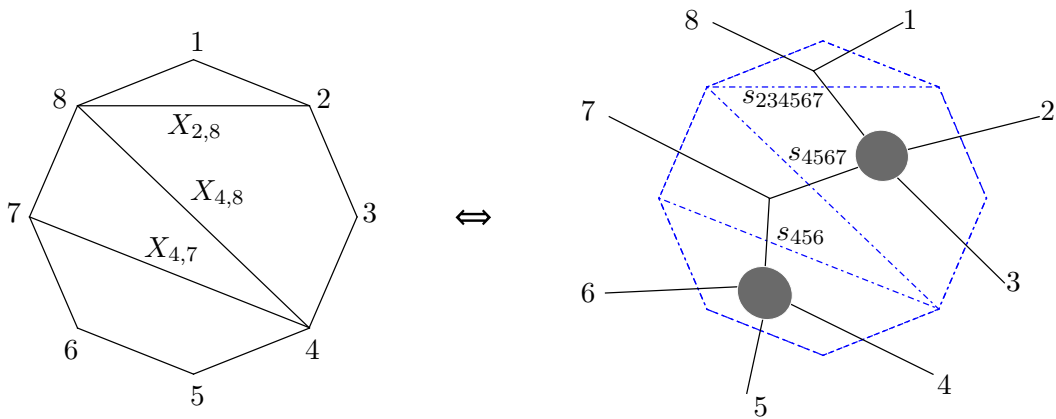

We now focus on kinematic variables that are particularly useful for cyclically ordered particles. For the {\it standard ordering} $(1,2,\ldots, n)$, we define {\it planar variables} with manifest cyclic symmetry:
\be\label{eq:planar_var}
X_{i,j} := s_{i,i{+}1,\ldots, j{-}1}
\ee
for any pair of indices $1\leq i<j\leq n$. Note that $X_{i,i{+}1}$ and $X_{1,n}$ vanish. Given a convex $n$-gon with cyclically ordered vertices, the variable $X_{i,j}$ can be visualized as the diagonal between vertices $i$ and $j$, as in Figure~\ref{fig:triang_cuts}\;(left).

The Mandelstam variables in particular can be expanded in terms of these variables, by the easily verified identity:
\be\label{eq:sXXXX}
s_{ij}=X_{i,j{+}1}+X_{i{+}1,j}-X_{i,j}-X_{i{+}1,j{+}1}
\ee
It follows that the non-vanishing planar variables form a spanning set of kinematic space. However, they also form a basis, since there are exactly $\dim \K_n = n(n{-}3)/2$ of them. It is rather curious that the number of planar variables is precisely the dimension of kinematic space. Examples of the basis include $\{s:= X_{1,3}, t:=X_{2,4}\}$ for $n{=}4$ particles and $\{s_{12}=X_{1,3}, s_{23}=X_{2,4}, s_{34}=X_{3,5}, $ $s_{123}=X_{1,4}, s_{234}=X_{2,5}\}$ for $n{=}5$.


More generally, for an ordering $\alpha:=(\alpha(1),\ldots, \alpha(n))$ of the external particles, we define {\it $\alpha$-planar variables}
\be
X_{\alpha(i),\alpha(j)} := s_{\alpha(i),\alpha(i{+}1),\ldots, \alpha(j{-}1)}
\ee
for any pair $i<j$ modulo $n$. As before, $X_{\alpha(i),\alpha(i{+}1)}$ and $X_{\alpha(1),\alpha(n)}$ vanish, and the non-vanishing variables form a basis of kinematic space. Also, each variable can be visualized as a diagonal of a convex $n$-gon whose vertices are cyclically ordered by $\alpha$. 



\subsection{The Planar Scattering Form}
\label{sec:planar_scatter_form}
We now move on to our main task of defining the {\it planar scattering form}. Let $g$ denote a (tree) cubic graph with propagators $X_{i_a,j_a}$ for $a=1,\ldots, n{-}3$. For each ordering of these propagators, we assign a value $\sign(g)\in \{\pm 1\}$ to the graph with the property that swapping two propagators flips the sign. Then, we assign to the graph a $d\log$ form:
\be
\sign (g)\bigwedge_{a=1}^{n{-}3}d\log X_{i_a,j_a}
\ee
where the $\sign(g)$ is evaluated on the ordering in which the propagators appear in the wedge product. There are of course two sign choices for each graph.
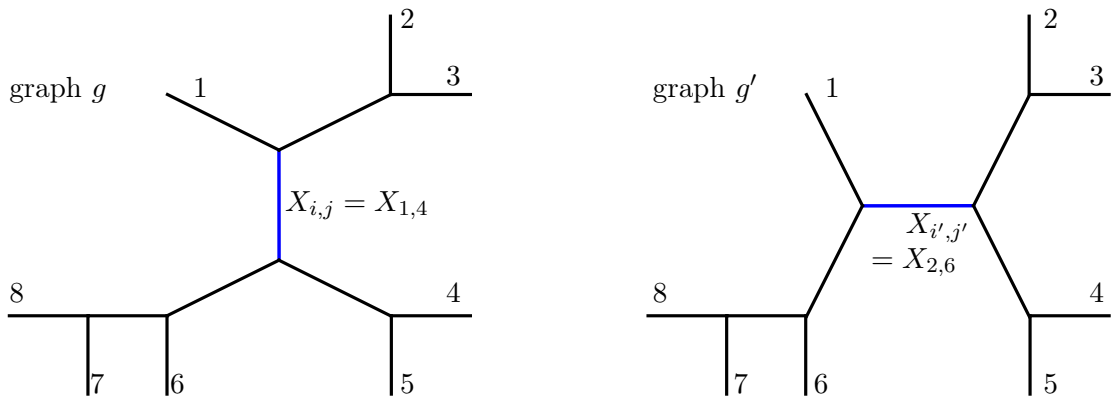
\begin{figure}
\centering
\begin{overpic}[width=\linewidth]{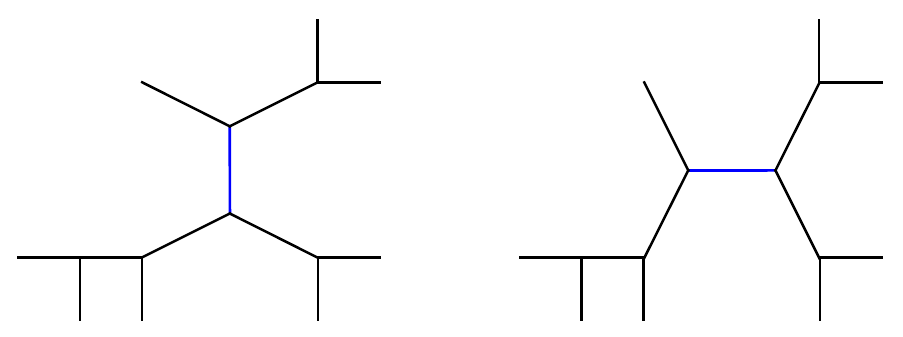}
\put(2,28){graph $g$}
\put(58,28){graph $g'$}
\put(18,28){$1$}\put(36,34){$2$}\put(40,29){$3$}\put(40,10){$4$}
\put(73,28){$1$}\put(92,34){$2$}\put(96,29){$3$}\put(96,10){$4$}
\put(36,2){$5$}\put(16,2){$6$}\put(9,2){$7$}\put(2,10){$8$}
\put(92,2){$5$}\put(72,2){$6$}\put(65,2){$7$}\put(58,10){$8$}
\put(26,18){$X_{i,j}=X_{1,4}$}\put(80,16){$X_{i',j'}$}\put(77,13){$=X_{2,6}$}
\end{overpic}
\caption{Two planar graphs related by a mutation given by an exchange of channel $X_{i,j}\rightarrow X_{i',j'}$ in a four point subgraph}
\label{fig:mutation}
\end{figure}

Finally, we introduce the {\it planar scattering form} of rank $(n{-}3)$:
\begin{equation}\label{eq:planar_form}
\Omega^{(n{-}3)}_n:= \sum_{{\textrm{planar } g} } \sign(g) \bigwedge_{a=1}^{n{-}3} d\log X_{i_a,j_a}
\end{equation}
where we sum over a $d\log$ form for every planar cubic graph $g$. Note that a particle ordering is implicitly assumed by the construction, so we also denote the form as $\Omega^{(n{-}3)}[1,\ldots, n]$ when we wish to emphasize the ordering. For $n{=}3$, we define $\Omega^{(0)}_{n{=}3}:=\pm 1$.

Since there are two sign choices for each graph, this amounts to many different scattering forms. However, there is a natural choice (unique up to overall sign) obtained by making the following requirement:
\be
\text{The planar scattering form is {\it projective}.} \nonumber
\ee
In other words, we require the form to be invariant under {\it local} $\GL(1)$ transformations $X_{i,j}\rightarrow \Lambda(X)X_{i,j}$ for any index pair $(i,j)$, or equivalently $s_I \to \Lambda (s) s_I$ for any index set $I$. This fixes the scattering form up to an overall sign which we ignore.

Moreover, this gives a simple {\it sign-flip rule} which we now describe. We say that two planar graphs $g,g'$ are related by a {\it mutation} if one can be obtained from the other by an exchange of channel in a four-point sub-graph (See Figure~\ref{fig:mutation}). Let $X_{i,j},X_{i',j'}$ denote the mutated propagators, respectively, and let $X_{i_b,j_b}$ for $b=1,\ldots, n{-}4$ denote the shared propagators. Under a local $\GL(1)$ transformation, the $\Lambda$-dependence of the scattering form becomes:
\begin{equation}
\left(\sign(g)+\sign(g')\right)d\log \Lambda\wedge \left(\bigwedge_{b=1}^{n{-}4}d\log X_{i_b,j_b}\right)+\cdots
\end{equation}
where we have only written the terms involving the $d\log$ of all shared propagators of $g$ and $g'$. Here $\sign(g')$ is evaluated on the same propagator ordering as $\sign(g)$ but with $X_{i,j}$ replaced by $X_{i',j'}$. The form is projective if the $\Lambda$-dependence disappears, {\it i.e.} when we have
\be\label{eq:sign_flip}
\sign(g)=-\sign(g')
\ee
for each mutation.

The sign flip rule has several immediate consequences. For instance, it ensures that the form is cyclically invariant up to a sign:
\be
i \to {i{+}1}\;\;\;\;\;\Rightarrow\;\;\;\;\;\Omega^{(n{-}3)}_n \to (-1)^{n{-}3}~\Omega^{(n{-}3)}_n
\ee
since it takes $(n{-}3)$ mutations (mod 2) to achieve the cyclic shift. The sign flip rule also ensures that the form factorizes correctly. Indeed, it suffices to consider the channel $X_{1,m} \to 0$ for any $m=3,\ldots, n{-}1$ for which
\begin{equation}\label{eq:scatter_fac}
\Omega^{(n{-}3)}(1,2,\ldots, n) \xrightarrow{X_{1,m}\to 0} \Omega^{(m{-}3)} (1,2,\ldots, m{-}1, I)\wedge 
\frac{d X_{1,m}}{X_{1,m}} \wedge \Omega^{(n{-}m{-}1)}(I^-, m, \ldots, n)\,,
\end{equation}
where $p_I=-\sum_{i=1}^{m{-}1} p_i$ is the on-shell internal particle. General channels can be obtained via cyclic shift. 

Projectivity is equivalent  to the natural statement that the form only depends on {\it ratios} of Mandelstam variables, as we can explicitly see in some simple examples for $n{=}4,5$:
\be\label{eq:planar_form_4}
\Omega^{(1)}
(1,2,3,4)=d \log s - d \log t
=\frac{ d s}{s}- \frac {d t}{t}=d\log\left(\frac{s}{t}\right)=d\log \left(\frac{X_{1,3}}{X_{2,4}}\right)
\ee
\begin{align}\label{eq:planar_form_5}
\Omega^{(2)}
(1,2,3,4,5)
=&~~~d\log X_{1,4} \wedge d\log X_{1,3}+d\log X_{1,3} \wedge d\log X_{3,5}+d\log X_{3,5} \wedge d\log X_{2,5}\nonumber\\
&+d\log X_{2,5} \wedge d\log X_{2,4}+d\log X_{2,4} \wedge d\log X_{1,4}\nonumber\\
=&~~~d\log \frac{X_{1,3}}{X_{2,4}} \wedge d\log \frac{X_{1,3}}{X_{1,4}} + d\log \frac{X_{1,3}}{X_{2,5}} \wedge d\log \frac{X_{3,5}}{X_{2,4}}\,
\end{align}
where we have written on the last expression for each example the form in terms of ratios of $X$'s only. For $n{=}6$, the form is given by summing over 14 planar graphs which can be expressed as ratios in the following way:
\begin{align}
\Omega^{(3)}_{n=6}=&~~~~d\log \frac{X_{2,4}}{X_{1,3}} \wedge d\log \frac{X_{1,4}}{X_{4,6}} \wedge d\log \frac{X_{1,5}}{X_{4,6}}+ d\log \frac{X_{2,6}}{X_{1,3}} \wedge d\log \frac{X_{3,6}}{X_{1,3}} \wedge d\log \frac{X_{4,6}}{X_{3,5}}\nonumber\\
&-d\log \frac{X_{2,6}}{X_{1,5}} \wedge d\log \frac{X_{2,5}}{X_{3,5}} \wedge d\log \frac{X_{2,4}}{X_{3,5}}- d\log \frac{X_{2,4}}{X_{1,3}} \wedge d\log \frac{X_{4,6}}{X_{3,5}} \wedge d\log \frac{X_{2,6}}{X_{1,5}}\,.\nonumber
\end{align}

Finally, for a general ordering $\alpha$ of the external particles, we define the scattering form $\Omega^{(n{-}3)}[\alpha]$ by making index replacements $i\rightarrow \alpha(i)$ on $\Omega^{(n{-}3)}_n$, which is equivalent to replacing Eq.~\eqref{eq:planar_form} with a sum over $\alpha$-planar graphs. Recall that a cubic graph is called {\it $\alpha$-planar} if it is planar when external legs are ordered by $\alpha$; alternatively, we say that the graph is {\it compatible} with the order. Furthermore, the form is projective.

We emphasize that projectivity is a rather remarkable property of the scattering form which is not true for each Feynman diagram separately. Indeed, no proper subset of Feynman diagrams provides a projective form---only the sum over all the diagrams (satisfying the sign flip rule) is projective. This foreshadows something we will see much more explicitly later on in connection to the positive geometry of the associahedron: the Feynman diagram expansion provides just one type of triangulation of the geometry, which introduces a spurious ``pole at infinity'' that cancels only in the sum over all terms. But other triangulations that are manifestly projective term-by-term are also possible, and often lead to even shorter expressions.




\section{The Kinematic Associahedron}

We introduce the {\it associahedron} polytope~\cite{Tamari, Stasheff_1,Stasheff_2} and discuss its connection to the bi-adjoint scalar theory. We begin by reviewing the combinatorial structure of the associahedron before providing a novel construction of the associahedron in kinematic space. We then argue that the tree level amplitude is a geometric invariant of the kinematic associahedron called its {\it canonical form} as review in Appendix~\ref{app:pos}, thus establishing the associahedron as the ``amplituhedron'' of the (tree) bi-adjoint theory.

\subsection{The Associahedron from Planar Cubic Diagrams}
\label{sec:assoc}
There exist many beautiful, combinatorial ways of constructing associahedra; an excellent survey of the subject, together with comprehensive references to the literaure, is given by~\cite{Ziegler}. In this section, we discuss one of the most fundamental descriptions of the associahedron which is also most closely related to scattering amplitudes. 
We begin by clarifying some terminology regarding polytopes.


A {\it boundary} of a polytope refers to a boundary of any codimension. A $k$-boundary is a boundary of dimension $k$. A {\it facet} is a codimension 1 boundary. Given a convex $n$-gon, a {\it diagonal} is a straight line between any two non-adjacent vertices. A {\it partial triangulation} is a collection of mutually non-crossing diagonals. A {\it full triangulation} or simply a {\it triangulation} is a partial triangulation with maximal number of diagonals, namely $(n{-}3)$.

\begin{figure}
\begin{subfigure}{.5\textwidth}
\centering
\begin{overpic}[width=8cm]
{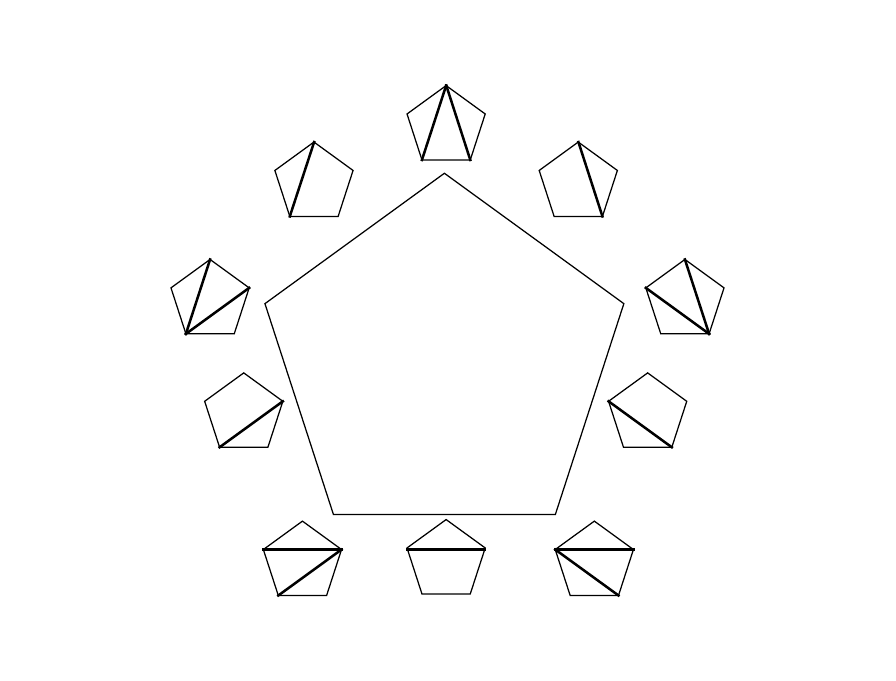}
\end{overpic}
\end{subfigure}
\begin{subfigure}{.56\textwidth}
\centering
\begin{overpic}[width=6cm]
{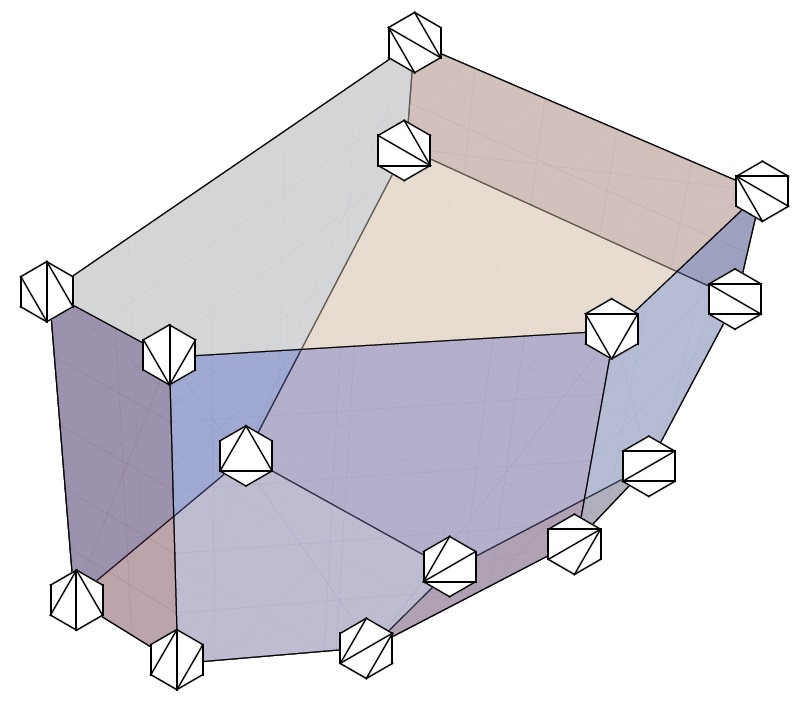}
\end{overpic}
\end{subfigure}

\caption{Combinatorial structure of the $n{=}5$ associahedron (left) and the $n{=}6$ associahedron (right). For simplicity, only vertices are labeled for the latter. } 
\label{fig:comb}
\end{figure}

For any $n{\geq} 3$, consider a convex polytope of dimension $(n{-}3)$ with the following properties:
\begin{enumerate}
    \item
    For every $d=0,1,\ldots,n{-}3$, there exists a one-to-one correspondence between the codimension $d$ boundaries and the $d$-diagonal partial triangulations of a convex $n$-gon.
    \item
    A codimension $d$ boundary $F_1$ and a codimension $d{+}k$ boundary $F_2$ are adjacent if and only if the partial triangulation of $F_2$ can be obtained by addition of $k$ diagonals to the partial triangulation of $F_1$.

\end{enumerate}
In particular, the triangulation with no diagonals corresponds to the polytope's interior, and:
\be
\text{The vertices correspond to the full triangulations.}
\ee
A classic result in combinatorics says that the number of full triangulations, and hence the number of vertices of our polytope, is the Catalan number $C_{n-2}$~\cite{catalan}. Any polytope $\A_n$ satisfying these properties is an {\it associahedron}. See Figure~\ref{fig:comb} for examples.

Before establishing a precise connection to scattering amplitudes, we make a few observations that provide some of the guiding principles. Let us order the edges of the $n$-gon cyclically with $1,\ldots, n$, and recall that:
\ba
\text{$d$-diagonal partial triangulations of the $n$-gon are in one-to-one correspondence}\nonumber\\ 
\text{with $d$-cuts on $n$-particle planar cubic diagrams. (See Figure~\ref{fig:triang_cuts})}\;\;\;\;\;\;
\ea
The edges of the $n$-gon correspond to external particles, while the diagonals correspond to cuts. 

Furthermore, the associahedron {\it factorizes combinatorially}. That is, consider a facet $F$ corresponding to some diagonal that subdivides the $n$-gon into a $m$-gon and a $(n{-}m{+}2)$-gon (See Figure~\ref{fig:polygon_fac}). The two lower polygons provide the combinatorial properties for two lower associahedra $\A_m$ and $\A_{n{-}m{+}2}$, respectively, and the facet is combinatorially identical to their direct product:
\be\label{eq:fac_comb}
F\cong \A_m\times \A_{n{-}m{+}2}
\ee
We show in Section~\ref{sec:fac} that this implies the factorization properties of amplitudes. 

Finally, we observe that the associahedron is a {\it simple} polytope, meaning that each vertex is adjacent to precisely $\dim\A_n=(n{-}3)$ facets. Indeed, given any associahedron vertex and its corresponding triangulation, the adjacent facets correspond to the $(n{-}3)$ diagonals.

\subsection{The Kinematic Associahedron}

We now show that there is an associahedron naturally living in the kinematic space for $n$ particles. The construction depends on an ordering for the particles which we take to be the standard ordering for simplicity.


We first define a region $\Delta_n$ in kinematic space by imposing the inequalities
\be\label{eq:simplex}
X_{i,j}\geq 0\;\;\;\text{for all $1\leq i<j\leq n$}
\ee
Recall that $X_{i,i{+}1}$ and $X_{1n}$ are trivially zero and therefore do not provide conditions. Since the number of non-vanishing planar variables is exactly the dimension of kinematic space, it follows that $\Delta_n$ is a simplex with a facet at infinity. This leads to an obvious problem. The associahedron $\A_n$ should have dimension $(n{-}3)$, which for $n>3$ is lower than the kinematic space dimension. We resolve this by restricting to a $(n{-}3)$-subspace $H_n\subset \K_n$ defined by a set of constants:
\ba\label{eq:cXXXX}
\text{Let $c_{ij}:=X_{i,j}+X_{i{+}1,j{+}1}-X_{i,j{+}1}-X_{i{+}1,j}$ be a {\it positive constant }}\nonumber\\
\text{for every pair of {\it non-adjacent} indices $1\leq i<j\leq n{-}1$}
\ea
Note that we have deliberately omitted $n$ from the index range. 
Also, Eq.~\eqref{eq:sXXXX} implies the following simple identity:
\be\label{eq:cs}
c_{ij}=-s_{ij}
\ee
The condition Eq.~\eqref{eq:cXXXX} is therefore equivalent to requiring $s_{ij}$ to be a {\it negative constant} for the same index range. 
Counting the number of constraints, we find the desired dimension:
\begin{equation}
\dim H_n = \dim \K_n - \frac{(n-2)(n-3)}{2} = n-3
\end{equation}
Finally, we let $\A_n:= H_n\cap \Delta_n$ be a polytope.
\begin{figure}
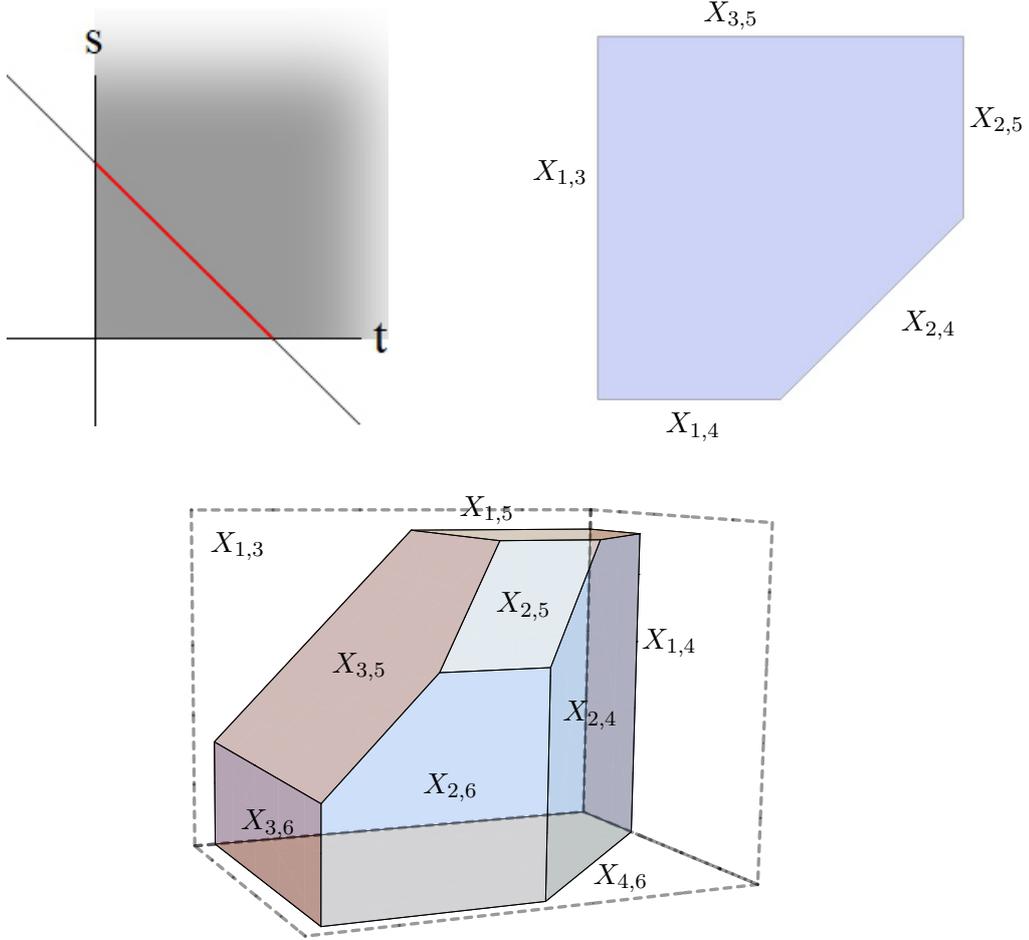

\begin{subfigure}{.5\textwidth}
\centering
\begin{overpic}[width=5cm]
{intro111.jpg}
\end{overpic}
\end{subfigure}
\begin{subfigure}{.5\textwidth}
\centering
\begin{overpic}[width=5cm]
{5pt.jpg}
\put(20,-7){$X_{1,4}$}
\put(-15,60){$X_{1,3}$}
\put(30,102){$X_{3,5}$}
\put(100,74){$X_{2,5}$}
\put(82,20){$X_{2,4}$}
\end{overpic}
\end{subfigure}
\linebreak
\linebreak
\linebreak
\linebreak
\begin{subfigure}{\textwidth}
\centering
\begin{overpic}[width=8cm]
{6ptreg.pdf}
\put(5,66){$X_{1,3}$}
\put(76,50){$X_{1,4}$}
\put(68,11){$X_{4,6}$}
\put(46,72){$X_{1,5}$}
\put(25,46){$X_{3,5}$}
\put(52,56){$X_{2,5}$}
\put(40,26){$X_{2,6}$}
\put(10,20){$X_{3,6}$}
\put(63,38){$X_{2,4}$}
\end{overpic}
\end{subfigure}
\caption{Kinematic associahedra for $n{=}4$ (top left), $n{=}5$ (top right) and $n{=}6$ (bottom).}\label{fig:kin_assoc}
\end{figure}
We claim that {\it $\A_n$ is an associahedron of dimension $(n{-}3)$}. See Figure~\ref{fig:kin_assoc} for examples. Recall from Section~\ref{sec:assoc} that the associahedron factorizes combinatorially, meaning that each facet is combinatorially the direct product of two lower associahedra as in Eq.~\eqref{eq:fac_comb}. In Section~\ref{sec:fac}, we show that the same property holds for the kinematic polytope $\A_n$, thereby implying our claim.

\begin{figure}
\centering
\begin{overpic}[width=6cm]
{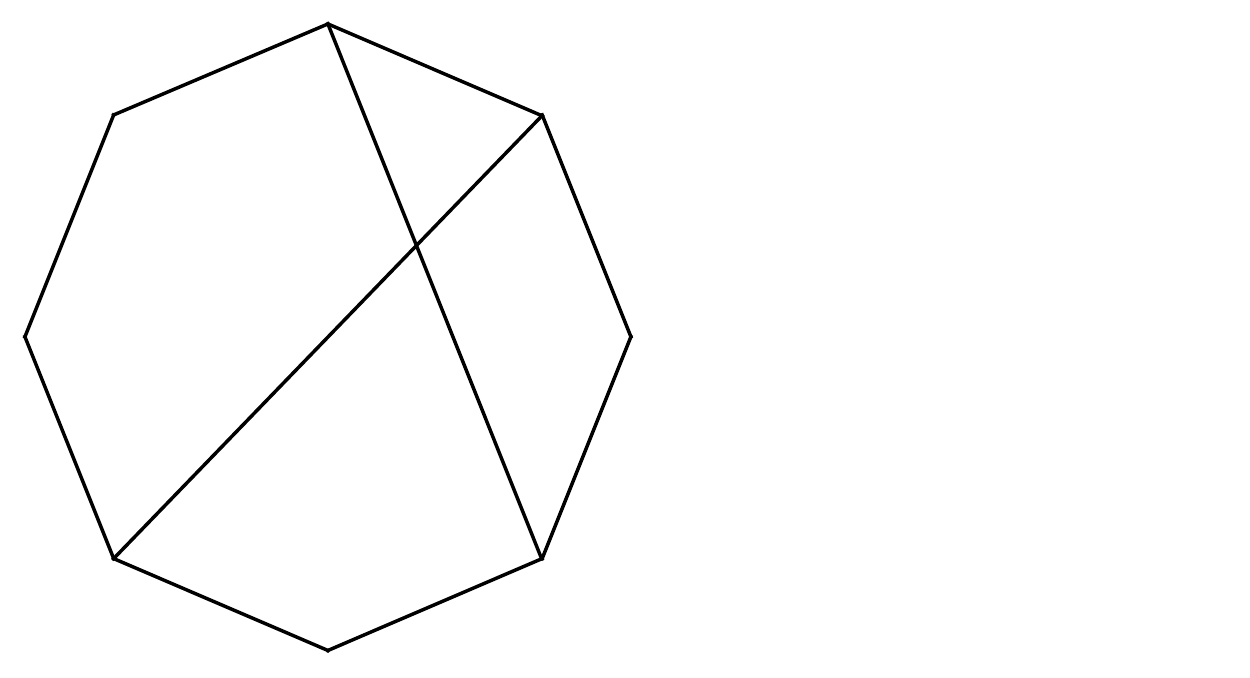}
\put(60,25){Not allowed!}
\end{overpic}
\caption{Planar variables $X_{i,j}$ corresponding to crossing diagonals cannot be simultaneously set to zero.}
\label{fig:cross}
\end{figure}

Here we highlight the key observation needed for showing factorization and hence the associahedron structure. Note that the boundaries are enforced by the positivity conditions $X_{i,j}\geq  0$, so that we can reach any codimension 1 boundary by setting some particular $X_{i,j} \to 0$. But then, to reach a lower dimensional boundary, we cannot set $X_{k,l} \to 0$ for any diagonal $(k,l)$ that {\it crosses} $(i,j)$ (See Figure~\ref{fig:cross}). Indeed, if we begin with the basic identity Eq.~\eqref{eq:cXXXX} with $(i,j)$ replaced by $(a,b)$ and sum $a,b$ over the range $i\leq a<j$ and $k\leq b<l$, the sums telescope and we find
\be\label{eq:XXXXc}
X_{j,k}+X_{i,l}=X_{i,k}+X_{j,l}-\sum_{\substack{i\leq a<j\\k\leq b<l}} c_{ab}
\ee
for any $1\leq i<j<k<l\leq n$. Now consider a situation like Figure~\ref{fig:diag_flip} (top) where the diagonals $X_{i,k}=0$ and $X_{j,l}=0$ cross, then
\be
X_{j,k}+X_{i,l}=-\sum_{\substack{i\leq a<j\\k\leq b<l}}c_{ab}
\ee
which is a contradiction since the left side is {\it nonnegative} while the right side is {\it strictly negative}. Geometrically, this means that every boundary of $\A_n$ is labeled by a set of non-crossing diagonals ({\it i.e.} a partial triangulation), as expected for the associahedron.

Let us do some quick examples. For $n{=}4$, the kinematic space with variables $(s,t,u)$ satisfies the constraint $s+t+u=0$ and is 2-dimensional. However, the kinematic associahedron is given by the line segment $0<s<-u$ where $u<0$ is a constant, as shown in Figure~\ref{fig:kin_assoc} (top left).
For $n{=}5$, the kinematic space is 5-dimensional, but the subspace $H_{n=5}$ is 2-dimensional defined by three constants $c_{13},c_{14},c_{24}$. If we parameterize the subspace in the basis $(X_{1,3},X_{1,4})$, then the associahedron $\A_{n=5}$ is a pentagon with edges given by:
\ba\label{eq:5_basis}
X_{1,3} &\geq & 0\\
X_{3,5} &=& -X_{1,4}+c_{14}+c_{24}\geq 0\\
X_{2,5} &=& -X_{1,3}+c_{13}+c_{14}\geq 0\\
X_{2,4} &=& X_{1,4}-X_{1,3}+c_{13} \geq 0\\
X_{1,4} &\geq& 0
\ea
where the edges are given in clockwise order (See Figure~\ref{fig:kin_assoc} (top right)). The $n{=}6$ example is given in Figure~\ref{fig:kin_assoc} (bottom).

The associahedron $\A_n$ in kinematic space is only one step away from scattering amplitudes, as we now show.

\subsection{Bi-adjoint $\phi^3$ Amplitudes}
\label{sec:amplitudes}

We now show the connection between the kinematic associahedron $\A_n$ and scattering amplitudes in bi-adjoint scalar theory. The discussion here applies to tree amplitudes with a pair of standard ordering, which we denote by $m_n$. We generalize to arbitrary ordering pairs $m[\alpha|\beta]$ in Section~\ref{sec:all_orderings}. This section relies on the concept of {\it positive geometries} and {\it canonical forms}, for which a quick review is given in Appendix~\ref{app:pos}. For readers unfamiliar with the subject, Appendices~\ref{app:pos_def},~\ref{app:pos_poly} and~\ref{app:pos_simple} suffice for the discussion in this section. A much more detailed discussion is given in~\cite{Arkani-Hamed:2017tmz}.

We make two claims in this section:
\begin{enumerate}
\item
The pullback of the cyclic scattering form $\Omega^{(n{-}3)}_n$ to the subspace $H_n$ is the canonical form of the associahedron $\A_n$.
\item
The canonical form of the associahedron $\A_n$ determines the tree amplitude of the bi-adjoint theory with identical ordering.

\end{enumerate}

Recall that the associahedron is a simple polytope (See end of Section~\ref{sec:assoc}), and the canonical form of a simple polytope (See Eq.~\eqref{eq:canon_simple}) is a sum over its vertices. For each vertex $Z$, let $X_{i_a,j_a}= 0$ denote its adjacent facets for $a=1,\ldots, n{-}3$. Furthermore, for each ordering of the facets, let $\sign(Z)\in\{\pm 1\}$ denote its orientation relative to the inherited orientation. The canonical form is therefore
\begin{equation}\label{eq:canon_assoc}
\Omega(\A_n)=\sum_{\text{vertex }Z} \sign(Z)\bigwedge_{a=1}^{n{-}3}d\log X_{i_a,j_a}
\end{equation}
where $\sign(Z)$ is evaluated on the ordering of the facets in the wedge product. Since the form is defined on the subspace $H_n$, it may be helpful to express the $X_{i,j}$ variables in terms of a basis of $(n{-}3)$ variables like Eq.~\eqref{eq:Y_basis}.

We argue that Eq.~\eqref{eq:canon_assoc} is equivalently the pullback of the scattering form Eq.~\eqref{eq:planar_form} to the subspace $H_n$. Since there is a one-to-one correspondence between vertices $Z$ and planar cubic graphs $g$, it suffices to show that the pullback of the $g$ term is the $Z$ term. This is true by inspection since $g$ and its corresponding $Z$ have the same propagators $X_{i_a,j_a}$. The only subtlety is that the $\sign(Z)$ appearing in Eq.~\eqref{eq:canon_assoc} is defined geometrically, while the $\sign(g)$ appearing in Eq.~\eqref{eq:planar_form} is defined by local $\GL(1)$ invariance. We now argue equivalence of the two by showing that $\sign(Z)$ satisfies the sign flip rule.

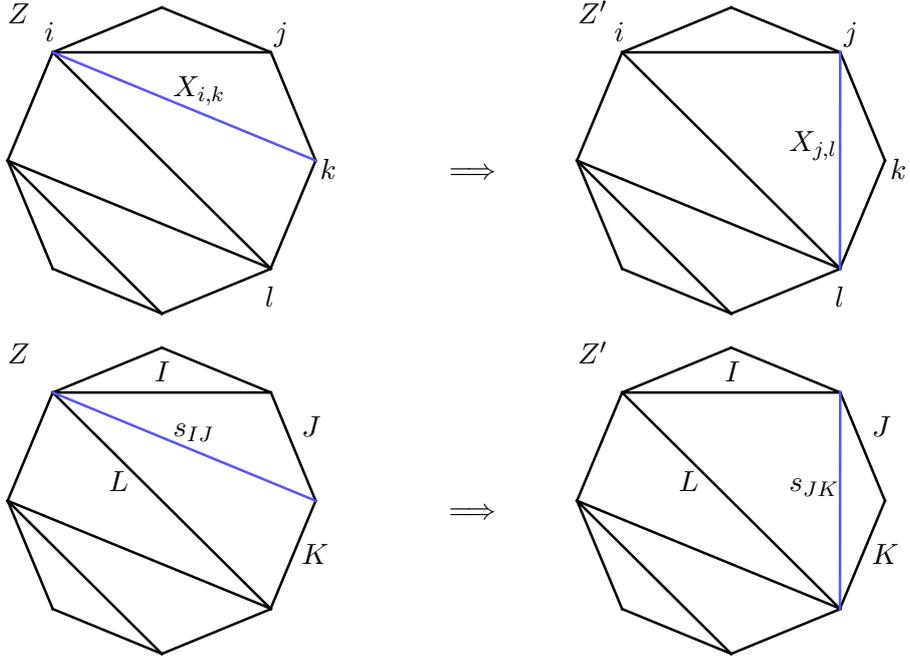
\begin{figure}
\centering
\begin{overpic}[width=0.8\linewidth]{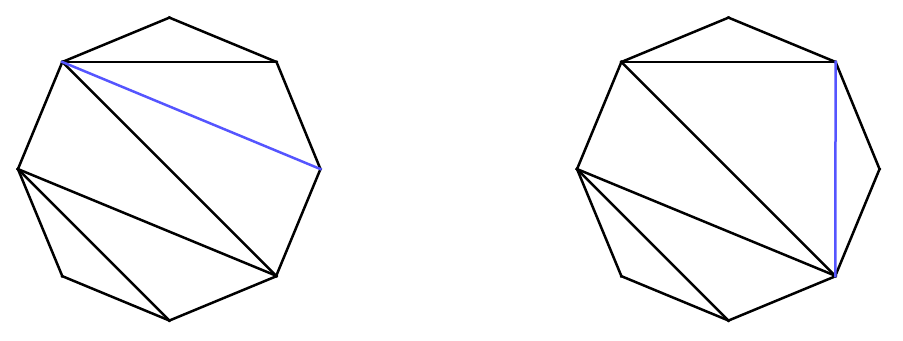}
\put(6,32){$i$}
\put(31,32){$j$}
\put(36,17){$k$}
\put(30,3){$l$}
\put(68,32){$i$}
\put(93,32){$j$}
\put(98,17){$k$}
\put(92,3){$l$}
\put(2,34){$Z$}\put(64,34){$Z'$}
\put(20,26){$X_{i,k}$}
\put(87,20){$X_{j,l}$}
\put(50,17){$\Longrightarrow$}
\end{overpic}
\begin{overpic}[width=0.8\linewidth]{ggst_gr5.eps}
\put(2,34){$Z$}\put(64,34){$Z'$}\put(18,32){$I$}\put(34,26){$J$}\put(34,12){$K$}\put(13,20){$L$}
\put(20,26){$s_{IJ}$}
\put(87,20){$s_{JK}$}
\put(80,32){$I$}\put(96,26){$J$}\put(96,12){$K$}\put(75,20){$L$}
\put(50,17){$\Longrightarrow$}
\end{overpic}
\caption{Two triangulations related by a mutation $X_{i,k}\rightarrow X_{j,l}$ (top) or equivalently $s_{IJ}\rightarrow s_{JK}$ (bottom).}\label{fig:diag_flip}
\end{figure}
Suppose $Z,Z'$ are vertices whose triangulations are related by a mutation. While mutations are defined as relations between planar cubic graphs (See Figure~\ref{fig:mutation}), they can equivalently be interpreted from the triangulation point of view. Indeed, two triangulations are related by a mutation if one can be obtained from the other by exchanging exactly one diagonal. For example, the two triangulations of a quadrilateral are related by mutation. For a generic triangulation of the $n$-gon, every mutation can be obtained by identifying a quadrilateral in the triangulation and exchanging its diagonal. In Figure~\ref{fig:diag_flip} (top), we show an example where a mutation is applied to the quadrilateral $(i,j,k,l)$ with the diagonal $(i,k)$ in $Z$ exchanged for the diagonal $(j,l)$ in $Z'$. Note that we have implicitly assumed $1\leq i<j<k<l\leq n$. Furthermore, taking the exterior derivative of the kinematic identity Eq.~\eqref{eq:XXXXc} gives us
\begin{equation}\label{eq:mutation}
dX_{j,k}+dX_{i,l}=dX_{i,k}+dX_{j,l}\,.
\end{equation}
Note that the two propagators on the left appear in both diagrams, while the two propagators on the right are related by mutation. It follows that 
\be\label{eq:jac}
\bigwedge_{a=1}^{n{-}3}dX_{i_a,j_a} =- \bigwedge_{a=1}^{n{-}3}dX_{i_a',j_a'}
\ee
The crucial part is the minus sign, which implies the sign flip rule:
\be
\sign(Z)=-\sign(Z')
\ee
We can therefore identify $\sign(Z)=\sign(g)$. Furthermore, an important consequence of \eqref{eq:jac} is that the following quantity is independent of $g$ on the pullback:
\begin{equation}\label{eq:dX}
d^{n{-}3}X:=\sign(g)\bigwedge_{a=1}^{n{-}3}d X_{i_a,j_a}
\end{equation}
Substituting into Eq.~\eqref{eq:canon_assoc} gives
\begin{equation}\label{eq:amp_form}
\Omega(\A_n)=\left(\sum_{\text{planar }g}\frac{1}{\prod_{a=1}^{n{-}3}X_{i_a,j_a}}\right)d^{n{-}3}X=m_n d^{n{-}3}X
\end{equation}
which gives the expected amplitude $m_n$, thus completing the argument for our second claim.
For convenience we sometimes denote the item in parentheses as $\aOmega(\A_n)$, called the {\it canonical rational function}. Thus,
\be\label{eq:canon_amp}
\aOmega(\A_n)=m_n
\ee

Let us do a quick and informative example for $n{=}4$. We use the usual Mandelstam variables $(s,t,u):=(X_{1,3},X_{2,4},-X_{1,3}-X_{2,4}=-c_{13})$. Here $u$ is a negative constant, and the associahedron is simply the line segment $0\leq s\leq -u$ in Figure~\ref{fig:kin_assoc} (top left), whose canonical form is
\be\label{eq:canon_4}
\Omega(\A_{n=4})=\left(\frac{1}{s}-\frac{1}{s+u}\right)ds=\left(\frac{1}{s}+\frac{1}{t}\right)ds
\ee
which of course is also the desired amplitude up to the $ds$ factor. Now consider pulling back the planar scattering form Eq.~\eqref{eq:planar_form_4}. Since $u$ is a constant on $H_{n=4}$ and $s+t+u=0$, hence $ds=-dt$ on the pullback. It follows that
\be
\Omega^{(1)}_{n=4}|_{H_{n=4}} = \left(\frac{1}{s}+\frac{1}{t}\right)ds
\ee
which is equal to Eq.~\eqref{eq:canon_4}.
We also demonstrate an example for $n=5$ where the associahedron is a pentagon as shown in Figure~\ref{fig:kin_assoc} (top right). We argue that the pullback of Eq.~\eqref{eq:planar_form_5} determines the 5-point amplitude by showing that the numerators have the expected sign on the pullback, namely $dX_{1,4}dX_{1,3}=dX_{1,3}dX_{3,5}=dX_{3,5}dX_{2,5}=dX_{2,5}dX_{2,4}=dX_{2,4}dX_{1,4}$. For instance, the identity $X_{3,5}=-X_{1,4}+c_{14}+c_{24}$ implies $\partial(X_{1,4},X_{1,3})/\partial(X_{1,3},X_{3,5})=1$, leading to the first equality. We leave the rest as an exercise for the reader. It follows that the pullback determines the corresponding amplitude.
\be\label{eq:amp_5pt}
\Omega^{(2)}_{n{=}5}|_{H_{n=5}}=\left(\frac{1}{X_{1,3}X_{1,4}}+
\frac{1}{X_{3,5}X_{1,3}}+
\frac{1}{X_{1,4}X_{2,4}}+
\frac{1}{X_{2,5}X_{3,5}}+
\frac{1}{X_{2,4}X_{2,5}}
\right)d^2X
\ee
Of course, this is also the canonical form of the pentagon.

\subsection{All Ordering Pairs of Bi-adjoint $\phi^3$ Amplitudes}
\label{sec:all_orderings}
We now generalize our results to every ordering pair of the bi-adjoint theory. Given an ordering pair $\alpha,\beta$, the amplitude is given by the sum of all cubic diagrams {\it compatible} with both orderings, with an overall sign from the trace decomposition~\cite{Cachazo:2013iea} that we postpone to Section \ref{sec:trace} and more specifically Eq.~\eqref{eq:standard_pull_on_beta}. Here we ignore the overall sign and simply define $m[\alpha|\beta]$ to be the sum over the cubic graphs.

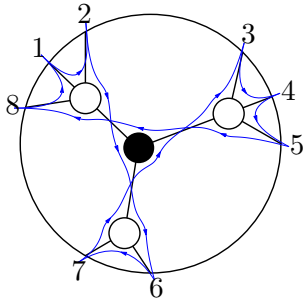
\begin{figure}
\centering
\begin{overpic}[width=0.25\linewidth]
{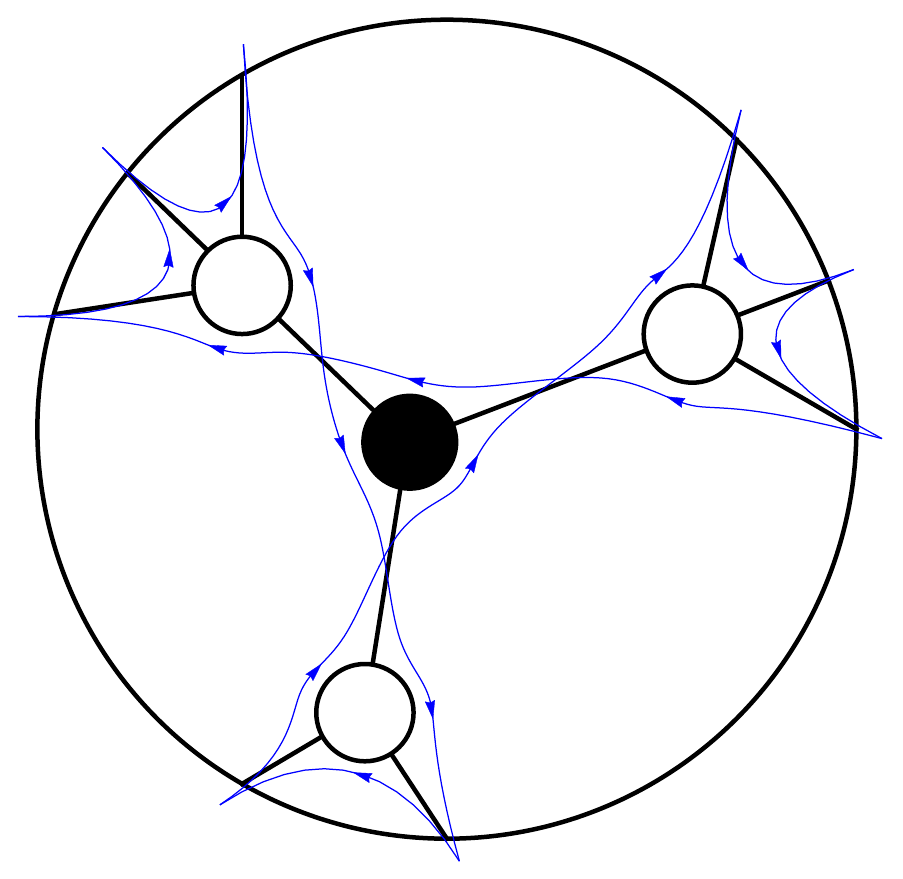}
\put(8,80){$1$}
\put(-1,60){$8$}
\put(24,92){$2$}
\put(22,2){$7$}
\put(49,-4){$6$}
\put(98,46){$5$}
\put(95,65){$4$}
\put(81,85){$3$}
\end{overpic}
\caption{This mutual cut diagram gives rise to $(\alpha,\beta)=(12345678,81267354)$ by the described rules.}
\label{dbodbodbo}
\end{figure}

We first review a simple diagrammatic procedure~\cite{Cachazo:2013iea} for obtaining all the graphs appearing in $m[\alpha|\beta]$ as illustrated in Figure~\ref{13090885}:
\begin{figure}[H]
\centering
\includegraphics[width=\linewidth]{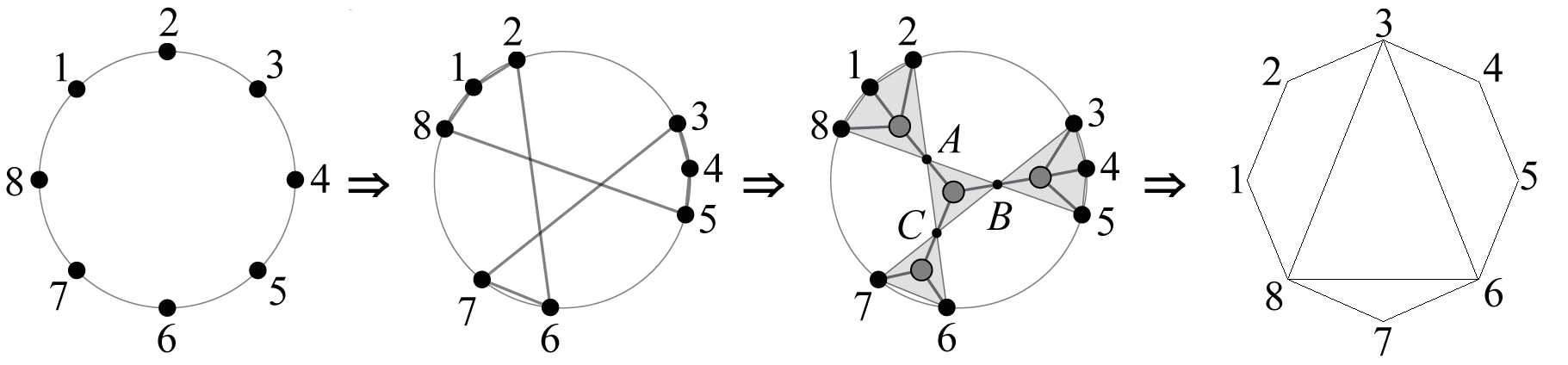}
\caption{Step-by-step procedure for obtaining the mutual cuts (3\textsuperscript{rd} picture) and the mutual partial triangulation (4\textsuperscript{th}) for $(\alpha,\beta)=(12345678\vert81267354)$. The first three pictures are found in~\cite{Cachazo:2013iea}.}
\label{13090885}
\end{figure}
\begin{enumerate}
\item
Draw $n$ points on the boundary of a disk ordered cyclically by $\alpha$. 
\item
Draw a closed path of line segments connecting the points in order $\beta$. These line segments enclose a set of polygons, forming a polygon decomposition.
\item
The internal vertices of the decomposition correspond to cuts on cubic graphs called {\it mutual cuts}.
\item
The cuts correspond to diagonals of the $\alpha$-ordered $n$-gon, forming a {\it mutual partial triangulation}.
\end{enumerate}
The cubic graphs compatible with both orderings are precisely those that admit all the mutual cuts. Equivalently, they correspond to all triangulations of the $\alpha$-ordered $n$-gon containing the mutual partial triangulation. Conversely, given a graph of mutual cuts or equivalently a mutual partial triangulation, we can reverse engineer the ordering $\beta$ up to dihedral transformation as follows:
\begin{enumerate}
\item
Color each vertex of the graph white or black like Figure~\ref{dbodbodbo} so that no two adjacent vertices have the same color.
\item
Draw a closed path that winds around white vertices clockwise and black vertices counterclockwise.
\item
The path gives the ordering $\beta$ up to cyclic shift. Changing the coloring corresponds to a reflection.
\end{enumerate}
The path gives the $\beta$ up to cyclic shift. Swapping the colors reverses the particle ordering. It follows that $\beta$ can be obtained up to dihedral transformations.

We are now ready to construct the kinematic polytope for an arbitrary ordering pair. We break the symmetry between the two orderings by using planar variables $X_{\alpha(i),\alpha(j)}$ discussed at the end of Section~\ref{sec:planar_var}. In analogy with Eq.~\eqref{eq:simplex}, we define a simplex $\Delta[\alpha]$ in kinematic space by requiring that:
\be
\text{$X_{\alpha(i),\alpha(j)}\geq 0$ for all $1\leq i<j\leq n$.}
\ee
Similar to before, $X_{\alpha(i),\alpha(i{+}1)}$ and $X_{\alpha(1),\alpha(n)}$ vanish and therefore do not provide conditions. We can visualize the variable $X_{\alpha(i),\alpha(j)}$ as the diagonal $(\alpha(i),\alpha(j))$ of a regular $n$-gon whose vertices are labeled by $\alpha$. Furthermore, we construct a $(n{-}3)$-subspace $H[\alpha|\beta]$ of kinematic space by making the following requirements:
\begin{enumerate}
\item
For each diagonal $(\alpha(i),\alpha(j))$ that crosses at least one diagonal in the mutual partial triangulation, we require $b_{\alpha(i),\alpha(j)}:=X_{\alpha(i),\alpha(j)}>0$ to be a positive constant.
\item
The mutual triangulation (assuming $d$ diagonals) subdivides the $n$-gon into $(d{+}1)$ sub-polygons, and we impose the non-adjacent constant conditions Eq.~\eqref{eq:cXXXX} to each sub-polygon.
\end{enumerate}
For the last step, it is necessary to omit an edge from each sub-polygon when imposing the non-adjacent constants. By convention, we omit edges corresponding to the diagonals of the mutual triangulation as well as edge $n$ of the $n$-gon so that no two sub-polygons omit the same element. A moment's thought reveals that there is only one way to do this. Finally, we define the kinematic polytope $\A[\alpha|\beta]:=H[\alpha|\beta]\cap\Delta[\alpha]$. In particular, for the standard ordering $\alpha=\beta=(1,\ldots, n)$, we recover $(\Delta[\alpha],H[\alpha|\beta],\A[\alpha|\beta])=(\Delta_n,H_n,\A_n)$. 


Let us get some intuition for the shape of the kinematic polytope. Clearly $\A[\alpha|\alpha]$ is just the associahedron with boundaries relabeled by $\alpha$. For general $\alpha,\beta$, we can think of the mutual partial triangulation (with $d$ diagonals) as a partial triangulation corresponding to some codimension $d$ boundary of the associahedron $\A[\alpha|\alpha]$. Now imagine ``zooming in'' on the boundary by pushing all non-adjacent boundaries to infinity. The non-adjacent boundaries precisely correspond to partial triangulations of the $\alpha$-ordered $n$-gon that cross at least one diagonal of the mutual partial triangulation. This provides the correct intuition for the ``shape'' of the kinematic polytope $\A[\alpha|\beta]$. Said in another way, the polytope $\A[\alpha|\beta]$ is again an associahedron but with incompatible boundaries pushed to infinity.

\begin{figure}
\centering
\begin{subfigure}{0.33\linewidth}
\centering
\includegraphics[width=3cm]{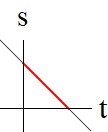}
\caption{$(1234)$\\ $u<0$ const \\ $d\log (s/t)$}
\end{subfigure}
\begin{subfigure}{0.3\linewidth}
\centering
\includegraphics[width=3cm]{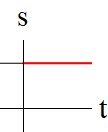}
\caption{$(1324)$\\ $s>0$ const \\ $d\log t$}
\end{subfigure}
\begin{subfigure}{0.3\linewidth}
\centering
\includegraphics[width=3cm]{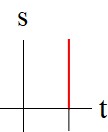}
\caption{$(2134)$\\ $t>0$ const \\ $d\log s$}\end{subfigure}
\caption{Three orderings for the $n{=}4$ kinematic polytopes. We assume the same $\alpha=(1234)$ but different $\beta$ (displayed above). Furthermore, we present the constant and canonical form for each geometry.}
\label{fig:4pt_orderings}
\end{figure}

\begin{figure}
\centering
\begin{overpic}[width=5cm]{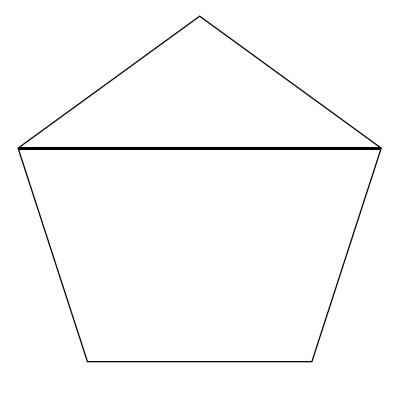}
\put(16,0){$1$}
\put(-5,55){$2$}
\put(45,100){$3$}
\put(100,55){$4$}
\put(80,0){$5$}
\put(45,53){$X_{2,4}$}
\end{overpic}
\qquad\qquad
\begin{overpic}[width=5cm]{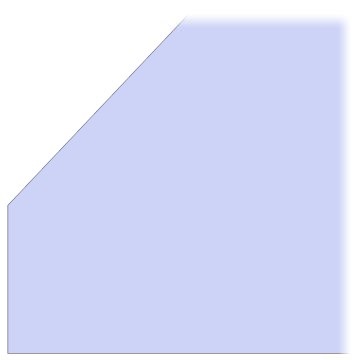}
\put(-15,20){$X_{2,4}$}
\put(15,80){$X_{1,4}$}
\put(40,-10){$X_{2,5}$}
\end{overpic}
\linebreak
\linebreak
\caption{The mutual partial triangulation for $(\alpha,\beta)=(12345,13245)$ (left) and its kinematic polytope (right). The faded area corresponds to the boundary at infinity. The two vertices correspond to the two cubic graphs compatible with both orderings.}
\label{fig:5pt_13245}
\end{figure}
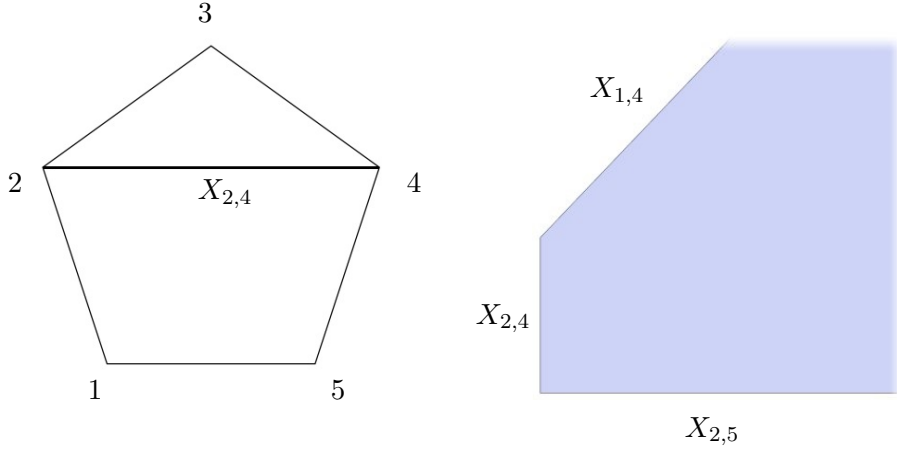

For $n{=}4$, the three distinct kinematic polytopes are shown in Figure~\ref{fig:4pt_orderings}. For $n{=}5$, consider the case $(\alpha,\beta)=(12345,13245)$. The mutual partial triangulation consists of the regular pentagon with the single diagonal $(2,4)$ (See Figure~\ref{fig:5pt_13245} (left)) with two compatible cubic graphs corresponding to the channels $(X_{2,4},X_{2,5})$ and $(X_{2,4},X_{1,4})$. The constants are given by
\ba
b_{1,3}&:=&X_{1,3}>0\\
b_{3,5}&:=&X_{3,5}>0\\
c_{14}&:=&X_{1,4}+X_{2,5}-X_{2,4}>0
\ea
and the inequalities are given by
\ba
X_{2,4}&\geq& 0\\
X_{2,5}&\geq& 0\\
X_{1,4}&\geq& 0
\ea
Finally we plot this region in the basis $(X_{2,4},X_{2,5})$ as shown in Figure~\ref{fig:5pt_13245} where the first two inequalities simply give the positive quadrant while the last inequality gives the diagonal boundary $X_{1,4}=c_{14}-X_{2,5}+X_{2,4}\geq 0$.

Having constructed the kinematic polytope $\A[\alpha|\beta]$, we now discuss its connection to bi-adjoint tree amplitude $m[\alpha|\beta]$ (omitting the overall sign). We make the following two claims in analogy to the two claims made near the beginning of Section~\ref{sec:amplitudes}:
\begin{enumerate}
\item
The pullback of the cyclic scattering form $\Omega^{(n{-}3)}[\alpha]$ to the subspace $H[\alpha|\beta]$ is the canonical form of the kinematic polytope $\A[\alpha|\beta]$. That is,
\be\label{eq:pullback_alpha_beta}
\Omega^{(n{-}3)}[\alpha]|_{H[\alpha|\beta]}=\Omega(\A[\alpha|\beta])
\ee
\item
The canonical form of the kinematic polytope $\A[\alpha|\beta]$ determines the amplitude $m[\alpha|\beta]$. That is,
\be\label{eq:canon_alpha_beta}
\aOmega(\A[\alpha|\beta])=m[\alpha|\beta]
\ee
\end{enumerate}
The derivation is not substantially different than what we have seen before, so we simply highlight a few subtleties. For the first claim, recall that the scattering form is a sum over all $\alpha$-planar graphs:
\be
\Omega^{(n{-}3)}[\alpha]=\sum_{\alpha\text{-planar }g}\sign(g)\bigwedge_{a=1}^{n{-}3}d\log X_{\alpha(i_a),\alpha(j_a)}
\ee
We claim that on the pullback to the subspace $H[\alpha|\beta]$, the numerator is identical and non-zero for every $(\alpha,\beta)$-planar graph $g$ and zero otherwise:
\be
\sign(g)\bigwedge_{a=1}^{n{-}3}dX_{\alpha(i_a),\alpha(j_a)}=\begin{cases}
d^{n{-}3}X  &\text{ if $g$ is $\beta$-planar}\\
0  &\text{ otherwise}
\end{cases}
\ee
The pullback therefore sums all the $\beta$-planar diagrams and destroys all other diagrams, thus giving the desired amplitude $m[\alpha|\beta]$:
\be
\Omega^{(n{-}3)}[\alpha]|_{H[\alpha|\beta]}
=\left(\sum_{(\alpha,\beta)\text{-planar }g}\frac{1}{\prod_{a=1}^{n{-}3}X_{\alpha(i_a),\alpha(j_a)}}\right)d^{n{-}3}X=m[\alpha|\beta]d^{n{-}3}X
\ee

As before, it can be shown that this is also the canonical form of the kinematic polytope $\A[\alpha|\beta]$. The canonical forms for the $n{=}4$ examples are given in Figure~\ref{fig:4pt_orderings}. The canonical form for the $n{=}5$ example in Figure~\ref{fig:5pt_13245} is
\ba
\Omega(\A[12345|13245])&=&d\log X_{2,5}d\log X_{2,4}+d\log X_{2,4}d\log X_{1,4}\nonumber\\
&=&\left(\frac{1}{X_{2,5}X_{2,4}}+\frac{1}{X_{2,4}X_{1,4}}\right)d^2X
\ea
where we used the fact that $dX_{2,5}dX_{2,4}=dX_{2,4}dX_{1,4}$ on the pullback, which follows from the identity $X_{1,4}=c_{14}-X_{2,5}+X_{2,4}$.

\subsection{The Associahedron is the Amplituhedron for Bi-adjoint $\phi^3$ Theory}
Let us summarize the story so far for the bi-adjoint $\phi^3$ theory. We have an obvious kinematic space $\K_n$ parametrized by the $X_{i,j}$ which is $n(n-3)/2$-dimensional. We also have a scattering form $\Omega^{(n-3)}_n$ of rank $(n{-}3)$ defined on this space, which for $n>3$ is of lower than top rank. This scattering form is fully determined by its association with a positive geometry living in the kinematic space defined in the following way. First, there is a top-dimensional ``positive region'' in the kinematic space given by $X_{i,j}\geq 0$ whose boundaries are associated with all the poles of the planar graphs. Next, there is a family of $(n{-}3)$-dimensional linear subspaces defined by $X_{i,j} + X_{i+1, j+1} - X_{i ,j+1} - X_{i+1, j} = c_{ij}$. With appropriate positivity constraints on the constants $c_{ij}>0$, this subspace intersects the ``positive region'' in a positive geometry---the kinematic associahedron $\A_n$. Furthermore, the scattering form $\Omega^{(n-3)}_n$ on the full kinematic space is fully determined by the property of pulling back to the canonical form of the associahedron on this family of subspaces. Hence, the physics of on-shell tree-level bi-adjoint $\phi^3$ amplitudes are completely determined by the positive geometry not in any auxiliary space but directly in kinematic space.

Furthermore, there is a striking similarity between this description of bi-adjoint $\phi^3$ scattering amplitudes and the description of planar $\mathcal{N}=4$ super Yang-Mills (SYM) with the amplituhedron as the positive geometry~\cite{Arkani-Hamed:2017vfh}. Indeed the general structure is identical. There is once again a kinematic space, which for planar ${\cal N} = 4$ SYM is given by the momentum-twistor variables $Z_i\in \mathbb{P}^3(\mathbb{R})$ for $i=1,\ldots, n$, and a differential form $\Omega_n^{(4 k)}$ of rank $4\times k$ (for N$^k$MHV) on kinematic space that is fully determined by its association with a positive geometry. We again begin with a ``positive region'' in the kinematic space which enforces positivity of all the poles of planar graphs via $\langle Z_i Z_{i+1} Z_j Z_{j+1} \rangle \geq 0$; however, also required is a set of topological ``winding number'' conditions enforced by a particular ``binary code''  of sign-flip patterns for the momentum-twistor data. This is a top-dimensional subspace of the full kinematic space. There is also a canonical $4 \times k$ dimensional subspace of the kinematic space, corresponding to an affine translation of a given set of external data ${\bf }Z_*$ in the direction of a fixed $k$-plane $\Delta$ in $n$ dimensions; this subspace is thus specified by a $(4+k) \times n$ matrix ${\cal Z} := (Z_*, \Delta)^T$. Provided the condition that all ordered $(4{+}k)\times(4{+}k)$ minors of ${\cal Z}$ are positive, this subspace intersects the ``positive region'' in a positive geometry---the (tree) amplituhedron. The form $\Omega_n^{(4 k)}$ on the full space is fully determined by the property of pulling back to the canonical form of the amplituhedron found on this family of subspaces. Once again this connection between scattering forms and positive geometry is seen directly in ordinary momentum-twistor space, without any reference to the auxiliary Grassmannian spaces where amplituhedra were originally defined to live. 

The nature of the relationship between ``kinematic space'', ``positive region'', ``positive family of subspaces'' and ``scattering form'' is literally identical in the two stories. We say therefore that ``the associahedron is the amplituhedron for bi-adjoint $\phi^3$ theory''. 

Of course there are some clear differences as well. Most notably, the scattering form $\Omega_n^{(4k)}$ is directly the super-amplitude with the differentials $dZ^I_i$ interpreted as Grassmann variables $\eta_i^I$, whereas for the bi-adjoint $\phi^3$ theory we have forms on the space of Mandelstam variables with no supersymmetric interpretation. While the planar ${\cal N}=4$ scattering forms are unifying different helicities into a single natural object, what are the forms in Mandelstam space doing? As we have already seen in the bi-adjoint example, and with more to come in later sections, these forms are instead {\it geometrizing color factors}, as established in Section~\ref{sec:color}.

\section{Factorization and ``Soft'' Limit}

We now derive two important properties of amplitudes by exploiting geometric properties of the associahedron:
\begin{enumerate}
\item
The amplitude factorizes on physical poles.
\item
The amplitude vanishes in a ``soft'' limit.
\end{enumerate}
We emphasize that both properties follow from geometric arguments. While amplitude factorization is familiar, here it emerges from the ``geometry factorization'' of the associahedron; and the vanishing in the ``soft limit'' is a property of the amplitude that is made more manifest by the geometry than Feynman diagrams.

\subsection{Factorization}
\label{sec:fac}

Recall from Section~\ref{sec:assoc} that the associahedron factorizes combinatorially, {\it i.e.} each facet is combinatorially identical to a product of two lower associahedra (See Eq.~\eqref{eq:fac_comb}). We now demonstrate this explicitly for the kinematic polytope $\A_n$, thus giving a simple derivation of the fact that $\A_n$ is indeed an associahedron. While Eq.~\eqref{eq:fac_comb} is a purely combinatorial statement, we go further in this section and find explicit geometric constructions for the two lower associahedra. We therefore say that $\A_n$ {\it factorizes geometrically}. Furthermore, we argue that {\it geometrical factorization} of $\A_n$ directly implies {\it amplitude factorization}, so that locality and unitarity of the amplitude are {\it emergent properties} of the geometry.

\begin{figure}
\centering
\begin{overpic}[width=0.3\linewidth]{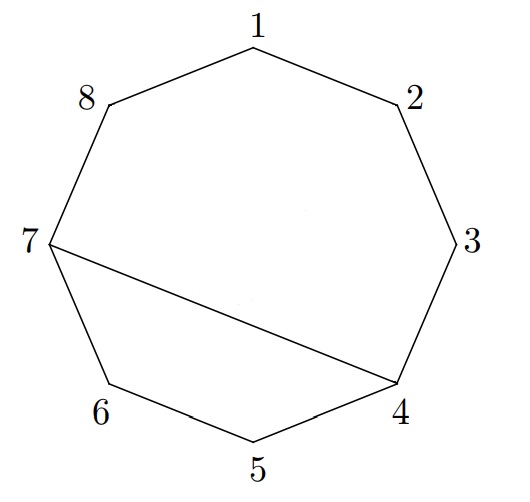}
\put(40,38){ $X_{4,7}$}
\end{overpic}
\caption{The diagonal $(4,7)$ subdivides the 8-gon into a 4-gon (on the ``left'') and a 6-gon (on the ``right''), suggesting that the facet $X_{4,7}=0$ of the associahedron $\A_{n{=}8}$ is combinatorially identical to $\A_{n{=}4}\times \A_{n{=}6}$.}
\label{fig:polygon_fac}
\end{figure}

We rewrite the kinematic associahedron $\A_n$ as $\A(1,2,\ldots, \bar{n})$ to emphasize the particle labels and their ordering; we put a bar over index $n$ to emphasize that the subspace $H_n$ is defined with non-adjacent indices omitting $n$ (See Eq.~\eqref{eq:cXXXX}). 
We make the following observations:
\begin{enumerate}
\item
{\it Geometric factorization}: The facet $X_{i,j}=0$ is equivalent to a product polytope
\be\label{eq:kin_fac}
\left.\A_n\right|_{X_{i,j}=0} \cong \A_L\times \A_R
\ee
where
\ba\label{eq:left_right}
\A_L&:=&\A(i,i{+}1,\ldots, j{-}1,\bar{I})\nonumber\\
\A_R&:=&\A(1,\ldots, i{-}1,I,j,j{+}1,\ldots,\bar{n})
\ea
and $I$ denotes the intermediate particle. The cut can be visualized as the diagonal $(i,j)$ on the convex $n$-gon (See Figure~\ref{fig:polygon_fac}).
\item
{\it Amplitude factorization}: The residue of the canonical form along the facet $X_{i,j}=0$ factors:
\be\label{eq:unit}
\Res_{X_{i,j}=0}\Omega(\A_n)=
\Omega(\A_L)
\wedge
\Omega(\A_R)
\ee
This implies factorization of the amplitude.
\end{enumerate}
We first construct the ``left associahedron'' $\A_L$ and the ``right associahedron'' $\A_R$ by Eq.~\eqref{eq:left_right} as {\it independent} associahedra living in {\it independent} kinematic spaces. The indices appearing in the construction are nothing more than well-chosen labels at this point. To emphasize this, we use independent planar variables for $\A_L$ and $\A_R$:
\ba
\A_L\;&:&\;L_{a,b}\text{ for }i\leq a<b< j\\
\A_R\;&:&\;R_{a,b}\text{ for }1\leq a<b<n\text{ except }i\leq a<b< j
\ea
The index ranges can be visualized as Figure~\ref{fig:polygon_fac} 
where the ``left'' planar variables $L_{a,b}$ correspond to diagonals of the ``left'' subpolygon, and likewise for the ``right''. Furthermore, the two associahedra come with positive non-adjacent constants $l_{ab}$, $r_{ab}$, respectively. For $l_{ab}$ the indices consist of all non-adjacent pairs $a,b$ in the range $i\leq a<b<j$. For $r_{ab}$ they consist of all non-adjacent pairs $a,b$ in the range $(1,\ldots, i{-}1,I,j,j{+}1,\ldots, n{-}1)$.

We now argue that there exists a one-to-one correspondence:
\be
\A_L\times \A_R\cong \A_n|_{X_{i,j}=0}
\ee
We begin by picking a kinematic basis for $\A_L$ consisting of $L_{a,b}$ variables corresponding to some triangulation of the left subpolygon in Figure~\ref{fig:polygon_fac}, and similarly for the $R_{a,b}$ variables. The two triangulations combine to form a partial triangulation of the $n$-gon with the diagonal $(i,j)$ omitted. Each diagonal corresponds to a planar variable, thus providing a basis for the subspace $H_n|_{X_{i,j}=0}$. Furthermore, we assume that the non-adjacent constants match so that $c_{ab}=l_{ab}$ for all $l_{ab}$. As for $r_{ab}$, we assume that $c_{ab}=r_{ab}$ for all $r_{ab}$ where $a,b\neq I$. Furthermore, $r_{aI}=\sum_{k\in I}c_{ak}$ for all $r_{aI}$.

We then write down the most obvious map $\A_L\times \A_R\rightarrow H_n|_{X_{i,j}=0}$ given by:
\ba
X_{a,b} &=& L_{a,b}\text{ for all left basis variables $L_{a,b}$}\\
X_{a,b} &=& R_{a,b}\text{ for all right basis variables $R_{a,b}$}
\ea
Since the $X_{a,b}$ variables in the image form a basis for $\A_n|_{X_{i,j}=0}$, this completely defines the map. We observe that $X_{a,b}=L_{a,b}$ holds not just for left basis variables, but for all left variables $L_{a,b}$. The idea is to rewrite $L_{a,b}$ in terms of basis variables and non-adjacent constants. Since the same formula holds for $X_{a,b}$, and the constants match by assumption, therefore the desired result must follow. Similarly, $X_{a,b}=R_{a,b}$ holds for all right variables $R_{a,b}$.

Now we argue that the image of the embedding lies in the facet $\A_n|_{X_{i,j}=0}$, which requires showing that all planar propagators $X_{a,b}$ are positive under the embedding except for $X_{i,j}=0$. This is trivially true for propagators whose diagonals do not cross $(i,j)$, since either $X_{a,b}=L_{a,b}$ or $X_{a,b}=R_{a,b}$. Now consider a crossing diagonal $(k,l)$ satisfying $1\leq i<k<j<l\leq n$. Applying Eq.~\eqref{eq:XXXXc} with indices $j,k$ swapped and setting $X_{i,j}=0$ gives
\be
X_{k,l}=X_{k,j}+X_{i,l}+\sum_{\substack{i\leq a<k\\j\leq b<l}} c_{ab}
\ee
Since $X_{k,j}$ is a diagonal of the left subpolygon and $X_{i,l}$ is a diagonal of the right, they are both positive. It follows that the right hand side is term-by-term positive, hence our crossing term $X_{k,l}$ must also be positive, as claimed. We emphasize that $X_{k,l}$ is actually strictly positive, implying that it cannot be cut. This is important because cutting crossing propagators simultaneously would violate the planar graph structure of the associahedron. Finally, it is easy to see that this is a one-to-one map, thus completing our argument for the first assertion Eq.~\eqref{eq:kin_fac}.

As an example, consider the $n{=}6$ kinematic associahedron shown in Figure~\ref{fig:kin_assoc} (bottom).  Let us consider the facet $X_{2,5}=0$ which by geometric factorization is a product of 4-point associahedra ({\it i.e.} a product of line segments) and must therefore be a quadrilateral. This agrees with Figure~\ref{fig:kin_assoc} (bottom) by inspection. The same is true for the facets $X_{1,4}=0$ and $X_{3,6}=0$. In contrast, the facet $X_{3,5}=0$ is given by the product of a point with a pentagon, and is therefore also a pentagon. The same holds for the remaining 5 facets.

The second assertion Eq.~\eqref{eq:unit} follows immediately from the first:
\be
\Res_{X_{i,j}=0}\Omega(\A_n)=\Omega(\A_n|_{X_{i,j}=0})=\Omega(\A_L\times\A_R)=\Omega(\A_L)\wedge\Omega(\A_R)
\ee
where the first equality follows from the residue property Eq.~\eqref{eq:canon_res}, the second from the first assertion Eq.~\eqref{eq:kin_fac} and the third from the product property Eq.~\eqref{eq:canon_prod}. This provides a geometric explanation for the factorization of the amplitude first discussed in Eq.~\eqref{eq:scatter_fac}.




\subsection{``Soft'' Limit}
\label{sec:soft}
The associahedron geometry suggests a natural ``soft limit'' where the polytope is ``squashed'' to a lower dimensional one, whereby the amplitude obviously vanishes.

Consider the associahedron $\A_n$ which lives in the subspace $H_n$ defined by non-adjacent constants $c_{ij}$. Let us consider the ``soft'' limit where the non-adjacent constants $c_{1i}\rightarrow 0$ go to zero for $i=3,\ldots, n{-}1$. It follows from kinematic constraints that
\be
X_{1,3}+X_{2,n}=s_{12}+s_{1n}=-\sum_{i=3}^{n{-}1}s_{1i}=\sum_{i=3}^{n{-}1}c_{1i}\rightarrow 0
\ee
But since both terms on the left are nonnegative $X_{1,3},X_{2,n}\geq 0$ inside the associahedron, the limit ``squashes'' the geometry to a lower dimension where $X_{1,3}=X_{2,n}=0$. The canonical form must therefore vanish everywhere on $H_n$, implying that the amplitude is identically zero. Note that if we restrict kinematic variables to the interior of the associahedron, then $p_1\cdot p_i\rightarrow 0$ for every $i$, yielding the true soft limit $p_1\rightarrow 0$. A similar argument can be given to show that the canonical form vanishes in the ``soft'' limit where $c_{i,n{-}1}\rightarrow 0$ for every $i=1,\ldots, n{-}3$. And by cyclic symmetry, the amplitude must vanish under every ``soft'' limit given by $s_{ij}\rightarrow 0$ for some fixed index $i$ and every index $j\neq i{-}1,i{+}1$. 

Furthermore, given any triangulation of the associahedron $\A_n$ of the kind discussed in Section~\ref{sec:assoc_triang}, every piece of the triangulation is squashed by the ``soft'' limit. It follows that the canonical form of each piece must vanish individually.

The fact that the amplitude $m_n$ vanishes in this limit is rather non-trivial from a physical point of view. While the geometric argument we provided is straightforward, there does not appear to be any obvious physical reason for it. It is another feature of the amplitude made obvious by the associahedron geometry.

As an example, the $n{=}5$ amplitude Eq.~\eqref{eq:amp_5pt} vanishes in the limit $c_{13},c_{14}\rightarrow 0$, which can be seen by substituting the equivalent limits $X_{1,3}\rightarrow X_{1,4}-X_{2,4}$ and $X_{2,5}\rightarrow X_{2,4}-X_{1,4}$ directly into the amplitude Eq.~\eqref{eq:amp_5pt}.

\section{Triangulations and Recursion Relations}


Since the scattering forms pull back to the canonical form on our associahedra, it is natural to expect that concrete expressions for the scattering amplitudes correspond to natural triangulations of the associahedron. This connection between triangulations of a positive geometry and various physical representations of amplitudes has been vigorously explored in the context of the positive Grassmannian/amplituhedron, with various triangulations of spaces and their duals corresponding to BCFW and ``local'' forms for scattering amplitudes.
In the present case of study for bi-adjoint $\phi^3$ theories, we encounter a lovely surprise: one of the canonical triangulations of the associahedron literally reproduced the Feynman diagram expansion! Ironically this representation also introduces spurious poles (at infinity!) that only cancel in the full sum over all diagrams; also, other properties of the amplitude, such as the vanishing in the ``soft'' limit discussed in Section~\ref{sec:soft}, are also not manifest term-by-term in this triangulation. We also explore a number of other natural triangulations of the geometry that make manifest the features hidden by the Feynman diagram triangulation. Quite surprisingly, some triangulations lead to even more compact expressions for these familiar and already very simple amplitudes!
Finally, we introduce a novel recursion relation for amplitudes based on the factorization properties discussed in Section~\ref{sec:fac}.

\subsection{The Dual Associahedron and Its Volume as the Bi-adjoint Amplitude}

Recall that every convex polytope $\A$ has a {\it dual polytope} $\A^*$ which we review in Appendix~\ref{app:pos_poly} where some notation is established. An important fact also explained in Appendix~\ref{app:pos_poly} says that the canonical form of any polytope $\A$ is determined by the volume of its dual $\A^*$:
\be\label{eq:canon_vol}
\aOmega(\A)=\text{Vol}(\A^*)
\ee
This identity has many implications for both physics and geometry. We refer the reader to~\cite{Arkani-Hamed:2017tmz} for a more thorough discussion.

Applying Eq.~\eqref{eq:canon_vol} to our discussion implies that the canonical form of the associahehdron $\A_n$ is determined by the volume of the {\it dual associahedron} $\A_n^*$:
\be\label{eq:canon_vol_assoc}
\aOmega(\A_n) = \text{Vol}(\A_n^*)
\ee
But in the same way, the canonical form is determined by the amplitude $m_n$ via Eq.~\eqref{eq:canon_amp}, thus suggesting that the amplitude is the volume of the dual:
\be
m_n=\text{Vol}(\A_n^*)
\ee
This leads to yet another geometric interpretation of the bi-adjoint amplitude. For the remainder of this section, we describe the construction of the dual associahedron in more detail, and provide the example for $n{=}5$.

Following the discussion in Appendix~\ref{app:pos_poly}, we embed the subspace $H_n$ in projective space $\Proj^{n{-}3}(\R)$, and we choose a basis $X_{i_1',j_1'},\ldots,X_{i_{n{-}3}',j_{n{-}3}'}$ of Mandelstam variables to denote coordinates on the subspace:
\be\label{eq:Y_basis}
Y=(1,X_{i_1',j_1'},\ldots, X_{i_{n{-}3}',j_{n{-}3}'})\in \mathbb{P}^{n{-3}}(\mathbb{R})
\ee
Here we have introduced a zeroth component ``1'' since the coordinates are embedded projectively. Any other basis can be obtained via a $GL(n{-}2)$ transformation.

\begin{figure}
\centering
\begin{overpic}[width=0.4\linewidth]
{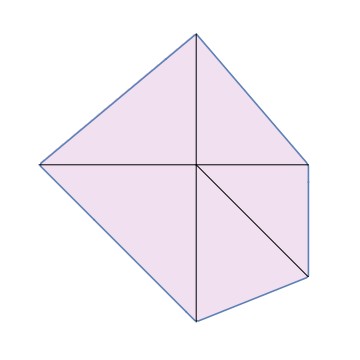}
\put(42,44){$W_*$}
\put(-5,53){$W_{1,3}$}
\put(48,95){$W_{3,5}$}
\put(87,53){$W_{2,5}$}
\put(87,15){$W_{2,4}$}
\put(48,4){$W_{1,4}$}
\end{overpic}
\;\;\;\;
\begin{overpic}[width=0.4\linewidth]
{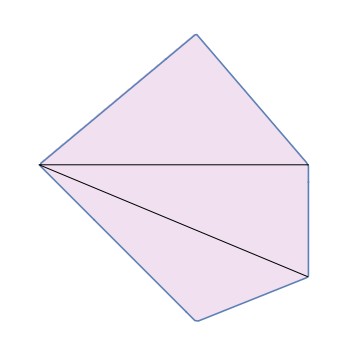}
\put(-5,53){$W_{1,3}$}
\put(48,95){$W_{3,5}$}
\put(87,53){$W_{2,5}$}
\put(87,15){$W_{2,4}$}
\put(48,4){$W_{1,4}$}
\end{overpic}
\caption{Two triangulations of the dual associahedron $\A_{n{=}5}^*$}
\label{fig:dual_pentagon}
\end{figure}
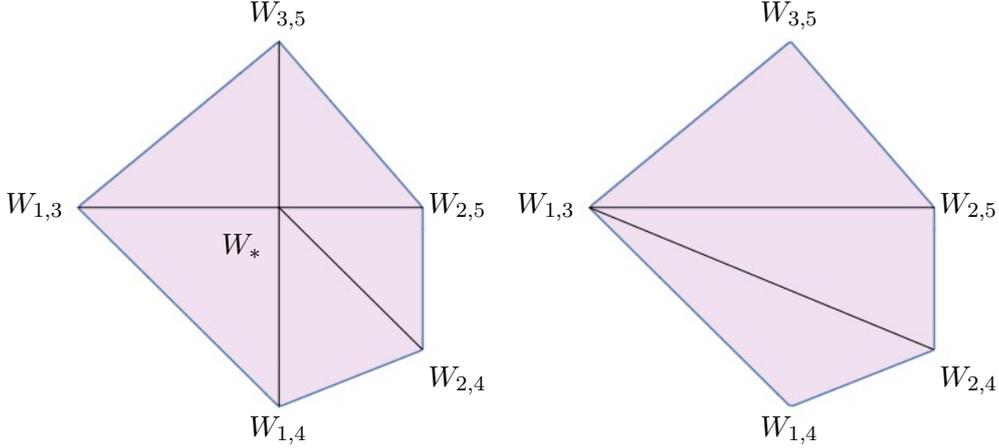

Furthermore, we denote the facets of the associahedron in projective coordinates. Recall that every facet of $\A_n$ is of the form $X_{i,j}= 0$. We rewrite this in the form $W_{i,j}\cdot Y=0$ for some dual vector $W_{i,j}$. For example, consider $n=5$ in the basis $Y=(1,X_{1,3},X_{1,4})$. Then
\be
Y\cdot W_{2,5}=
X_{2,5}=c_{13}+c_{14}-X_{1,3} = (c_{13}+c_{14},-1,0)\cdot Y
\ee
which implies that $W_{2,5}=(c_{13}+c_{14},-1,0)$. More generally, the components of any $W_{i,j}$ can be read off from the expansion of $X_{i,j}$ in terms of basis variables $X_{i_a',j_a'}$ and non-adjacent constants. Here we present all the dual vectors for the $n=5$ pentagon in Figure~\ref{fig:kin_assoc} (top right):
\ba\label{eq:WWWWW}
W_{1,3}&=&(0,1,0)\nonumber\\
W_{3,5}&=&(c_{14}+c_{24},0,-1)\nonumber\\
W_{2,5}&=&(c_{13}+c_{14},-1,0)\nonumber\\
W_{2,4}&=&(c_{13},-1,1)\nonumber\\
W_{1,4}&=&(0,0,1)
\ea
Once the coordinates for the dual vectors $W_{i,j}$ are computed, they can be thought of as {\it vertices} of the dual associahedron $\A_n^*$ in the dual projective space. For $n{=}5$, the dual associahedron is a pentagon whose vertices are Eq.~\eqref{eq:WWWWW} (See Figure~\ref{fig:dual_pentagon}).

\subsection{Feynman Diagrams as a Triangulation of the Dual Associahedron Volume}

We now compute the volume of $\A^*$ by triangulation and summing over the volume of each piece. We make use of the fact that $\A_n^*$ is a {\it simplicial} polytope, meaning that each facet is a simplex. This is equivalent to $\A_n$ being a simple polytope. In this case the dual is easily triangulated by the following method:
\begin{enumerate}
\item
Take a reference point $W_*$ on the interior of the dual polytope.
\item
For each facet of the dual, take the convex hull of the facet with $W_*$ which gives a simplex.
\item
The union of all such simplices forms a triangulation of the dual.
\end{enumerate}
Let $Z$ denote a {\it facet} of the dual $\A_n^*$. Then $Z$ is adjacent to some vertices $W_{i_1,j_1},\ldots,$ $W_{i_{n{-}3},j_{n{-}3}}$ corresponding to propagators $X_{i_1,j_1},\ldots, X_{i_{n{-}3},j_{n{-}3}}$, respectively. By taking the convex hull of the facet $Z$ with $W_*$, and taking the union over all facets, we get a triangulation of the dual associahedron whose volume is the sum over the volume of each simplex. Recalling the formula for the volume of a simplex Eq.~\eqref{eq:simplex_vol}, we find
\ba\label{eq:dual_vol}
\text{Vol}(\A_n^*)&=&
\sum_{\text{vertex }Z} \text{Vol}(W_*,W_{i_1,j_1},\ldots, W_{i_{n{-}3},j_{n{-}3}})\nonumber\\
&=&
\sum_{\text{vertex }Z}
\frac{\sign(Z)\left<W_*W_{i_1,j_1}\cdots W_{i_{n{-}3},j_{n{-}3}} \right>}{(Y\cdot W_*)\prod_{a=1}^{n{-}3}(Y\cdot W_{i_a,j_a})}
\ea
where $\sign(Z)$ is the orientation of the adjacent vertices $W_{i_1,j_1},\ldots, W_{i_{n{-}3},j_{n{-}3}}$ (in that order) relative to the inherited orientation. Note that the antisymmetry of $\sign(Z)$ is compensated by the antisymmetry of the determinant $\left<\cdots\right>$ in the numerator, and the sum is independent of the choice of reference point $W_*$. Furthermore, the $\sign(Z)$ here is equivalent to the $\sign(Z)$ appearing in Eq.~\eqref{eq:canon_assoc} where $Z$ denotes the corresponding {\it vertex} of $\A_n$. In fact, we now argue that for an appropriate choice of reference point $W_*$, the Feynman diagram expansion Eq.~\eqref{eq:amp_form} is term-by-term equivalent to the expression Eq.~\eqref{eq:dual_vol}, where each $Z$ is associated with its corresponding planar cubic graph $g$.

With the benefit of hindsight, we set the reference point to $W_*=(1,0\ldots,0)$, which is particularly convenient because the numerators in Eq.~\eqref{eq:dual_vol} are now equivalent for all $Z$. Indeed, since $X_{i_a,j_a}=Y\cdot W_{i_a,j_a}$, we have
\be 
\left<W_*W_{i_1,i_1}\cdots W_{i_{n{-}3},j_{n{-}3}}\right>
=
\frac{\partial(X_{i_1,j_1},\ldots, X_{i_{n{-}3},j_{n{-}3}})}
{\partial(X_{i_1',j_1'},\ldots, X_{i_{n{-}3}',j_{n{-}3}'})}
=\sign(Z)/\sign(Z')
\ee
where the primed variables form the basis we chose back in Eq.~\eqref{eq:Y_basis}, and the second equality follows from Eq.~\eqref{eq:jac}. This shows that all the numerators in Eq.~\eqref{eq:dual_vol} are equivalent to $\sign(Z')$, which we set to one. Finally, substituting $(Y\cdot W_*)=1$ and $(Y\cdot W_{i_a,j_a})=X_{i_a,j_a}$ into Eq.~\eqref{eq:dual_vol} and replacing $Z$ by $g$ gives
\be
\text{Vol}(\A_n^*)=\sum_{\text{planar }g}\frac{1}{\prod_{a=1}^{n{-}3}X_{i_a,j_a}}
\ee
which is precisely the Feynman diagram expansion Eq.~\eqref{eq:amp_form} for the amplitude. It follows that the amplitude is the volume of the dual associahedron
\be
\text{Vol}(\A_n^*)=\aOmega(\A_n)=m_n
\ee
of which the Feynman diagram expansion is a particular triangulation.

We point out that the Feynman diagram expansion introduces a spurious vertex $W_*$, which term-by-term gives rise to a pole at infinity that cancels in the sum. From the point of view of the original associahedron, this corresponds to a ``signed'' triangulation of $\A_n$ with overlapping simplices, whereby every simplex consists of all the facets that meet at a vertex together with the boundary at infinity. The presence of bad poles at infinity in individual Feynman diagrams that only cancel in the sum over all diagrams bears striking resemblance to the behavior of Feynman diagrams under BCFW shifts in gauge theories and gravity. There too, individual Feynman diagrams  have poles at infinity, even though the final amplitude does not, and this surprising vanishing at infinity is critically related to the magical properties of amplitudes in these theories. Indeed, the absence of poles at infinity in Yang-Mills theory finds a deeper explanation in terms of the symmetry of dual conformal invariance. It is thus particularly amusing to see an analog of this hidden symmetry even for something as innocent-seeming as bi-adjoint $\phi^3$ theory! Furthermore, the scattering form in the full kinematic space is projectively invariant, a symmetry invisible in individual diagrams. And the pullback of the forms to the associahedron subspaces are {\it also} projectively invariant, with no pole at infinity. In Yang-Mills theories, we have discovered representations (such as those based on BCFW recursion relations) that make the dual conformal symmetry manifest term-by-term, and these were much later seen to be associated with triangulations of the amplituhedron. Similarly, we now turn to other natural triangulations of the associahedron which do not introduce new vertices and thus have no spurious poles at infinity, thus making manifest term-by-term the analogous feature of bi-adjoint $\phi^3$ amplitudes that is hidden in Feynman diagrams.

\subsection{More Triangulations of the Dual Associahedron}
Returning to Eq.~\eqref{eq:dual_vol}, a different choice of $W_*$ would have led to alternative triangulations, and hence novel formulas for the amplitude. For instance, for $n{=}5$, we can take the limit $W_*\rightarrow W_{13}$. This kills two volume terms and gives a three-term triangulation as shown in Figure~\ref{fig:dual_pentagon} (right):
\ba
m_{n{=}5}=\frac{X_{1,3}+X_{2,5}}
{X_{1,3}X_{3,5}X_{2,5}}
+\frac{X_{1,3}+X_{2,5}}
{X_{1,3}X_{2,5}X_{2,4}}
+\frac{X_{1,3}-X_{1,4}+X_{2,4}}
{X_{1,3}X_{2,4}X_{1,4}}
\ea
Note that we have re-written the non-adjacent constants $c_{ij}$ in terms of planar variables via Eq.~\eqref{eq:cXXXX}. The sum of these three volumes gives the volume of the dual associahedron, and hence the amplitude. Furthermore, since no spurious vertices are introduced, the result makes manifest term-by-term the absence of poles at infinity. This contrasts the Feynman diagram expansion where spurious poles appear term-by-term. Finally, this method of setting $W_*$ to one of the vertices can be repeated for arbitrary $n$, and in general produces fewer terms than with Feynman diagrams.


\subsection{Direct Triangulations of the Kinematic Associahedron}
\label{sec:assoc_triang}
Recall that canonical forms are {\it triangulation independent}, hence the canonical form of a polytope can be obtained by triangulation and summation over the canonical form of each piece. A brief review is given in Appendix~\ref{app:pos_triang}. We now exploit this property to compute the canonical form of the associahedron, thus establishing another method for computing amplitudes.

\begin{figure}
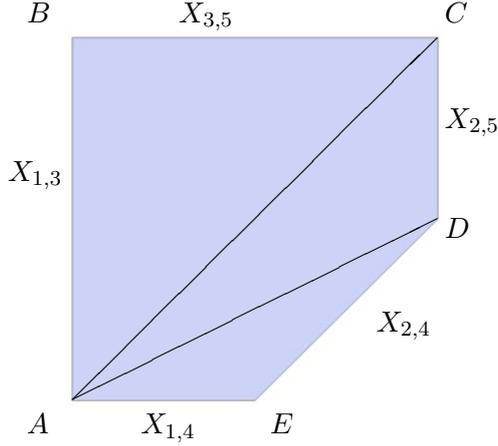

\centering
\begin{overpic}[width=5cm]
{5pt.jpg}
\put(20,-7){$X_{1,4}$}
\put(-15,60){$X_{1,3}$}
\put(30,102){$X_{3,5}$}
\put(100,74){$X_{2,5}$}
\put(82,20){$X_{2,4}$}
\put(-10,-7){$A$}
\put(-10,102){$B$}
\put(100,102){$C$}
\put(100,45){$D$}
\put(54,-7){$E$}
\put(2,2){\color{black}\line(1,1){96}}
\put(2,2){\color{black}\line(2,1){96}}
\end{overpic}
\linebreak
\linebreak
\caption{A triangulation of the associahedron $\A_{n{=}5}$}
\label{fig:pentagon_triang}
\end{figure}

We wish to compute the $n{=}5$ amplitude for which the associahedron is a pentagon. We choose the basis $Y=(1,X_{13},X_{14})$, and triangulate the associahedron as the union of three triangles $ABC$, $ACD$ and $ADE$ (See Figure~\ref{fig:pentagon_triang}). It follows that
\be\label{eq:pentagon_triang_3}
\Omega(\A_{n=5})=\Omega(ABC)+\Omega(ACD)+\Omega(ADE)
\ee
Note that the triangles must be oriented in the same way as the associahedron (clockwise in this case). Getting the wrong orientation would cause a sign error. The boundaries of the triangles are given by $W\cdot Y=0$ for:
\ba
W_{AB} = (0,1,0) \;\; \;\;
W_{BC} =(c_{14}+c_{24},0,-1)\nonumber\\\ 
W_{CD} =(c_{13}+c_{14},-1,0)\;\;\;\;
W_{DE} = (c_{13},-1,1)\;\;\;\;
W_{AE} = (0,0,1)\nonumber\\
W_{AC} = (0,-c_{14}-c_{24},c_{13}+c_{14})\;\;\;\;
W_{AD} = (0, -c_{14},c_{13}+c_{14})
\ea
Recalling the canonical form for a simplex Eq.~\eqref{eq:canon_simplex}, we get
\ba
\Omega(ABC) &=& \frac{
(X_{1,3}+X_{2,5})
(X_{1,4}+X_{3,5})
d^2X}
{X_{1,3}
X_{3,5}
(X_{1,4}X_{2,5}-X_{1,3}X_{3,5})
}\nonumber\\
\Omega(ACD) &=& \frac{
(X_{1,3}+X_{2,5})^2(X_{2,4}-X_{2,5}+X_{3,5})
d^2X}{
X_{2,5}
(-X_{1,4}X_{2,5}-X_{1,3}X_{2,4}+X_{1,3}X_{2,5})
(X_{1,4}X_{2,5}-X_{1,3}X_{3,5})
}\nonumber\\
\Omega(ADE) &=& \frac{
(X_{1,3}-X_{1,4}+X_{2,4})(-X_{2,4}+X_{1,4}+X_{2,5})
d^2X}
{X_{1,4}
X_{2,4}
(-X_{1,4}X_{2,5}-X_{1,3}X_{2,4}+X_{1,3}X_{2,5})
}\nonumber\\ \nonumber\\
\Omega(\A_{n=5})&=&\Omega(ABC)+\Omega(ACD)+\Omega(ADE)
\nonumber
\ea
where again we have rewritten the non-adjacent constants $c_{ij}$ in terms of planar variables via Eq.~\eqref{eq:cXXXX}. The sum of these three quantities determines the amplitude. This expansion is fundamentally different in character from the Feynman diagram expansion due to the appearance of (non-linear) spurious poles that occur in the presence of spurious boundaries $AC$ and $AD$.

This approach can be extended to all $n$ provided that a triangulation is known. Two important properties of the bi-adjoint amplitude, which are obscured by individual Feynman diagrams, become manifest in this triangulation. First, unlike that for each Feynman diagram, the form for each piece of the triangulation is projective, which means it only depends on the ratio of $X$ variables. Moreover, geometrically it is obvious that the vanishing ``soft'' limit also works term-by-term, which is certainly not the case for each Feynman diagram.

\subsection{Recursion Relations for Bi-adjoint $\phi^3$ Amplitudes}
\label{sec:rec}
We propose a simple recursion relation for computing the amplitude $\Omega(\A_n)$ as a form. Our derivation applies the recursion relations from Appendix~\ref{app:pos_rec} and the factorization properties from Section~\ref{sec:fac}. The result is reminiscent of {\it BCFW triangulation} for the amplituhedron~\cite{Arkani-Hamed:2013jha,Bai:2014cna
}. While it is not obvious from the field theory point of view, the recursion follows naturally from the geometric picture.

We begin by picking a kinematic basis
\be\label{eq:rec_basis}
Y=(1,X_{i_1,j_1},\ldots,X_{i_{n{-}3},j_{n{-}3}})
\ee
For simplicity let $Z_*=(1,0,\ldots, 0)$ denote the reference point appearing in Eq.~\eqref{eq:rec_shift}. Furthermore, for any facet given by $X_{i,j}=W_{i,j}\cdot Y=0$ corresponding to some dual vector $W_{i,j}$, we let
\be 
X_{i,j}^0 := W_{i,j}\cdot Z_* \qquad X_{i,j}' := W_{i,j}\cdot Y-W_{i,j}\cdot Z_*
\ee
Equivalently, we can expand the propagator $X_{i,j}$ by
\be
X_{i,j} = X_{i,j}^0 + X_{i,j}'
\ee
where $X_{i,j}^0$ is a linear combination of non-adjacent constants while $X_{i,j}'$ is a linear combination of the basis variables. This expansion is basis-dependent, but unique for each basis. Furthermore, the deformation Eq.~\eqref{eq:rec_shift} $Y\rightarrow \hat{Y}$ is given by
\ba
\hat{Y} &=& Y-\left(\frac{W_{i,j}\cdot Y}{W_{i,j}\cdot Z_*}\right)Z_*\\
&=& \left(\frac{X_{i,j}'}{-X_{i,j}^0},X_{i_1,j_1},\ldots, X_{i_{n{-}3},j_{n{-}3}}\right)\\
&=&\left(1,\left(\frac{-X_{i,j}^0}{X_{i,j}'}\right)X_{i_1,j_1},\ldots,
\left(\frac{-X_{i,j}^0}{X_{i,j}'}\right)X_{i_{n{-}3},j_{n{-}3}}
\right)
\ea
where in the last step we rescaled the vector by an overall factor to put it in the same form as Eq.~\eqref{eq:rec_basis}, which is possible since the vector is projective. This gives us the deformations $\hat{X}_{i_a,j_a}=\left(-X_{i,j}^0/X_{i,j}'\right) X_{i_a,j_b}$ for every basis variable. We caution the reader that this deformation is only applied on the basis variables, not on all kinematic variables $X_{k,l}$. The non-adjacent constants are invariant under the deformation $\hat{c}_{kl}=c_{kl}$, and the deformation for any other kinematic variable can be obtained by expanding it in terms of basis variables and non-adjacent constants. In particular, the deformation for $X_{i,j}$ vanishes:
\be
\hat{X}_{i,j} = \hat{X}_{i,j}'+\hat{X}^0_{i,j} = \left(\frac{-X_{i,j}^0}{X_{i,j}'}\right)X_{i,j}'+X^0_{i,j}=0
\ee
which is expected since the deformation is a projection onto the cut.

From Eq.~\eqref{eq:rec} we propose that the canonical form of the associahedron can be obtained from the canonical form of each facet:
\be
\Omega(\A_n) = \sum_{\text{facet }F_{i,j}}D_{{i,j}} \hat{\Omega}(F_{i,j})
\ee
where $F_{i,j}$ denotes the facet along $X_{i,j}=0$, and we sum over all facets. The hat operator denotes a pullback via the deformation $X_{kl}\rightarrow \hat{X}_{kl}$, and the $D_{{i,j}}$ operator denotes the ``numerator replacement'' rule (See Eq.~\eqref{eq:num_replace}):
\be
\left<Xd^{n{-}4}X\right>\rightarrow 
\left(\frac{X_{i,j}^0}{X_{i,j}}\right)d^{n{-}3}X
\ee
where $X$ denotes the vector $Y$ with the initial component chopped off, and the angle brackets denote the determinant $\left<X d^{n{-}4}X\right>:=\det(X,dX,\ldots, dX)$.
Finally, recall from Section~\ref{sec:fac} that each $F_{i,j}$ factorizes into a product of lower associahedra like $F_{i,j}\cong\A_L\times \A_R$. It follows that
\be
\Omega(\A_n) = \sum_{\text{facet }F_{i,j}}D_{i,j}\left(\hat{\Omega}(\A_L)\wedge \hat{\Omega}(\A_R)\right)
\ee
This provides a recursion relation for the amplitude because $\Omega(\A_L)$ and $\Omega(\A_R)$ are determined by lower point amplitudes. The existence of such a recursion for bi-adjoint amplitudes is not expected from the usual field-theory point of view, but here we have seen that it follows directly from the geometry. 

We now do an example for $n{=}5$ (See Figure~\ref{fig:pentagon_triang}). We pick the basis $Y=(1,X_{1,3},X_{1,4})$, and we consider the contribution from the facet $X_{2,5}=c_{13}+c_{14}-X_{1,3}$ which implies $X_{2,5}^0=c_{13}+c_{14}$ and $X_{2,5}'=-X_{1,3}$. The deformations are given by
\ba\label{eq:def_5}
\hat{X}_{1,3} &=& c_{13}+c_{14}\\
\hat{X}_{1,4} &=& \frac{c_{13}+c_{14}}{X_{1,3}}X_{1,4}\\
\hat{X}_{3,5} &=& -\frac{c_{13}+c_{14}}{X_{1,3}}X_{1,4}+c_{14}+c_{24}\\
\hat{X}_{2,5}&=&0\\
\hat{X}_{2,4}&=&\frac{c_{13}+c_{14}}{X_{1,3}}X_{1,4}-c_{1,4}
\ea
And the required numerator replacement is given by
\be\label{eq:num_replace_X25}
\left<XdX\right>\rightarrow
\left(\frac{c_{13}+c_{14}}{X_{2,5}}\right)d^2X
\ee
On the cut $X_{2,5}=0$, the associahedron factorizes into the product $\A_L\times\A_R$ given by
\ba
\A_L &=& \A(2,3,4,\bar{I})\\
\A_R &=& \A(1,I,\bar{5})
\ea
where $I$ is the intermediate particle. See the discussion around Eq.~\eqref{eq:left_right} for more details. Recalling the 4- and 3-point amplitudes, we have
\ba
\Omega(\A_L) &=& d\log X_{2,4}-d\log X_{3,5}\\
\Omega(\A_R) &=& 1
\ea
Then the pullback $\hat{\Omega}(\A_L)\wedge\hat{\Omega}(\A_R)$ gives
\ba
&&d\log \hat{X}_{2,4}-d\log\hat{X}_{3,5} \\
&=&\frac{(c_{13}+c_{14})c_{24}\left< XdX\right>}
{(c_{14}X_{1,3}-c_{13}X_{1,4}-c_{14}X_{1,4})
(c_{14}X_{1,3}+c_{24}X_{1,3}-c_{13}X_{1,4}-c_{14}X_{1,4}
)}\nonumber
\ea
Applying the numerator replacement Eq.~\eqref{eq:num_replace_X25} and rewriting the non-adjacent variables in terms of planar variables via Eq.~\eqref{eq:cXXXX} gives
\be
\frac{
(X_{1,3}+X_{2,5})^2(X_{2,4}-X_{2,5}+X_{3,5})
d^2X}{
X_{2,5}
(-X_{1,4}X_{2,5}-X_{1,3}X_{2,4}+X_{1,3}X_{2,5})
(X_{1,4}X_{2,5}-X_{1,3}X_{3,5})
}
\ee
But this is precisely the $\Omega(ACD)$ term appearing in Eq.~\eqref{eq:pentagon_triang_3}, which is the canonical form of the triangle $ACD$ in Figure~\ref{fig:pentagon_triang}. This confirms the discussion in Appendix~\ref{app:pos_rec} where we expected to find the canonical form of the triangle given by the convex hull of $Z_*=A$ and the facet $CD$, which is precisely $ACD$.

Similarly, the contribution from $X_{3,5}$ and $X_{2,4}$ give $\Omega(ABC)$ and $\Omega(ADE)$, respectively. The contributions from the remaining cuts $X_{1,3}$ and $X_{1,4}$ vanish because they intersect the reference point $Z_*$ and hence the geometry is degenerate. It follows that the recursion provides a triangulation of the associahedron with reference point $Z_*=(1,0,0)$ identical to Eq.~\eqref{eq:pentagon_triang_3}.

More generally, given a choice of basis and reference point $Z_*$, the recursion gives a triangulation of the associahedron with a reference point. Again, we emphasize that this ``BCFW-like'' representation of bi-adjoint amplitudes is very different from Feynman diagrams, and it is not obvious how to derive it from a field-theory argument. 

\section{The Worldsheet Associahedron}
\label{sec:worldsheet}

We have seen that scattering amplitudes are better thought of as differential forms on the space of kinematic variables that pullback to the canonical forms of associahedra in kinematic space. This is a deeply satisfying connection. After all, the associahedron is perhaps the most fundamental and primitive object whose boundary structure embodies ``factorization'' as a combinatorial and geometric property.

Furthermore, string theorists have long known of the fundamental role of the associahedron for the open string. After all, the boundary structure of the open string moduli space---the moduli space of $n$ ordered points on the boundary of a disk---also famously ``factorizes'' in the same way. In fact, it is well-known that the Deligne-Mumford compactification~\cite{DM,1998math} of this space has precisely the same boundary structure as the associahedron. The implications of this ``worldsheet associahedron'' for aspects of stringy physics have also been explored in {\it e.g.} \cite{Hanson:2006zc, Mizera:2017cqs}.

Moreover, from general considerations of positive geometries we know that there should also be a ``worldsheet canonical form'' associated with this worldsheet associahedron, which turns out to be the famous ``worldsheet Parke-Taylor form''~\cite{strings} (for related discussions see {\it e.g.} \cite{Mizera:2017cqs, Mainz}), an object whose importance has been highlighted in Nair's observation~\cite{Nair} and Witten's twistor string~\cite{Witten:2003nn}, and especially in the story of scattering equations and the CHY formulas for scattering amplitudes~\cite{Cachazo:2013gna,Cachazo:2013hca,Cachazo:2013iaa,Cachazo:2013iea}.

But how is the worldsheet associahedron related to the kinematic associahedron? 
This simple question has a striking answer: The scattering equations act as a {\it diffeomorphism} from the worldsheet associahedron to the kinematic associahedron! From general grounds, it follows that the kinematic scattering form is the pushforward of the worldsheet Parke-Taylor form under the scattering equation map. This gives a beautiful raison d’etre to the scattering equations, and a quick geometric derivation of the bi-adjoint CHY formulas. We now explain these ideas in more detail.


\subsection{Associahedron from the Open String Moduli Space}

Recall that the moduli space of genus zero $\M_{0,n}$ is the space of configurations of $n$ distinct punctures on the Riemann sphere $\mathbb{CP}^1$ modulo ${\rm SL}(2,\mathbb{C})$. The real part $\M_{0,n}(\R)$ is the {\it open-string moduli space} consisting of all distinct points $\sigma_i\;(i=1,\ldots, n)$ on the real line (and infinity) modulo ${\rm SL}(2,\mathbb{R})$. While there are $n!$ ways of ordering the $\sigma_i$ variables, any pair of orderings related by dihedral transformation are ${\rm SL}(2,\mathbb{R})$ equivalent. It follows that the real part is tiled by $(n{-}1)!/2$ distinct regions given by inequivalent orderings of the $\sigma_i$ variables~\cite{1998math}. The region given by the standard ordering is called the {\it positive part} of the open string moduli space or more simply the {\it positive moduli space}
\be
\M_{0,n}^+:=\{\sigma_1<\sigma_2<\cdots<\sigma_n\}/{\rm SL}(2,\mathbb{R})
\ee
where the ${\rm SL}(2,\mathbb{R})$ redundancy can be ``gauge fixed'' in the standard way by setting fixing three variables $(\sigma_1,\sigma_{n{-}1},\sigma_{n})=(0,1,\infty)$ in which case $\M_{0,n}^+=\{0<\sigma_2<\cdots<\sigma_{n{-}2}<1\}$. Sometimes we also denote the space by $\M_{0,n}^+(1,2,\ldots,n)$ to emphasize the ordering. Furthermore, recall that $\M_{0,n}^+$ can also be constructed as the (strictly) positive Grassmannian ${\rm G}_{>0}(2,n)$ modded out by the torus action $\mathbb{R}_{>0}^n$. More precisely, we consider the set of all $2\times n$ matrices $(C_1,\ldots, C_n)$ with positive Pl\"{u}cker coordinates $(ab):=\det(C_a,C_b)>0$ for $1\leq a<b\leq n$, modded out by ${\rm GL}(2)$ action and column rescaling.

In analogy to what we did for the kinematic polytope, we make two claims for the positive moduli space:
\begin{enumerate}
\item
The (compactified) positive moduli space is an associahedron which we call the {\it worldsheet associahedron}. 
\item
The canonical form of the worldsheet associahedron is the Parke-Taylor form,
\be\label{PTform}
\omega_n^{\rm WS}:=\frac{1}{{\rm vol~[SL}(2)]}~\prod_{a=1}^n~\frac{ d\sigma_a}{\sigma_a - \sigma_{a+1}}
=
\frac{1}{{\rm vol~[SL}(2){\times}{\rm GL}(1)^n]}\prod_{a=1}^n~\frac{d^2 C_a}{(a\, a{+}1)}
\ee
where in the last expression we rewrote the form in Pl\"{u}cker coordinates.
\end{enumerate}
More precisely, the process of compactification provides the positive moduli space $\M_{0,n}^+$ with boundaries of all codimensions, and here we present a natural compactification called the {\it $u$-space compactification} that produces the boundary structure of the associahedron. Of course, the associahedron structure of the positive moduli space is well-known~\cite{DM,1998math}, but the discussion we present here is instructive for later sections.

\begin{figure}
\centering
\begin{overpic}[width=6cm]
{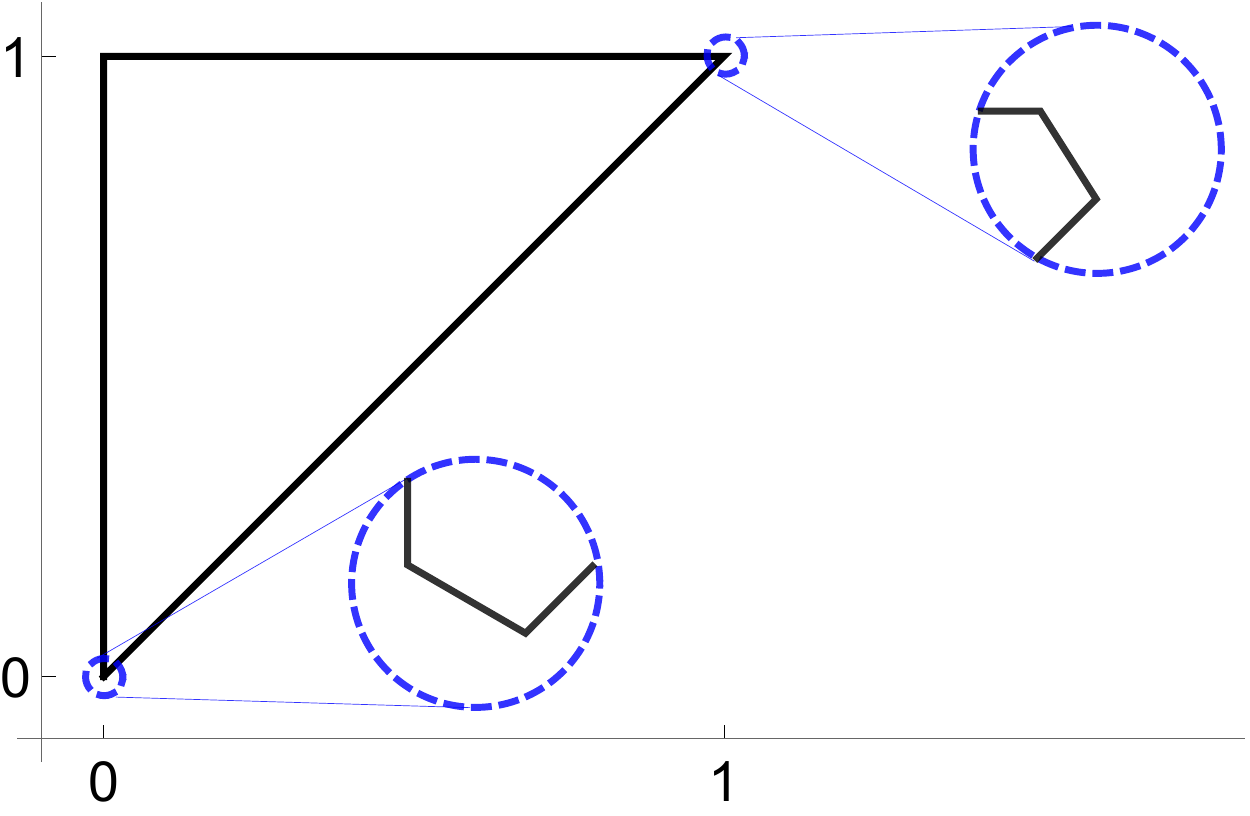}
\put(-2,68){$\sigma_3$}
\put(68,1){$\sigma_2$}
\end{overpic}
\caption{A blowup of the $n{=}5$ worldsheet associahedron showing all boundaries.}
\label{blowupup}
\end{figure}

The compactification is very subtle in $\sigma_i$ variables because our naive gauge choice fails to make all boundaries manifest. Nonetheless, all the boundaries can be visualized via a ``blowup'' procedure. Consider the case $n{=}5$ where only three of the five boundaries are manifest in the standard gauge as shown in Figure~\ref{blowupup}. The two ``hidden'' boundaries can be recovered by introducing a blowup at the vertices $(\sigma_2,\sigma_3)=(0,0)$ and $(1,1)$ as shown in Figure~\ref{blowupup}. A similar procedure applies for all $n$. We will come back to this picture when we discuss the canonical form, but now we provide an explicit compactification that makes manifest all the boundaries.

We introduce the variables $u_{i,j}$ for $1\leq i<j{-}1<n$ which are constrained to the region $0\leq u_{i,j}\leq 1$. The $u_{i,j}$ is analogous to the planar kinematic variable $X_{i,j}$ introduced in Eq.~\eqref{eq:planar_var}, and can therefore be visualized as the diagonal $(i,j)$ of a convex $n$-gon with cyclically ordered labels like Figure~\ref{fig:triang_cuts} (left). There are of course $n(n{-}3)/2$ of these variables. Furthermore, we impose the {\it non-crossing identity}
\be\label{eq:non_cross}
u_{i,j}=1-\prod_{(k,l)\in(i,j)^c}u_{k,l}
\ee
for each diagonal $(i,j)$, where $(i,j)^c$ denotes the set of all diagonals that cross $(i,j)$. Only $(n{-}2)(n{-}3)/2$ of these $n(n{-}3)/2$ constraints are independent, so the space is of dimension $(n{-}3)$.

\begin{figure}
\centering
\begin{overpic}[width=5cm]
{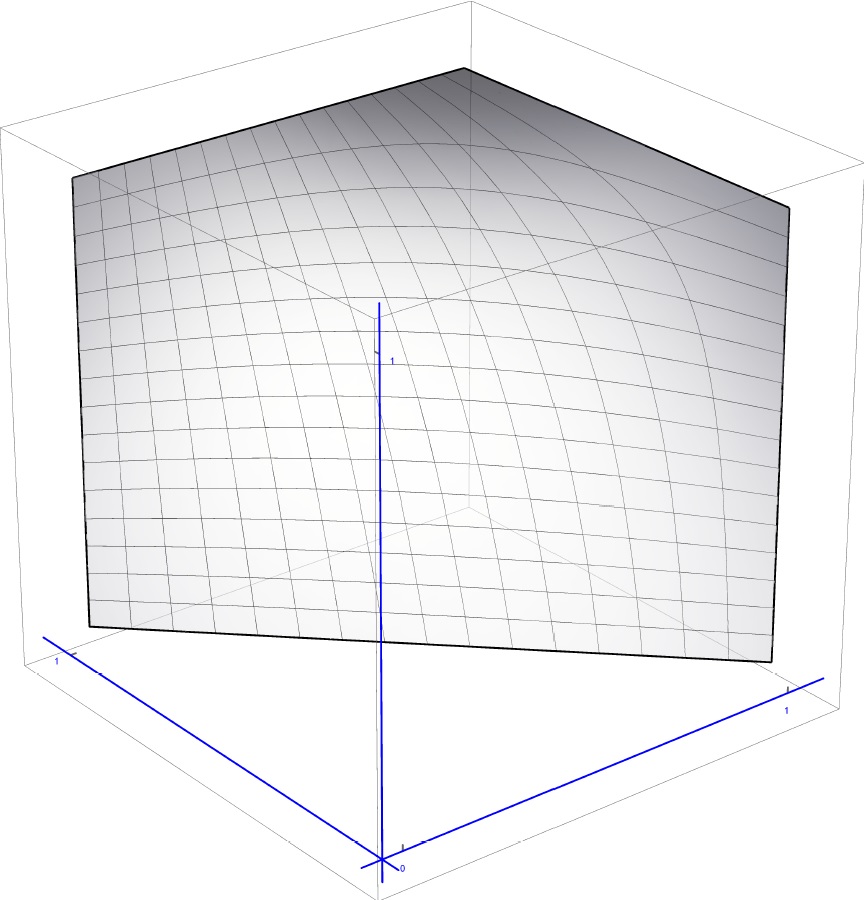}
\put(100,22){\textcolor{blue}{$u_{1,3}$}}
\put(37,70){\textcolor{blue}{$u_{3,5}$}}
\put(-10,26){\textcolor{blue}{$u_{2,5}$}}
\put(-5,52){$u_{1,3}$}
\put(90,52){$u_{2,5}$}
\put(50,23){$u_{3,5}$}
\put(20,90){$u_{1,4}$}
\put(70,90){$u_{2,4}$}
\end{overpic}
\caption{The worldsheet associahedron for $n{=}5$ presented in a coordinate chart where all boundaries are manifest. We caution the reader that some coordinate charts do not make manifest all the boundaries.}
\label{5ptassws}
\end{figure}
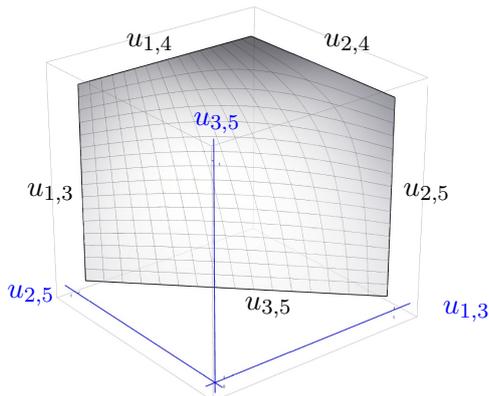

Let us consider some examples. For $n{=}4$ we have two variables with one constraint
\be
u_{1,3}=1-u_{2,4}
\ee
For $n{=}5$ we have five variables satisfying the constraint
\be\label{5ptconstraint}
u_{1,3}=1-u_{2,4}~u_{2,5}\,,
\ee
and four others related by cyclic shift; but only three constraints are independent, thus giving a 2-dimensional surface shown in Figure~\ref{5ptassws}. For $n{=}6$, there are two types of constraints corresponding to two types of diagonals of the hexagon. Here we present the constraints for the diagonals $(1,3)$ and $(1,4)$, and the rest are related via cyclic shift.
\be\label{6ptconstraint}
u_{1,3}=1-u_{2,4}~u_{2,5}~u_{2,6}\,\qquad u_{1,4}=1-u_{2,5}~u_{2,6}~u_{3,5}~u_{3,6}\,,
\ee
This gives $6+3=9$ constraints, but only six are independent.

The $u$-space provides an explicit compactification of the positive moduli space. To see this, we begin by constructing a map from the positive moduli space $\M_{0,n}^+$ to the interior of $u$-space via the following cross ratio formula:
\be\label{cross-ratio}
u_{i,j}=\frac{(\sigma_i-\sigma_{j{-}1})(\sigma_{i{-}1}-\sigma_j)}{(\sigma_i-\sigma_j)(\sigma_{i{-}1}-\sigma_{j{-}1})}
=\frac{(i\;j{-}1)(i{-}1\;j)}{(i\;j)(i{-}1\;j{-}1)}
\ee
which has already been studied extensively in the original dual resonance model ({\it c.f.}~\cite{Koba:1969kh}
and more recently in~\cite{Mafra:2011nw}). The map provides a diffeomorphism between the positive moduli space and the $u$-space interior. Taking the closure in $u$-space thereby provides the required compactification. Henceforth we denote $u$-space by $\oM_{0,n}^+$.

We now argue that the compactification $\oM_{0,n}^+$ is an associahedron. We begin by showing that there are exactly $n(n{-}3)/2$ codimension 1 boundaries given individually by $u_{i,j}=0$ for every diagonal $(i,j)$. We then show that every codimension 1 boundary ``factors'' like Eq.~\eqref{eq:fac_comb}, from which the desired conclusion follows.

Clearly the boundaries of the space are given by $u_{i,j}=0$ or $1$. However, if $u_{i,j}=1$ then by the non-crossing identity Eq.~\eqref{eq:non_cross} we must have $u_{k,l}=0$ for at least one diagonal $(k,l)\in (i,j)^c$. It therefore suffices to only consider $u_{i,j}=0$. We claim that every boundary $u_{i,j}=0$ ``factors'' geometrically into a product of lower-dimensional worldsheets:
\be\label{eq:fac_ws}
\partial_{(i,j)}\oM_{0,n}^+\cong \oM_{0,n_L}^+\times \oM_{0,n_R}^+
\ee
where
\ba
\oM_{0,n_L}^+&:=&\oM_{0,n_L}^+(i,\ldots, j{-}1,I)\\
\oM_{0,n_R}^+&:=&\oM_{0,n_R}^+(1,\ldots, i{-}1,I,j,\ldots,n)
\ea
with $I$ denoting an auxiliary label and $(n_L,n_R)=(j{-}i{+}1,n{+}i{-}j{+}1)$. Similar to the geometric factorization of the kinematic polytope discussed in Section~\ref{sec:fac}, we visualize the geometric factorization of the compactification as the diagonal $(i,j)$ that subdivides the convex $n$-gon into a ``left'' subpolygon and a ``right'' subpolygon as shown in Figure~\ref{fig:polygon_fac}. Furthermore, note that Eq.~\eqref{eq:fac_ws} immediately implies that the boundary is of dimension $(n-4)$ and hence codimension 1. From the $\sigma$-space point of view, the limit $u_{i,j}=0$ corresponds to the usual degeneration where the $\sigma_a$ for all $a=i,\ldots, j{-}1$ pinch together on the left subpolygon, and similarly the $\sigma_a$ for all $a=j,\ldots, n,1,\ldots, i{-}1$ pinch together on the right subpolygon.

To derive Eq.~\eqref{eq:fac_ws}, let $L,R$ denote the set of diagonals of the left and right subpolygons, respectively. Then in the limit $u_{i,j}=0$, we get $u_{k,l}=1$ for every diagonal $(k,l)$ that crosses $(i,j)$. It follows that the constraints Eq.~\eqref{eq:non_cross} split into two independent sets of constraints, one for each subpolygon:
\ba
\text{Left: }\qquad&&\left\{u_{k,l}=1-\prod_{(p,q)\in (k,l)^c\cap L} u_{p,q} \;\middle|\; (k,l)\in L\right\}\\
\text{Right: }\qquad&&\left\{u_{k,l}=1-\prod_{(p,q)\in (k,l)^c\cap R} u_{p,q} \;\middle|\; (k,l)\in R\right\}
\ea
These provide precisely the constraints for the left and right factors $\oM_{0,n_L}^+$ and $\oM_{0,n_R}^+$, thereby implying Eq.~\eqref{eq:fac_ws}. We conclude therefore that the compactified space $\oM_{0,n}^+$ is an associahedron. As an example, the $n{=}5$ worldsheet associahedron is shown in Figure~\ref{5ptassws}.


We now compute the canonical form. Since the worldsheet associahedron has the same boundary structure as the kinematic associahedron, therefore its canonical form should take on a similar form as Eq.~\eqref{eq:canon_assoc}. Indeed, let us work in the standard gauge $(\sigma_1,\sigma_{n{-}1},\sigma_n)=(0,1,\infty)$ where the moduli space interior is the simplex $0<\sigma_2<\sigma_3<\cdots<\sigma_{n{-}2}<1$. We now blow up the boundaries of the simplex to form an associahedron polytope, in the manner discussed earlier. We assume that our blowup is small of order $\epsilon$, with boundaries given by $B_{i,j}(\epsilon;\sigma)\geq 0$ corresponding to the diagonals $(i,j)$ of the $n$-gon. The exact expression for $B_{i,j}$ is not unique; however, since the boundary $(i,j)$ corresponds to the limit where $\sigma_{i},\sigma_{i{+}1},\ldots, \sigma_{j{-}1}$ pinch, it is thereby necessary that $\lim_{\epsilon\rightarrow 0} B_{i,j}(\epsilon,\sigma)=\sigma_{i,j{-}1}$. Now, we compute the canonical form by substituting $X_{i,j}\rightarrow B_{i,j}$ into Eq.~\eqref{eq:canon_assoc}, then removing the blowup by taking the limit $\epsilon\rightarrow 0$:
\ba\label{bigformWS}
\Omega\left(\oM_{0,n}^+\right)&=&\lim_{\epsilon\rightarrow 0}\sum_{{\rm planar}~g} {\rm sign}(g)~\bigwedge_{a=1}^{n{-}3} d\log B_{i_a,j_a}(\epsilon;\sigma)\\
&=&\sum_{{\rm planar}~g} {\rm sign}(g)~\bigwedge_{a=1}^{n{-}3} d\log \sigma_{i_a,j_a{-}1}
\ea
where we sum over all planar cubic graphs $g$, and for every $g$ the $(i_a,j_a)$ for $a=1,\ldots, n{-}3$ are the diagonals of the corresponding triangulation. The $\sign(g)$ is defined by the sign flip rule Eq.~\eqref{eq:sign_flip} as before. We caution the reader that the naive substitution $X_{i,j}\rightarrow u_{i,j}$ is incorrect; since the $u_{i,j}$ variables are constrained by non-linear equations ({\it i.e.} the non-crossing identities Eq.~\eqref{eq:non_cross}), hence there is no known dual polytope with boundaries $u_{i,j}\geq 0$ whose volume takes the form Eq.~\eqref{eq:canon_assoc}. 

Furthermore, since the $\epsilon\rightarrow 0$ limit reduces to a simplex, the canonical form must also reduce to the form for that simplex, which we recognize as the Parke-Taylor form Eq.~\eqref{PTform}:
\be\label{pullbackWS}
\Omega\left(\oM_{0,n}^+\right)=-\frac{d^{n{-}3}\sigma}{\sigma_2(\sigma_2{-}\sigma_3)\cdots(\sigma_{n{-}2}-1)}
\ee
While Eq.~\eqref{bigformWS} and Eq.~\eqref{pullbackWS} look very different, their equivalence is guaranteed by the geometric argument provided. In fact, the former can be thought of as a triangulation (with overlapping pieces that ``cancel'') of the latter. 

Finally, we present Eq.~\eqref{bigformWS} in a SL(2) invariant way:
\be
\Omega\left(\oM_{0,n}^+\right)=\sum_{{\rm planar}~g} {\rm sign}(g)~\bigwedge_{a=1}^{n{-}3} d\log \left(\frac{
\sigma_{i_a,j_a{-}1}\;\sigma_{1,n}\;\sigma_{n{-}1,n}}{
\sigma_{1,n{-}1}\;\sigma_{i_a,n}\;\sigma_{j_a{-}1,n}
}
\right)
\ee




\subsection{Scattering Equations as a Diffeomorphism Between Associahedra}

We have now seen two associahedra: the kinematic associahedron $\A_n$ in kinematic space $\mathcal{K}_n$ and the worldsheet associahedron $\oM_{0,n}^+$ in moduli space $\M_{0,n}$. Furthermore, recall that the scattering equations~\cite{Cachazo:2013gna} relate points in moduli space to points in kinematic space. It is therefore natural to expect that the same equations should relate the two associahedra. We begin by reinterpreting the scattering equations as a map from moduli space to kinematic space, giving the {\it scattering equation map}. We then make the striking observation that the scattering equation map acts as a {\it diffeomorphim} between the two associahedra.
\ba
\M_{0,n}\qquad \xrightarrow[\text{as a map}]{\text{scattering equations}}
\qquad
\mathcal{K}_n\\
\oM_{0,n}^+\qquad\xrightarrow[\text{as a diffeomorphism}]{\text{scattering equations}}
\qquad
\A_n
\ea
This has immediate consequences for amplitudes, including a novel derivation of the CHY formula for bi-adjoint scalars and much more. But before jumping ahead, let us establish the map.

Recall that the scattering equations~\cite{Cachazo:2013gna} read
\be
E_i:=\sum_{j=1;j\neq i}^n\frac{s_{ij}}{\sigma_{i,j}}=0\;\;\text{ for $i=1,\ldots, n$}
\ee
where $\sigma_{i,j}:=\sigma_i-\sigma_j$, and only $(n{-}3)$ equations are independent due to SL(2) redundancy. It is convenient to first send $\sigma_n \to \infty$ so that by adding all $E_1,E_2,...,E_c$ together we find
\be\label{twoparticle}
s_{c,c{+}1}=-\sum_{\substack{1\le i\le c \\ c+1\le j\le n-1\\ (i,j)\neq(c,c+1)}} \sigma_{c,c{+}1} \frac{s_{ij}}{\sigma_{i,j}}\,.
\ee
for the range $1\leq c\leq n{-}2$. 
Combining variables $s_{c,c{+}1}$ that have adjacent indices and variables $s_{ij}$ that have non-adjacent indices ({\it i.e.} $j{-}i>1$) gives us a formula for every planar variable $X_{a,b}$:
\be
X_{a,b}=-\sum_{\substack{1\leq i<a\\a<j<b}} \sigma_{a , j}~\frac{s_{ij}}{\sigma_{i,j}} - \sum_{\substack{a\leq i<b \\ b\leq j<n}} \sigma_{i , b{-}1}~\frac{s_{ij}}{\sigma_{i,j}} - \sum_{\substack{1\leq i<a \\ b\leq j<n}}\sigma_{a , b{-}1}~\frac{s_{ij}}{\sigma_{i,j}} \,,
\ee
whereby every index pair $i,j$ on the right hand side is non-adjacent with $i,j\neq n$. 
This provides a remarkable rewriting of the scattering equations because every Mandelstam variable on the right is a constant $s_{ij}=-c_{ij}$. Substituting the constants and recovering the SL(2) invariance by rewriting the $\sigma$ variables as cross-ratios of Pl\"{u}cker coordinates gives
\be\label{SE2}
X_{a,b}=\sum_{\substack{1\leq i<a\\a<j<b}}~\frac{(a\,j)(i\,n)}{(i\,j)(a\,n)}~c_{ij} + \sum_{\substack{a\leq i<b-1\\ b\leq j<n}}~\frac{(j\,n)(i\,b{-}1)}{(i\,j)(b{-}1\,n)}~c_{ij} + \sum_{\substack{1\leq i<a \\ b\leq j<n}}~\frac{(i\,n)(j\,n)(a\,b{-}1)}{(i\,j)(a\,n)(b{-}1\,n)}~c_{ij}\,.
\ee
Since the right hand side consists only of constants and $\sigma$ variables, this provides a map $\sigma\rightarrow X$ from moduli space to kinematic space (more specifically to the subspace $H_n$ when the $\sigma_i$ variables are real), thus providing the {\it scattering equation map} that we are after.

Let us look at the map more closely. First and foremost, every point $X_{a,b}$ on the image is manifestly positive when the $\sigma_{i}$ variables are ordered since the constants $c_{ij}>0$ are positive. It follows that Eq.~\eqref{SE2} maps the worldsheet associahedron $\oM_{0,n}^+$ into the kinematic associahedron $\A_n$.

Moreover, every boundary of the worldsheet associahedron (of any codimension) is mapped to the corresponding boundary of the kinematic associahedron. Indeed, consider a codimension 1 boundary $u_{a,b}\rightarrow 0$. In this limit, the variables $\sigma_a,\ldots, \sigma_{b{-}1}$ all pinch to a point so that $\sigma_{i,j}\rightarrow 0$ for all $a\leq i<j<b$. By direct inspection of Eq.~\eqref{SE2} we find that $X_{a,b}\rightarrow 0$ in this limit. It follows therefore that every boundary $u_{a,b}=0$ of the worldsheet associahedron is mapped to the corresponding boundary $X_{a,b}=0$ of the kinematic associahedron. An extended statement holds for boundaries of all codimensions. We say therefore that the scattering equation map preserves the associahedron boundary structure. Furthermore, this suggests that every point on the kinematic associahedron is reached by the map.

Finally, we make a numerical observation. For every point on the interior of the kinematic polytope, exactly one of the $(n{-}3)!$ solutions of the scattering equations lies on the interior of the worldsheet associahedron. In other words, provided that planar propagators $s_{a\cdots b{-}1}>0$ are positive and the non-adjacent constants $s_{ij}<0$ are negative, then there exists exactly one real ordered solution $\sigma_1<\cdots<\sigma_n$. We have checked this thoroughly up to $n{=}10$ for a substantial amount of data. Note that our kinematic inequalities are different from the ones introduced in~\cite{Cachazo:2016ror} where all solutions are {\it real}.

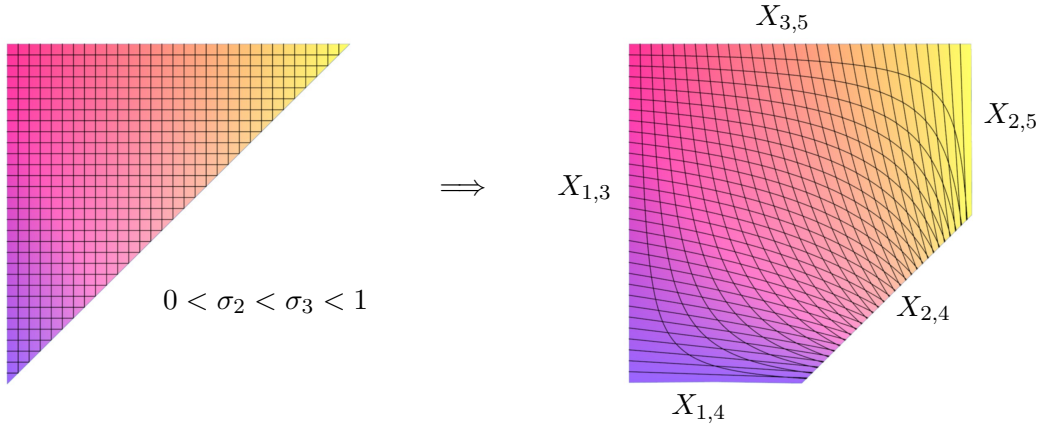
\begin{figure}
    \centering
    \begin{overpic}[width=5cm]
{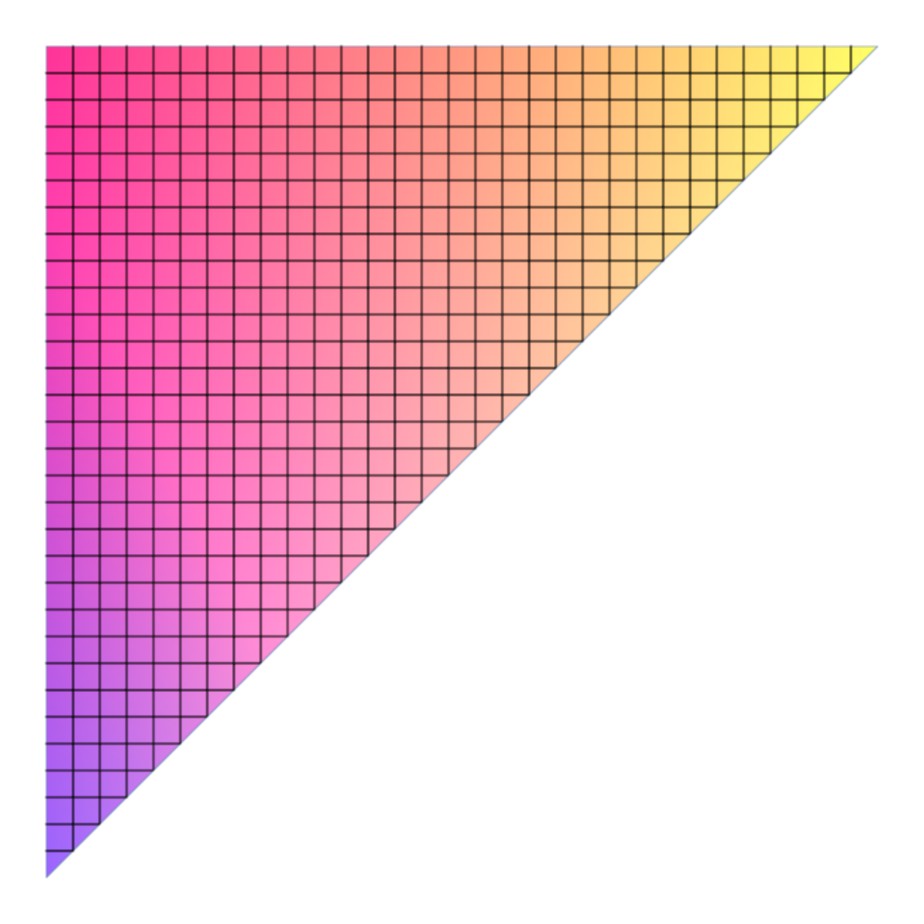}
\put(46,25){$0<\sigma_{2}<\sigma_3<1$}
\end{overpic}
    \qquad\qquad
    \qquad\qquad
    \begin{overpic}[width=5cm]
    {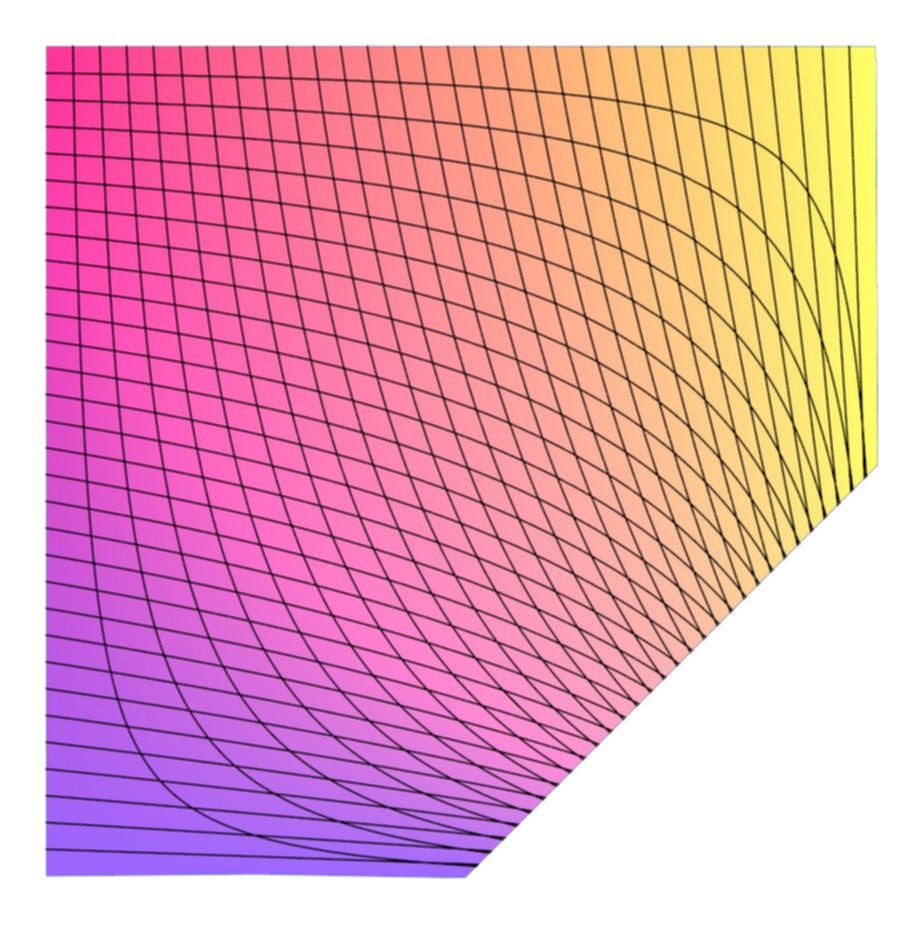}
    \put(-45,55){$\Longrightarrow$}
    \put(16,-3){$X_{1,4}$}
    \put(-14,55){$X_{1,3}$}
    \put(38,100){$X_{3,5}$}
    \put(98,74){$X_{2,5}$}
    \put(75,23){$X_{2,4}$}
    \put(-60, -15)

    \end{overpic}
    \newline
    \caption{Graphical evidence demonstrating that for $n{=}5$, the interior of the worldsheet associahedron is mapped diffeomorphically to the interior of the kinematic associahedron by the scattering equation map. Each contour line denotes a locus where one of $\sigma_2,\sigma_3$ is constant.}
    \label{fig:5pt_diffeo}
\end{figure}

We conjecture therefore that the scattering equation map is a {\it diffeomorphism} from the worldsheet associahedron to the kinematic associahedron. For $n{=}5$, the scattering equation map is given by
\ba\label{eq:5pt_map}
X_{1,3} &=& \frac{\sigma_2}{\sigma_3}(c_{13}+\sigma_3 c_{14})\\
X_{1,4} &=& \frac{1}{1-\sigma_2}((\sigma_3-\sigma_2)c_{24}+\sigma_3(1-\sigma_2)c_{14})
\ea
In Figure~\ref{fig:5pt_diffeo}, we present graphical evidence showing that these equations provide a diffeomorphism.

Diffeomorphisms play an important role in the theory of positive geometries and canonical forms. Recall from Appendix~\ref{app:pos_push} and more specifically Eq.~\eqref{eq:push_claim} that provided a diffeomorphism $\phi:\A\rightarrow\mathcal{B}$ between two positive geometries, the map pushes the canonical form of one to the other:
\ba
\A\qquad&\xrightarrow[]{\text{diffeomorphism } \phi}&\qquad\B\\
\Omega(\A)\qquad&\xrightarrow[]{\text{pushforward by } \phi}&\qquad\Omega(\B)
\ea
Applying this to our scenario, we find that the scattering equation map pushes the canonical form of the worldsheet associahedron to that of the kinematic associahedron.
\ba
\oM_{0,n}^+\qquad&\xrightarrow[\text{as diffeomorphism}]{\text{scattering equations}}&\qquad\A_n\\
\Omega\left(\oM_{0,n}^+\right)\qquad&\xrightarrow[\text{scattering equations}]{\text{pushforward by}}&\qquad \Omega(\A_n)
\ea
But Eq.~\eqref{pullbackWS} and Eq.~\eqref{eq:amp_form} imply
\be
\omega_n^{\rm WS}\qquad\xrightarrow[\text{scattering equations}]{\text{pushforward by}}\qquad m_n d^{n{-}3}X
\ee
It follows that the amplitude $m_n$ can be obtained by pushing forward the Parke-Taylor form via the scattering equations. Recalling the definition of the pushforward from Eq.~\eqref{eq:push_def}, we obtain the amplitude form by taking the Parke-Taylor form, substituting all roots of the scattering equations and summing over all roots.
\be\label{eq:push_n}
\sum_{{\rm sol.}\;\sigma} \omega_n^{\rm WS} = m_n d^{n{-}3}X
\ee
For a general ordering pair $\alpha,\beta$, this generalizes to the following statement
\be\label{eq:push_alpha_beta}
\sum_{{\rm sol.}\;\sigma} \omega_n^{\rm WS}[\alpha] = m[\alpha|\beta]d^{n{-}3}X
\ee
where $\omega_n^{\rm WS}[\alpha]$ denotes the Parke-Taylor form for the ordering $\alpha$, and the scattering equations are reinterpreted as a map $\M_{0,n}\rightarrow \K_n$ that restricts to a diffeomorphism $\oM_{0,n}^+[\alpha]\rightarrow \A[\alpha|\beta]$, where $\oM_{0,n}^+[\alpha]$ denotes the (compactified) $\alpha$-ordered part of the open string moduli space.

We caution the reader that the pullback of the right hand side in Eq.~\eqref{eq:push_n} does {\it not } produce the left hand side. Indeed, pulling back a canonical form does not necessarily produce another canonical form. For instance, pulling back $d\log y$ via $y=x^2$ gives $2d\log x$, which does not even have unit residue.

We observe that Eq.~\eqref{eq:push_alpha_beta} is reminiscent of the CHY formula for the bi-adjoint scalar. Indeed they are equivalent, as we now show. We begin by  rewriting our pushforward in delta function form:
\be\label{eq:m_push}
m_n=\int \omega_{n}(\sigma)\left[\prod_{a=1}^{n{-}3}\delta(X_{i_a,j_a}-\phi_a(\sigma))\right]
\ee
where the variables $X_{i_a,j_a}$ form a planar basis (corresponding to the diagonals of a triangulation), and $X_{i_a,j_a}=\phi_a(\sigma)$ is the scattering equation map Eq.~\eqref{SE2}. It is necessary that the basis variables appear with unit Jacobian in the delta functions, because $m_n$ is obtained from $\Omega(\A_n)$ by stripping away $\prod_{a=1}^{n{-}3}ds_{I_a}$. In other words, the delta functions must be normalized in the basis in which $m_n$ is obtained from $\Omega(\A_n)$. 

Now we claim that Eq.~\eqref{eq:m_push} is equivalent to the corresponding CHY formula:
\be\label{eq:CHY_push}
m_{n,{\rm CHY}}:=\int \omega_n^{\rm WS}[\sigma] \left[\frac{1}{\prod_{a=1}^n(\sigma_a-\sigma_{a{+}1})}{\prod_a}'\delta\left(\sum_{b\neq a}\frac{s_{ab}}{\sigma_a-\sigma_b}\right)\right]
\ee
Here we have deliberately isolated the Parke-Taylor form and grouped the other Parke-Taylor factor with the delta function. With a little bit of work, it can be shown that the square bracket expressions in Eq.~\eqref{eq:CHY_push} and Eq.~\eqref{eq:m_push} are equivalent. Thus, the second Parke-Taylor factor acts as a Jacobian factor for pushing forward onto the subspace $H_n$. More generally, a delta function dressed with an $\alpha$-ordered Parke-Taylor factor provides the pushforward onto the subspace $H[\alpha]$ defined in Eq.~\eqref{eq:H_beta} or equivalently $H[\alpha|\alpha]$ defined in Section~\ref{sec:all_orderings}. It follows that
\be
m_n=m_{n,\text{CHY}}
\ee
We have thus provided a novel derivation of the CHY formula for the bi-adjoint scalar. This derivation is purely geometric, and does not rely on the usual arguments involving factorization.

Finally, we make a brief comment about all ordering pairs. In Section~\ref{sec:all_orderings}, we obtained the partial amplitude $m[\alpha|\beta]$ from the pullback of the planar scattering form $\Omega^{(n{-}3)}[\alpha]$ to the subspace $H[\alpha|\beta]$. However, around Eq.~\eqref{eq:standard_pull_on_beta} we argue that the same amplitude can also be obtained by pulling back the same form to a different subspace $H[\beta]$. Hence, the amplitude can be expressed as the integral of the $\alpha$-ordered Parke-Taylor form over the delta function dressed with $\beta$-ordered Parke-Taylor factor, which is precisely the CHY formula. It follows that
\be
m[\alpha|\beta]=m_{\rm CHY}[\alpha|\beta]
\ee
for every ordering pair.

\section{``Big Kinematic'' Space and Scattering Forms}
So far we have considered scattering forms for amplitudes where there is some natural notion of an ordering, and with it, an associahedron geometry where an ordering is also important. In this section, we lay the groundwork for discussing scattering forms and positive geometries in much more general theories with no notion of ordering at all. Remarkably, this will be associated with a new ``projective'' understanding of color-kinematics relations, and as we will see in Section~\ref{sec:color}, even a geometrization of color itself!

In order to do this, we retrace our steps to the beginning, and think of kinematic space in a more fundamental way. Most treatments of the space of independent Mandelstam invariants simply posit that the natural variables are the $s_{ij}$ subject to the constraint $\sum_i s_{ij}=0$. Already in the case where we had a natural ordering, we found that this was not useful, and that a better set of independent variables---the planar variables $X_{a,b}$---was needed to expose the connection between physics and geometry. But why was this important? And how can we generalize to situations where we do not have an ordering?

A key realization is that there was never anything canonical about choosing $s_{ij}$ as co-ordinates for kinematic space---apart from being constrained, they are just a particular random collection of momentum dot products. On the other hand, something physical was gained by working with $X_{i,j}$ variables: the kinematic space is described by {\it physical propagators associated with cubic graphs} that directly encode all possible singularities of a local theory.

This motivates a new way of thinking about the kinematic space where the fundamental variables are {\it all} collections of possible propagators associated with cubic graphs. Of course as we will see this is a highly redundant set, and these objects satisfy certain relations. Nonetheless, we find an especially simple way of characterizing this space that makes the fundamental link between kinematics and color transparent.

We begin by constructing a higher dimensional {\it big kinematic space} $\K_n^*$ before reducing to the usual kinematic space $\K_n$ of Mandelstam variables which we henceforth refer to as {\it small kinematic space}. We find that the big space is important in its own right with connections to Jacobi relations. Furthermore, in Section~\ref{sec:scatter_form}, we discuss a large class of {\it scattering forms} beyond the planar scattering form of Section~\ref{sec:planar_scatter_form}, some of which have polytope interpretations and some have additional symmetries like permutation invariance. 

\subsection{The Big Kinematic Space}

We begin by constructing the {\it big kinematic space} $\K_n^*$. Consider a set of abstract variables $S_I$ indexed by all subsets $I\subset\{1,2,\ldots, n\}$ subject to two conditions,
\begin{itemize}
\item
$S_I=S_{\bar{I}}$ where $\bar{I}$ is the complement of $I$
\item
$S_I=0$ for $|I|=0,1,n{-}1,n$
\end{itemize}
For example, $\K_{n{=}4}^*$ is a 3-dimensional space spanned by the variables
\be
\{S_{12}=S_{34},\;\; S_{13}=S_{24},\;\; S_{14}=S_{23}\}
\ee
while $\K_{n{=}5}^*$ is a 10-dimensional space spanned by $S_{ij}$'s, and $\K_{n{=}6}^*$ is a 25-dimensional space spanned by 15 $S_{ij}$'s and 10 $S_{ijk}$'s. The dimension for general $n$ is given by
\be
\dim \K_n^* = 2^{n{-}1}{-}n{-}1
\ee
which for $n>3$ is higher than the dimension $n(n{-}3)/2$ of the small kinematic space $\K_n$. Nonetheless, the latter can be recovered by imposing a {\it 7-term identity} which we now describe.

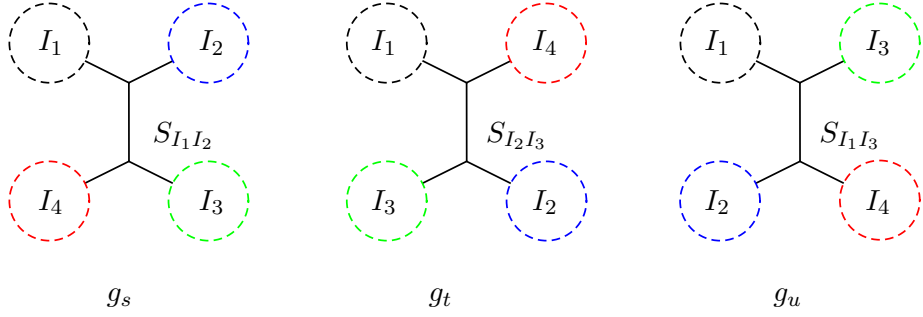
\begin{figure}
\begin{center}
\begin{overpic}[width=\linewidth]{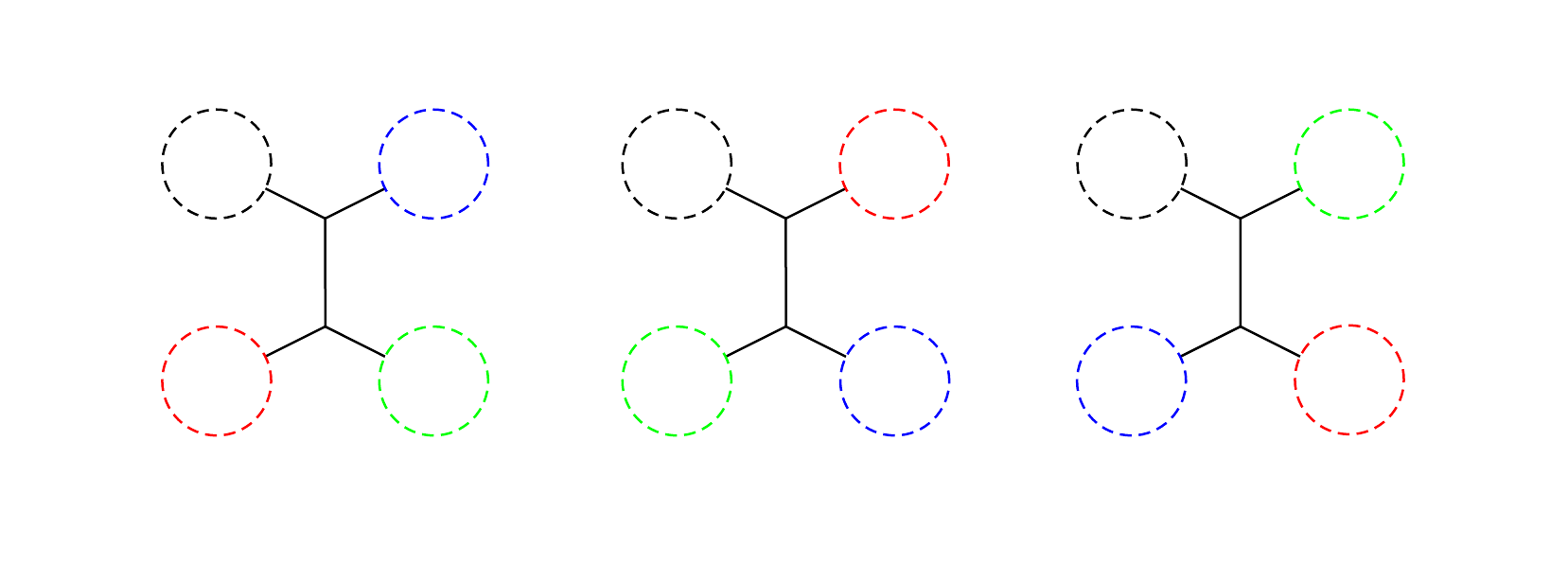}\label{ggst7}
\put(12,26){$I_1$}\put(26,26){$I_2$}\put(26,12){$I_3$}\put(12,12){$I_4$}
\put(41,26){$I_1$}\put(55,26){$I_4$}\put(55,12){$I_2$}\put(41,12){$I_3$}
\put(70,26){$I_1$}\put(84,26){$I_3$}\put(84,12){$I_4$}\put(70,12){$I_2$}

\put(22,18){$S_{I_1I_2}$}\put(51,18){$S_{I_2I_3}$}\put(80,18){$S_{I_1I_3}$}

\put(18,4){$g_s$}\put(46,4){$g_t$}\put(76,4){$g_u$}
\end{overpic}
\end{center}
\caption{A four set partition $I_1\sqcup I_2\sqcup I_3\sqcup I_4$ of the external labels and the three corresponding channels. The three graphs $g_s,g_t,g_u$ are identical except for a 4-point subgraph.}
\label{fig:4_pt_kinematics}
\end{figure}

For every partition of $n$ particles into four subsets 
\be
I_1 \sqcup I_2 \sqcup I_3 \sqcup I_4 = \{1,2,\cdots, n\}
\ee
we impose the following identity consisting of 7 terms (See Figure~\ref{fig:4_pt_kinematics}):
\be\label{7term}
S_{I_1I_2}+S_{I_2I_3}+S_{I_1I_3}=S_{I_1}+S_{I_2}+S_{I_3}+S_{I_4}
\ee
where $S_{IJ}:=S_{I\cup J}$. We can visualize this identity as a triplet of cubic graphs which are identical except for a four point subgraph, with the propagators on the left corresponding to the three channels of the subgraph, and the propagators on the right corresponding to the four legs of the subgraph. See Figure~\ref{fig:4_pt_kinematics} for an illustration. Moreover, recall that while Eq.~\eqref{7term} is usually presented as a derived property of 4-point kinematics, here we take a different point of view whereby the small kinematic space $\K_n$ is constructed by requiring Eq.~\eqref{7term} as an ``axiomatic identity'' from which the usual kinematic identities follow:
\be\label{7termresult}
S_I=\sum_{i<j;\;i,j\in I} S_{ij}\,\quad \text{for all } I; \quad \sum_{j=1; j\neq i}^n S_{ij}=0\, \quad \text{for all } i\,
\ee

We derive the first identity by induction on $m=|I|$, which is trivial for $m\leq 2$. Now assume that the assertion has been proven for $m<k$, and $|I|=k$ for some index set $I$. We first isolate two elements $a,b\in I$ and define $K:=I\backslash\{a,b\}$. Applying Eq.~\eqref{7term} to the partition $\bar{I}\sqcup K\sqcup \{a\}\sqcup \{b\}$ gives
\be
S_{ab}+S_{aK}+S_{bK}=S_K+S_{\bar{I}}
\ee
where we used $S_a=S_b=0$. It follows that
\be
S_{I}=S_{\bar{I}} = S_{ab}+S_{aK}+S_{bK}-S_K = \sum_{i<j;\;i,j\in I} S_{ij}
\ee
where for the last equality we applied the induction hypothesis to each of the four terms on the left hand side. This completes the derivation.

For the second identity in Eq.~\eqref{7termresult}, we apply the first identity to $\bar{I}$ for $I:=\{i\}$ which gives
\be\label{eq:all_but_one}
\sum_{a<b;\; a,b\neq i}S_{ab}=S_{\bar{I}}=S_I=0
\ee
Applying the first identity again to the full index set gives
\be\label{eq:all}
\sum_{a<b}S_{ab}=0
\ee
Subtracting Eq.~\eqref{eq:all_but_one} from Eq.~\eqref{eq:all} gives the desired result.

It follows therefore that the 7-term identity reduces the big kinematic space $\K_n^*$ to the small kinematic space $\K_n$, in which case the abstract variables can be identified with Mandelstam variables:
\be
S_I = s_I\;\;\;\text{for each $I$}
\ee
For some purposes, we find it useful to study geometries and differential forms directly in the big kinematic space prior to imposing the 7-term identity. 

\subsection{Scattering Forms and Projectivity}
\label{sec:scatter_form}

We introduce {\it scattering forms} as a generalization of the planar scattering forms from Section~\ref{sec:planar_scatter_form} to {\it all} cubic graphs. We then explore the implications of {\it projectivity} in this general framework and discover {\it Jacobi identities} for kinematic numerators as a direct consequence. 

Before defining the scattering forms, we establish the properties of cubic graphs from the point of view of the big space. Recall that a cubic graph $g$ consists of $(n{-}3)$ Mandelstam variables $s_{I_a}$ corresponding to the propagators of the graph. Then the corresponding big $S_{I_a}$ variables form a mutually compatible set, whereby any pair of variables $S_I$ and $S_J$ are said to be {\it compatible} if the index sets are either disjoint $I\cap J=\emptyset$ or one is contained in the other. Furthermore, we define an {\it ordered cubic graph} as a pair $(g|\alpha)$ consisting of a cubic graph $g$ and an ordering $\alpha$ for the external legs, assuming that $g$ is compatible with $\alpha$.

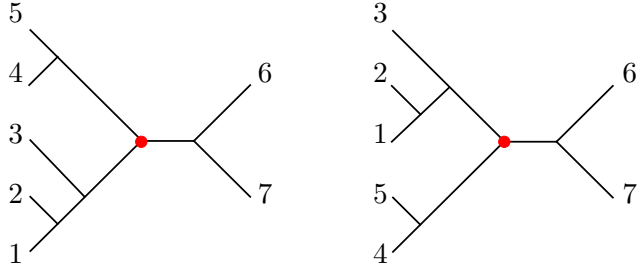
\begin{figure}
\begin{center}
\begin{overpic}[width=8cm]{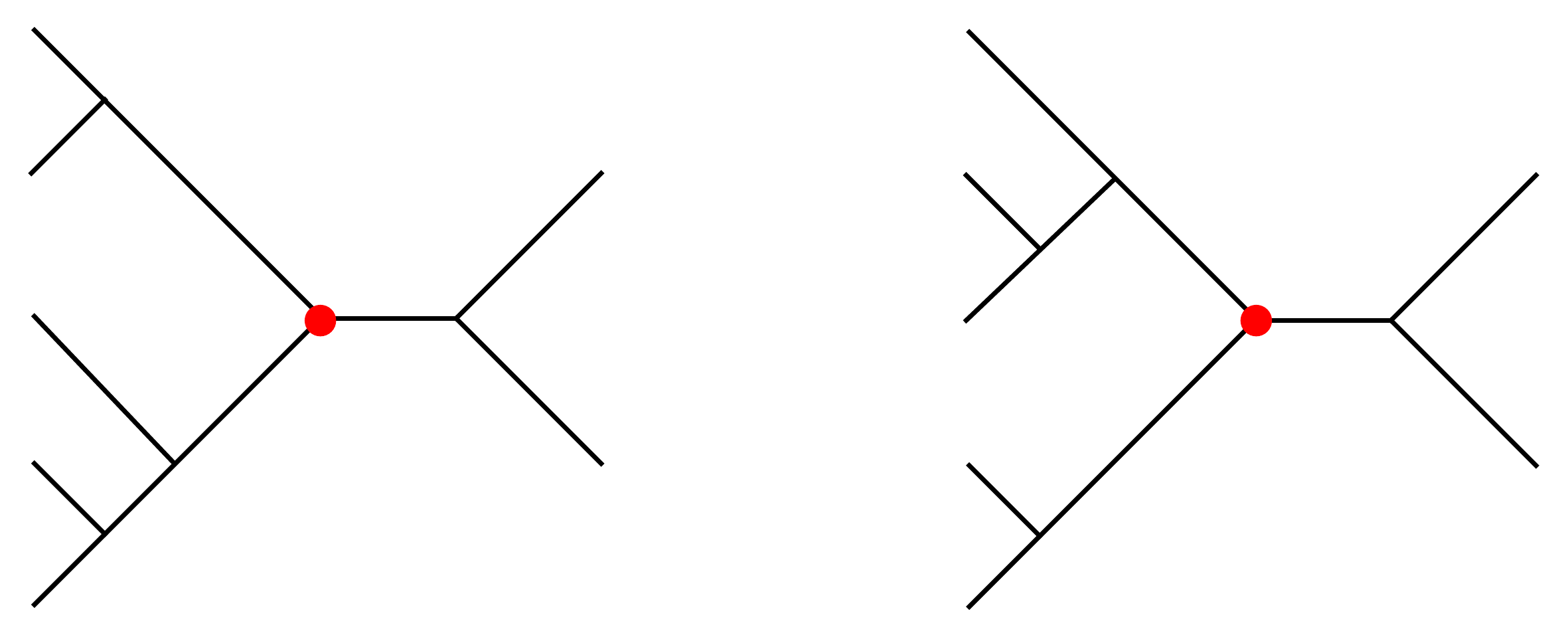}\label{YBCAFLIP}
\put(-3,0){$1$}
\put(-3,10){$2$}
\put(-3,20){$3$}
\put(-3,30){$4$}
\put(-3,40){$5$}
\put(38,10){$7$}
\put(38,30){$6$}
\put(57,0){$4$}
\put(57,10){$5$}
\put(57,20){$1$}
\put(57,30){$2$}
\put(57,40){$3$}
\put(98,10){$7$}
\put(98,30){$6$}
\end{overpic}
\end{center}
\caption{A vertex flip at the red vertex}
\label{fig:vertex_flip}
\end{figure}

For every ordered cubic graph $(g|\alpha)$ with propagators $S_{I_a}$, we define a $d\log$ form
\be\label{eq:Omega}
\Omega^{(n{-}3)}(g|\alpha):=\sign(g|\alpha)\bigwedge_{a=1}^{n{-}3}d\log S_{I_a}
\ee
where $\sign(g|\alpha_g)\in\{\pm 1\}$ depends not only on the ordered graph but also on the ordering of the propagators so that swapping two propagators changes the sign. The antisymmetry of the sign is of course compensated by the antisymmetry of the wedge product. Furthermore, we impose relations between the sign of different ordered cubic graphs via a {\it sign flip rule}. Recall that two graphs $g,g'$ with the same ordering $\alpha$ are related by a {\it mutation} if one can be obtained from the other by an exchange of channel in a 4-point subgraph like Figure~\ref{fig:mutation}. We assume that planarity in the ordering $\alpha$ is preserved by the mutation, so that only one mutation is possible in every 4-point subgraph of any cubic graph. Furthermore, we say that two orderings $\alpha,\alpha'$ for the same graph $g$ are related by a {\it vertex flip} if $(g|\alpha')$ can be obtained from $(g|\alpha)$ by exchanging two legs of a vertex (See Figure~\ref{fig:vertex_flip}). Finally, we define the sign flip rule by requiring a sign change for every mutation and every vertex flip.
\ba
\text{Mutation: }&\;\;\;&\sign(g|\alpha)=-\sign(g'|\alpha)\\
\text{Vertex flip: }&\;\;\;&\sign(g|\alpha)=-\sign(g|\alpha')
\ea
For a generic pair of ordered graphs $(g|\alpha), (g|\beta)$ related by a sequence of sign flips, let ${\rm flip}(\alpha,\beta)$ denote the number of flips involved (modulo 2) so that
\be
\sign(g|\alpha)=(-1)^{{\rm flip}(\alpha,\beta)}\sign(g|\beta)
\ee
If we restrict $\alpha$ to the standard ordering, then the vertex flip is irrelevant and we reduce to the sign flip rule for the planar scattering form Eq.~\eqref{eq:sign_flip}. More generally, we require the sign rule under vertex flip for any quantity $Q(g|\alpha)$ labeled by ordered cubic graphs. It follows that a product like $Q(g|\alpha)Q'(g|\alpha)$ of two such quantities is independent of the ordering and can therefore be written in a condensed form $Q(g)Q'(g)$. 

We now define the {\it scattering form} for $n$ particles as a rank $(n{-}3)$ form on $\K_n^*$ of the following form
\be\label{eq:scatter_form_S}
\Omega^{(n{-}3)}[N] = \sum_{\text{cubic}\;g}N(g|\alpha_g)\Omega^{(n{-}3)}(g|\alpha_g)
\ee
where we sum over all cubic graphs $g$, and to every cubic graph we assign an ordering $\alpha_g$ and a kinematic numerator $N(g|\alpha_g)$ which we assume to be independent of big $S$ variables. However, since every term is independent of the ordering $\alpha_g$, we can condense our notation as follows:
\be
\Omega^{(n{-}3)}[N] =\sum_{\text{cubic}\;g}N(g)\Omega^{(n{-}3)}(g)
\ee

Furthermore, we consider {\it projective scattering forms}, which are scattering forms that are invariant under {\it local ${\rm GL}(1)$ transformations} $S_I\rightarrow \Lambda(S)S_I$. This imposes constraints on the kinematic numerators which we now explain. Consider a triplet of cubic graphs $g_s,g_t,g_u$ like Figure~\ref{fig:4_pt_kinematics}. Under the transformation, the $\Lambda$-dependence of the scattering form becomes
\be\label{eq:triplet_sum}
\left(N(g_s|I_1I_2I_3I_4){+}N(g_t|I_1I_4I_2I_3){+}N(g_u|I_1I_3I_4I_2)\right)d\log\Lambda \wedge\left(\bigwedge_{b=1}^{n{-}4}d\log S_{J_b}\right){+}\cdots
\ee
where the $S_{J_b}$ denote the $(n{-}4)$ propagators shared by the triplet, and the $\cdots$ denotes similar expressions for all other triplets. Now, since the non-vanishing propagators are independent in the big kinematic space, therefore the $\Lambda$-dependence vanishes precisely if the coefficient of every triplet vanishes. This gives us $(2n{-}5)!!(n{-}3)/3$ identities (not all independent), one for each triplet, of the following form:
\be\label{Jacobi}
N(g_s|I_1I_2I_3I_4){+}N(g_t|I_1I_4I_2I_3){+}N(g_u|I_1I_3I_4I_2)=0
\ee
Note that we have explicitly written out the ordering for each graph which is important for making sure that the three terms add. We refer to Eq.~\eqref{Jacobi} as a {\it Jacobi identity} due to its similarity to the Jacobi identity for structure constants of Lie groups. It follows that the scattering form is projective if and only if its numerators satisfy Jacobi identities. 

We make a few comments before providing examples. Note that Eq.~\eqref{Jacobi} is derived without imposing the 7-term identity Eq.~\eqref{7term}. This is crucial, as imposing the identity would reduce us to the small kinematic space $\K_n$ where the set of all propagators no longer forms a basis (although the set of all {\it planar} propagators does), in which case we cannot require the coefficient of every triplet in Eq.~\eqref{eq:triplet_sum} to vanish. Furthermore, the ${\rm GL}(1)$ transformation does not act on the kinematic numerators, which may depend on usual kinematic quantities like $(p_i\cdot p_j)$, $(\epsilon_i\cdot p_j)$ and $(\epsilon_i\cdot \epsilon_j)$ that we assume to be independent of big $S$ variables. Nonetheless, we can define a similar local ${\rm GL}(1)$ transformation acting directly on the small space via $p_i\rightarrow \sqrt{\Lambda(p)}\;p_i$. It is straightforward then to show that ${\rm GL}(1)$ invariance in the big space directly implies {\it ${\rm GL}(1)$ covariance} in the small space, meaning $\Omega[N]^{(n{-}3)}(s)\rightarrow \Lambda^{D/2}\Omega^{(n{-}3)}[N](s)$ where $D$ is the mass dimension of the numerators.

Let us consider some examples of projective scattering forms. The simplest case is the {\it $\alpha$-planar scattering form}
\be\label{eq:planar_form_big}
\Omega_{\phi^3}^{(n{-}3)}[\alpha]=
\sum_{\alpha\text{-planar }g}\sign(g|\alpha)\bigwedge_{a=1}^{n{-}3}d\log S_{I_a}
\ee
where we sum over all cubic graphs $g$ compatible with the ordering $\alpha$. For the standard ordering this reduces to Eq.~\eqref{eq:planar_form} in the small kinematic space. In this case, the kinematic numerator $N(g|\alpha)$ vanishes for any graph incompatible with $\alpha$, and is $\pm 1$ otherwise. More specifically, for every triplet, either none of the three graphs is compatible, or exactly two are. For instance, if the first two of the triplet $g_s,g_t,g_u$ are compatible, then
\be
N(g_s|I_1I_2I_3I_4)=\pm 1\qquad N(g_t|I_1I_4I_2I_3)=\mp 1\qquad N(g_u|I_1I_3I_4I_2)=0
\ee

One way to generalize the planar scattering form without introducing any additional structures such as spin or color is to drop the planarity requirement and consider all projective scattering forms whose numerators are $0,\pm 1$. This provides a large class of scattering forms called {$d\log$ scattering forms} of which the planar case is only one. Furthermore, as the planar form is closely tied to the geometry of the associahedron, many of these other forms are also closely tied to polytopes of their own such as the permutohedron. We provide more details on this topic in the Outlook.

Furthermore, we point out that while planar forms have cyclic symmetry, it is also possible to construct projective forms with {\it permutation symmetry}. As will be discussed in the next section, such scattering forms can be obtained from {\it color-dressed amplitudes} that are permutation invariant, via an important connection between differential forms and color.  These include scattering forms for theories like Yang-Mills and Non-linear Sigma Model, which we discuss in more detail in Section~\ref{sec:YMNLSM}. 

Last but not least, we state an important property for any projective scattering form. Since planar scattering forms are projective, it follows that every linear combination of them is also projective:
\be
\Omega^{(n{-}3)}[C]=\sum_{\alpha\in S_n/Z_n} C(\alpha)\Omega^{(n{-}3)}_{\phi^3}[\alpha]
\ee
where the $C(\alpha)$ coefficients are independent of big $S$ variables. Remarkably, the converse is also true, {\it i.e.} every projective scattering form is a linear combination of planar scattering forms. We give a detailed derivation in Appendix~\ref{app:bcj}, and the upshot is that any projective scattering form can be expanded in terms of a basis of $(n{-}2)!$ planar forms,
\be\label{eq:expand_planar}
\Omega^{(n{-}3)}[C']=\sum_{\pi\in S_{n{-}2}}C'(\pi)\;\Omega_{\phi^3}^{(n{-}3)}[1,\pi(2),\ldots, \pi(n{-}1),n]\,.
\ee

\section{Color is Kinematics}
\label{sec:color}

\begin{figure}
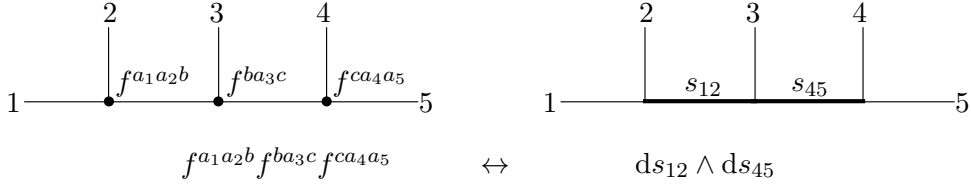

\centering
\begin{overpic}[width=5.4cm]
{CA5ptL.pdf}
\put(-2.5,20){$1$}
\put(21,42){$2$}
\put(47,42){$3$}
\put(73,42){$4$}
\put(98,20){$5$}
\put(24,25){$f^{a_1a_2b}$}
\put(51,25){$f^{b a_3 c}$}
\put(77,25){$f^{c a_4 a_5}$}
\put(40,5){$f^{a_1 a_2 b}f^{b a_3 c}f^{c a_4 a_5}$}
\put(113,5){$\leftrightarrow$}
\end{overpic}
\qquad\qquad
\begin{overpic}[width=5.4cm]
{CA5ptR.pdf}
\put(-2.5,20){$1$}
\put(21,42){$2$}
\put(47,42){$3$}
\put(73,42){$4$}
\put(98,20){$5$}
\put(32,25){$s_{12}$}
\put(58,25){$s_{45}$}
\put(20,5){$\mathrm{d}s_{12}\wedge\mathrm{d}s_{45}$}
\end{overpic}
\caption{An example of the duality between color factors and differential forms}
\label{fig:dualex}
\end{figure}
In the last section we have seen a striking relationship between {\it projective scattering forms} in the big kinematic space, and the {\it color-kinematics} connection for numerator factors. But this is just half of the story. In this section we see another related but distinct relationship between color and kinematics. Indeed we have become accustomed to speaking of ``color-kinematics duality'', but this relationship is even more basic from the scattering form point of view so that in a precise sense, ``Color is Kinematics!'' Temporarily ignoring the correct assignment of signs, the basic observation is extremely simple: any scattering form involves a sum over cubic graphs in kinematic space, and these all have a factor that is the wedge product of all the $d s$’s associated with the propagators. But quite beautifully, as a consequence of the 7-term identity Eq.~\eqref{7term}, these wedge-product factors associated with any cubic graph satisfy exactly the same Jacobi identities as the color factors associated with the same graphs! This leads naturally to a duality between color factors and differential forms, as suggested by Figure~\ref{fig:dualex}. We will see that this ``Color is Kinematics'' relation goes even deeper, with the {\it trace decomposition} of color factors directly equivalent to {\it subspace pullbacks} of the form. This connection allows us to geometrize color, and makes it possible to speak of colored theories, such as Yang-Mills theories and the Non-linear sigma model, purely in terms of scattering forms which can be freely exchanged for explicit color factors.

\subsection{Duality Between Color and Form}
\label{sec:dualitycolor}
We establish the {\it duality between color factors and  differential forms on kinematic space $\K_n$} by showing that the latter satisfy {\it Jacobi relations} similar to the usual Jacobi relations for structure constants. This leads naturally to a {\it duality between color-dressed amplitudes and scattering forms}.

We begin by reviewing the algebra of color. Given an ordered graph we define a color factor $C(g|\alpha)$ by first drawing $g$ as a planar graph whose external legs are ordered clockwise by $\alpha$ (See Figure~\ref{fig:dualex} (left)). Then, for each internal and external line we assign an index, and for each vertex $v$ we assign a structure constant $f^{a_vb_vc_v}$, where the indices $a_v,b_v,c_v$ correspond to the three adjacent lines in clockwise order. Finally, we obtain the color factor by multiplying the structure constants and contracting repeated indices (which occur along internal lines). Hence,
\be
C(g|\alpha)=\prod_v f^{a_vb_vc_v}
\ee
where index contraction is implicitly assumed. The antisymmetry of the structure constants implies the vertex flip sign rule Eq.~\eqref{eq:color_flip} while the usual Jacobi identities for the structure constants imply Jacobi identities for the color factors Eq.~\eqref{eq:color_jacobi} for any triple like Figure~\ref{fig:4_pt_kinematics}:
\be\label{eq:color_flip}
C(g|\alpha)=(-1)^{{\rm flip}(\alpha,\beta)}C(g|\beta)
\ee
\be
\label{eq:color_jacobi}
C(g_s|I_1I_2I_3I_4){+}C(g_t|I_1I_4I_2I_3){+}C(g_u|I_1I_3I_4I_2)=0
\ee

We now argue that a similar set of identities hold for differential forms on the kinematic space $\K_n$. For every ordered graph $(g|\alpha)$ with propagators $s_{I_a}$ for $a=1,\ldots, n{-}3$, we define the $(n{-}3)$-form
\be\label{eq:W_def}
W(g|\alpha)=\sign(g|\alpha)\bigwedge_{a=1}^{n{-}3}ds_{I_a}
\ee
We claim that the form satisfies the vertex flip sign rule Eq.~\eqref{eq:W_flip} and the Jacobi identity Eq.~\eqref{eq:W_jacobi} in perfect analogy with color factors.
\be\label{eq:W_flip}
W(g|\alpha)=(-1)^{{\rm flip}(\alpha|\beta)}W(g|\beta)
\ee
\be
\label{eq:W_jacobi}
W(g_s|I_1I_2I_3I_4){+}W(g_t|I_1I_4I_2I_3){+}W(g_u|I_1I_3I_4I_2)=0
\ee
The former follows from the $\sign(g|\alpha)$ factor in Eq.~\eqref{eq:W_def}. The latter follows from applying the 7-term identity Eq.~\eqref{7term} to the triplet $g_s,g_t,g_u$ from Figure~\ref{fig:4_pt_kinematics}:
\be
ds_{I_1I_2}+ds_{I_2I_3}+ds_{I_1I_3}=ds_{I_1}+ds_{I_2}+ds_{I_3}+ds_{I_4}
\ee
Note that on the left the propagators correspond to the three channels of the triplet, while on the right the propagators correspond to the legs of the 4-point subgraph. Moreover, let $s_{J_b}$ for $b=1,\ldots, n{-}4$ denote the propagators shared by the triplet. It follows that
\be
(ds_{I_1I_2}+ds_{I_2I_3}+ds_{I_1I_3})\wedge\bigwedge_{b=1}^{n{-}4}ds_{J_b}=0
\ee
where every term on the right hand side has vanished. In particular, external legs vanish by on-shell condition while internal legs vanish since they already appear in the product $\wedge_{b=1}^{n{-}4}ds_{J_b}$. The result is precisely the sought after Jacobi relation.

This implies a {\it duality between color factors and differential forms on kinematic space $\K_n$}:
\be\label{eq:color_dual}
C(g|\alpha)\qquad\leftrightarrow \qquad W(g|\alpha)
\ee
Hence ``Color is Kinematics''. We emphasize that the 7-term identity is absolutely crucial for this property to hold, thus providing one of the motivations for constructing kinematic space $\K_n$ by the 7-term identity directly.

We provide some examples for low $n$. For $n{=}4$, there are three color factors dual to 1-forms:
\ba
C_s=f^{a_1a_2b}f^{ba_3a_4}\qquad&\leftrightarrow&\qquad ds\nonumber\\
C_t=f^{a_1a_4b}f^{ba_2a_3}\qquad&\leftrightarrow&\qquad dt\nonumber\\
C_u=f^{a_1a_3b}f^{ba_4a_2}\qquad&\leftrightarrow&\qquad du\nonumber
\ea
For $n{=}5$, color factors are dual to 2-forms. Here we provide one example as illustrated in Figure~\ref{fig:dualex}.

Furthermore, the duality Eq.~\eqref{eq:color_dual} leads naturally to a {\it duality between color-dressed amplitudes and scattering forms}. Consider a colored-dressed amplitude ${\bf M}_n[N]$ with kinematic numerators $N$:
\be\label{eq:amp_M}
{\bf M}_n[N]=\sum_{\text{cubic }g}N(g|\alpha_g)C(g|\alpha_g)\prod_{I\in g}\frac{1}{s_I}
\ee
where we sum over all cubic graphs $g$, and $s_I$ for $I\in g$ denote the propagators in the graph. We now map this amplitude to a form on kinematic space by applying Eq.~\eqref{eq:color_dual} to each color factor individually, giving
\be\label{eq:scatter_form_s}
\Omega^{(n{-}3)}[N]=\sum_{\text{cubic }g}N(g|\alpha_g)W(g|\alpha_g)\prod_{I\in g}\frac{1}{s_I}
\ee
which we recognize as a scattering form Eq.~\eqref{eq:scatter_form_S} with $S_I\rightarrow s_I$. Likewise, we can return to the amplitude Eq.~\eqref{eq:amp_M} by applying Eq.~\eqref{eq:color_dual} backwards. Thus, the duality Eq.~\eqref{eq:color_dual} implies the duality Eq.~\eqref{eq:color_dual_amp}
\be\label{eq:color_dual_amp}
{\bf M}_n[N]\qquad\leftrightarrow\qquad \Omega^{(n{-}3)}[N]
\ee
We henceforth refer to both dualities as {\it color-form} duality. Note that for any permutation-invariant color-dressed amplitude, Eq.~\eqref{eq:color_dual_amp} gives a scattering form that is nicely permutation invariant. Furthermore, we comment on the role of projectivity. Recall that the numerators $N(g)$ satisfy Jacobi relations provided that the scattering form ${\Omega}^{(n{-}3)}[N](s)$ is derived from a projective form in the big kinematic space. The dual amplitude ${\bf M}_n[N]$ therefore admits an expansion with $N(g)$ as {\it BCJ numerators}, first proposed by Bern, Carrasco and Johansson in~\cite{Bern:2008qj}.

For the special case of bi-adjoint scalar with double color group ${\rm SU}(N)\times {\rm SU}(N)$. The scattering form is obtained by simply choosing $N(g)=C(g)$ for every graph $g$:
\be\label{eq:scalar_form}
\Omega^{(n{-}3)}_{\phi^3} = \sum_{\text{cubic }g} C(g|\alpha_g)\;\sign(g|\alpha_g)\bigwedge_{a=1}^{n{-}3} d\log s_{I_a}
\ee
which is both permutation invariant and projective. The corresponding double color-dressed amplitude is given by
\be
{\bf M}_{\phi^3,n}=\sum_{\text{cubic }g}\frac{C(g)\tilde{C}(g)}{\prod_{I\in g} s_I}
\ee

\subsection{Trace Decomposition as Pullback of Scattering Forms}
\label{sec:trace}
We explore color-form duality further by examining {\it partial amplitudes} and their interpretation from the differential form point of view. We find that {\it trace decomposition of color-dressed amplitudes are dual to pullbacks of the scattering form} to appropriate subspaces of dimension $(n{-}3)$.

Recall that for the color groups ${\rm U}(N)$ and ${\rm SU}(N)$, the color factors can be decomposed as traces from which partial amplitudes are obtained. More precisely, we have
\be\label{eq:C_trace}
C(g|\alpha) = \sum_{\beta\in O(g)/Z_n}(-1)^{{\rm flip}(\alpha,\beta)}{\rm Tr}({\beta(1)},\ldots, {\beta(n)})
\ee
where $O(g)/Z_n$ denotes all $2^{n{-}2}$ orderings compatible with the graph $g$ modulo cyclic transformations. 
In other words, out of all $(n{-}1)!$ distinct trace terms, the color factor $C(g|\alpha)$ is expanded precisely in terms of those traces whose ordering is compatible with the graph. 

Substituting Eq.~\eqref{eq:C_trace} into Eq.~\eqref{eq:amp_M} for every graph $g$ gives us the trace decomposition for the amplitude:
\be
{\bf M}_n[N]=\sum_{\beta\in S_n/Z_n}{\rm Tr}(\beta(1),\ldots, \beta(n))M_n[N;\beta]
\ee
where the {\it partial amplitude} $M_n[N;\beta]$ is given by a sum over $\beta$-planar graphs:
\be\label{eq:partial}
M_n[N;\beta] = \sum_{\beta-\text{planar }g}N(g|\beta)\prod_{I\in g}\frac{1}{s_I}
\ee
As an example, for $n{=}4$, the color factors decompose as
\ba
C_s{=}{\rm Tr}(1234)-{\rm Tr}(2134)-{\rm Tr}(1243)+{\rm Tr}(2143)\\
C_t{=}{\rm Tr}(1423)-{\rm Tr}(4123)-{\rm Tr}(1432)+{\rm Tr}(4132)\\
C_u{=}{\rm Tr}(1342)-{\rm Tr}(3142)-{\rm Tr}(1324)+{\rm Tr}(3124)
\ea
where both the $s$ and $t$ channels contribute to the ordering $\beta=(1234)$, thus giving
\be
M_4[N; 1234] = \frac{N(s|1234)}{s}+\frac{N(t|1234)}{t}
\ee

We now argue that the partial amplitude Eq.~\eqref{eq:partial} can be obtained by pulling back the scattering form Eq.~\eqref{eq:scatter_form_s} to an $(n{-}3)$-dimensional subspace $H[\beta]$ which we define by imposing $(n{-}2)(n{-}3)/2$ independent conditions:
\be\label{eq:H_beta}
H[\beta]:=\left\{s_{\beta(i)\beta(j)}\;\text{ is constant}\;\;|\;\; 1\leq i<j{-}1\leq n{-}2\right\}
\ee
This coincides with the subspace $H_n$ define in Eq.~\eqref{eq:cXXXX} if $\beta$ is the standard ordering and the constants $s_{\beta(i)\beta(j)}$ are negative. Now for any graph $g$ compatible with $\beta$, we define the pullback
\be\label{eq:def_V}
dV[\beta]:=W(g|\beta)|_{H[\beta]}
\ee
which is independent of the graph as shown around Eq.~\eqref{eq:dX} for the standard ordering. More generally, for a pair of orderings $\alpha,\beta$, we have
\be
W(g|\alpha)|_{H[\beta]}=
\begin{cases}
(-1)^{{\rm flips}(\alpha,\beta)}dV[\beta] & \text{if $g$ is compatible with $\beta$}\\
0 & \text{otherwise}
\end{cases}
\ee
where the first line follows immediately from the definition Eq.~\eqref{eq:def_V}, while the second line requires a proof for which we provide a sketch. Our strategy is to argue by induction on the number of particles, beginning with $n{=}4$ which can be verified directly. For higher $n$, suppose $g$ is a cubic graph that is compatible with $\alpha$ but not with $\beta$, and for simplicity let us assume that $\beta$ is the standard ordering. We observe that the graph must consist of at least one propagator of the form $s_{ij}$ where $i<j$ and $i,j \neq n$. If $i,j$ are non-adjacent, then $ds_{ij}=0$ on the pullback, and we are done. Otherwise, the propagator must be $s_{i,i+1}$, giving $W(g)=ds_{i,i+1}\wedge W'(g)$ for some form $W'(g)$. Since factors of $ds_{i,i+1}$ within $W'(g)$ do not contribute, we can therefore think of $W'(g)$ as the form for a reduced graph $g'$ obtained from $g$ by collapsing particles $i$ and $i{+}1$ into a single particle. But $W'(g)$ vanishes by induction, thus completing the argument. One subtlety of the last step is that the particle $(i,i+1)$ is generically off-shell with mass-squared given by $s_{i,i+1}$, which appears to violate the induction hypothesis. But since factors of $ds_{i,i+1}$ are effectively zero, the induction still holds. It follows that the pullback of the scattering form $\Omega^{(n{-}3)}[N]$ to the subspace $H[\beta]$ gives the partial amplitude $M_n[N;\beta]$:
\be
\Omega^{(n{-}3)}[N]|_{H[\beta]}=\left(\sum_{\beta\text{-planar }g}N(g|\beta)\prod_{I\in g}\frac{1}{s_I}\right)dV[\beta]=M_n[N;\beta]dV[\beta]
\ee

Applying this to the planar scattering form $\Omega_{\phi^3}^{(n{-}3)}[\alpha]$ for the bi-adjoint scalar gives us the double partial amplitude $m[\alpha|\beta]$:
\be\label{eq:standard_pull_on_beta}
\Omega_{\phi^3}[\alpha]|_{H[\beta]}=(-1)^{{\rm flip}(\alpha,\beta)}m[\alpha|\beta]dV[\beta]
\ee
This is very different from Eq.~\eqref{eq:pullback_alpha_beta} where the same amplitude was obtained by pulling back to a different subspace $H[\alpha|\beta]$. The advantage of the latter is that it provides a geometric interpretation for the amplitude (form) as the canonical form of a positive geometry as in Eq.~\eqref{eq:canon_alpha_beta}. The former, however, can be applied to trace decompose any colored tree amplitude.

\begin{figure}
\centering
\begin{overpic}[width=7cm]
{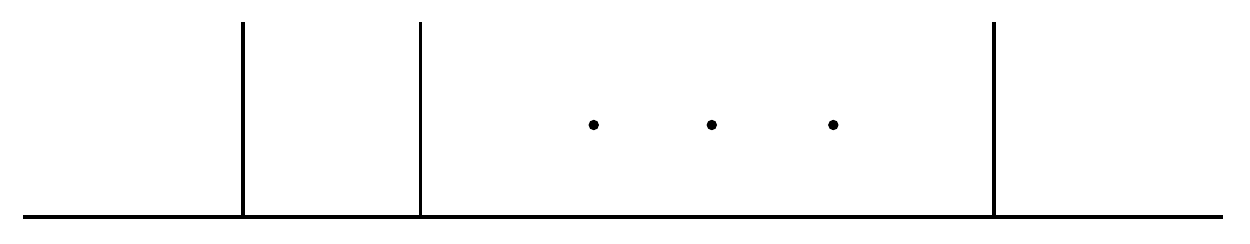}
\put(-2,0){$1$}
\put(100,0){$n$}
\put(13,20){$\pi(2)$}
\put(28,20){$\pi(3)$}
\put(72,20){$\pi(n-1)$}
\end{overpic}
\caption{Multi-peripheral graph with respect to $1$ and $n$ for the ordering $\pi\in S_{n-2}$}\label{mtpphgrphpi}
\end{figure}
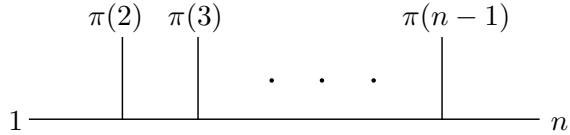

Finally, we discuss the role of some well-known amplitude relations. Recall the decomposition a la Del Duca, Dixon and Maltoni (DDM)~\cite{DelDuca:1999rs} given in Eq.~\eqref{eq:DDM_C}, where $g_\pi$ denotes the multi-peripheral graph with respect to $1$ and $n$ for the ordering $\pi\in S_{n{-}2}$ as shown in Figure~\ref{mtpphgrphpi}. 
\be\label{eq:DDM_C}
{\bf M}_n[N] = \sum_{\pi\in S_{n{-}2}}C(g_\pi|\pi)\;M_n[N;1,\pi(2),\ldots, \pi(n{-}1),n]
\ee
It follows that the color-dressed amplitude can be expanded in terms of only $(n{-}2)!$ partial amplitudes of the form given in Eq.~\eqref{eq:DDM_C}, which is more efficient than the $(n{-}1)!$-term expansion of the standard trace decomposition. This also follows from the Kleiss-Kuijf (KK)~\cite{Kleiss:1988ne} relations. Furthermore, applying the color-form duality to Eq.~\eqref{eq:DDM_C} gives an analogous identity for the scattering form
\be\label{eq:DDM_form}
\Omega^{(n{-}3)}[N] = \sum_{\pi\in S_{n{-}2}}W(g_\pi|\pi)\;M_n[N;1,\pi(2),\ldots, \pi(n{-}1),n]
\ee
Note that the expansion is unique both for the color-dressed amplitude and for the form, since the multi-peripheral graphs $g_\pi$ form a basis. Furthermore, we find that Bern-Carrasco-Johansson (BCJ) relations~\cite{Bern:2008qj} follow from requiring the scattering form to be projective, as shown in Appendix~\ref{app:bcj}.

Last but not least, as we have discussed around Eq.~\eqref{eq:expand_planar}, every projective form can be expanded in a basis of $(n{-}1)!$ planar scattering forms labeled by the orderings $\pi\in S_{n{-}2}$, and now we can spell out the coefficients. As shown in Appendix~\ref{app:bcj}, the coefficient for the $\pi$ term is nothing but the kinematic numerator $N(g_\pi|\pi)$:
\be\label{eq:dualbasis}
\Omega^{(n{-}3)}[N]=\sum_{\pi\in S_{n{-}2}} N (g_\pi|\pi)~\Omega^{(n{-}3)}_{\phi^3} (1, \pi(2), \cdots, \pi(n{-}1),n)\,.
\ee
Note that Eq.~\eqref{eq:DDM_form} and Eq.~\eqref{eq:dualbasis} are complementary to each other. By the color-form duality, the latter is equivalent to the well-known {\it dual-basis expansion}~\cite{Bern:2011ia} of the color-dressed amplitude:
\be
{\bf M}_n[N]=\sum_{\pi\in S_{n{-}2}} N (g_\pi|\pi) M_{n}^{\phi^3} [1,\pi(2),\ldots, \pi(n{-}1),n]
\ee

\section{Scattering Forms for Gluons and Pions}
\label{sec:YMNLSM}
There are two prime examples of permutation invariant forms on kinematic space: the scattering forms associated with the scattering of gluons in Yang-Mills theory, and of pions in the Non-linear Sigma Model. Let us stress again the central novelty of this claim: there is a differential form on the kinematic space, with coefficients that depend on either momenta and polarization vectors (for Yang-Mills) or Mandelstam variables (for the NLSM), which are fully permutation invariant with no $f^{abc}$ factors anywhere in sight. Nonetheless, the geometrization of color discussed in the previous sections tells us that these forms contain all the information about color-dressed amplitudes. 

In fact more is true: the scattering forms for gluons and pions are remarkably rigid objects. For gluons, we find that there is a  {\it unique} differential form with the usual minimal power-counting in momenta that is both gauge invariant and projectively invariant. In particular, the permutation invariance need not be stipulated but is derived. Similarly, the form for pions is the unique form where the requirement of gauge invariance for each leg is replaced with that of the Adler zero in the soft limit. Such ``uniqueness theorems'' have recently been established in \cite{Arkani-Hamed:2016rak}, for partial amplitudes from which the uniqueness of the full scattering form follows, provided the crucial extra requirement of projectivity. We also show that these forms have a natural pushforward origin from the worldsheet. 

\subsection{Gauge Invariance, Adler Zero, and Uniqueness of YM and NLSM Forms}
\label{sec:unique}
We establish general conditions under which scattering forms for gluons and (two-derivative-couple, massless) pions are unique. Consider general scattering forms $\Omega^{(n{-}3)}_{\rm gluon}$ and $\Omega^{(n{-}3)}_{\rm pion}$ for pure gluons and pure pions, respectively. For the gluons, we require the kinematic numerators to consist of contractions in momenta $p_i^\mu$ and polarizations $\epsilon_i^\mu$ with each polarization appearing exactly once; moreover we require the expected power counting, which suggests in particular that there can be no more than $(n{-}2)$ contractions like $(\epsilon_i\cdot p_j)$ in any term; finally, we require gauge invariance ({\it i.e.} invariance under the shift $\epsilon_i^\mu\rightarrow \epsilon_i^\mu+\alpha p_i^\mu$). For the pions, we require the numerators to be polynomials of Mandelstam variables with the right power counting ({\it i.e.} with degree $(n{-}2)$ in Mandelstams), and the Adler zero condition ({\it i.e}. vanishing under every soft limit $p_i^\mu\rightarrow 0$). Finally, we assume that the forms are projective. We claim that in both cases, the scattering form is unique up to an overall constant.

To derive these two claims, we decompose the scattering forms a la DDM Eq.~\eqref{eq:DDM_form}, and denote the partial amplitudes for the ordering $\pi\in S_{n{-}2}$ as $M^{\rm gluon}_{n}(\pi)$ and $M^{\rm pion}_{n}(\pi)$, respectively, which are given by Eq.~\eqref{eq:partial} with appropriate numerators. Given the linear-independence of the $W(g_\pi|\pi)$ factors, it is clear that each gluon partial amplitude inherits gauge invariance from $\Omega^{(n{-}3)}_{\rm gluon}$ while each pion partial amplitude inherits Adler zero from $\Omega^{(n{-}3)}_{\rm pion}$. However, the main result of~\cite{Arkani-Hamed:2016rak} states that any expression satisfying the assumptions of $M^{\rm gluon}_n(\pi)$ must be the Yang-Mills partial amplitude $M^{\rm YM}_n(\pi)$ up to a constant, and similarly any expression satisfying the assumptions of $M^{\rm pion}_n(\pi)$ must be the Non-linear Sigma Model partial amplitude $M^{\rm NLSM}_n(\pi)$. Hence, there exist constants $\alpha_\pi, \alpha'_\pi$ for every $\pi$ so that
\be\label{prop}
M_n^{\rm gluon}(\pi)=\alpha_\pi M_n^{\rm YM}(\pi)\qquad\qquad M_n^{\rm pion}(\pi)=\alpha_\pi' M_n^{\rm NLSM}(\pi)
\ee
Finally, recall that the partial amplitudes satisfy BCJ relations due to projectivity of the form. It follows that the constants $\alpha:=\alpha_\pi$ are identical for all $\pi$ and likewise for $\alpha':=\alpha_\pi'$ so that the scattering forms are unique up to a constant:
\be
\Omega^{(n{-}3)}_{\rm gluon}=\alpha~\Omega^{(n{-}3)}_{\rm YM}\,\quad\qquad \Omega^{(n{-}3)}_{\rm pion}=\alpha'~\Omega^{(n{-}3)}_{\rm NLSM}\,
\ee

Note that projectivity plays a crucial role without which we could have put arbitrary constants on the right hand side of Eq.~\eqref{eq:DDM_form}, thus leading to a $(n{-}2)!$-parameter family of solutions. 
Furthermore, permutation symmetry, unitarity and factorization all emerge as natural consequences of gauge invariance/Adler's zero and projectivity (and some technical constraints on the numerators), even though none was assumed. 

For all $n$, these forms can be obtained from the color-dressed amplitude by directly applying the relation Eq.~\eqref{eq:color_dual}, thus establishing their existence. Here we give explicit examples for $n{=}4$. The NLSM form reads:
\be
\Omega^{(1)}_{\rm NLSM}= s\,t\,d\log \left(\frac st\right)=t\,d\,s-s\,d\,t
\ee 
which also equals $(u\,d t- t\,d u)=(s\,d u-u\,d s)$ and is thus permutation invariant up to a sign. We can express the YM form as a combination of two $\phi^3$ forms:
\be \label{4ptYM}
\Omega^{(1)}_{\rm YM}=N_s\,d\log \left(\frac s t\right)+ N_u\,d\log \left(\frac u t\right)
\ee
where $N_s, N_u$ are BCJ numerators for the $s$ and $u$ channels (see {\it e.g.}~\cite{Mafra:2011kj}).



\subsection{YM and NLSM from the Worldsheet}
\label{sec:pushforward}
We now discuss the worldsheet origin of projective scattering forms with YM and NLSM as the primary examples. First we show that every projective scattering form $\Omega^{(n{-}3)}[N]$ on ${\cal K}_n$ can be obtained as the pushforward of an equivalence class of forms $\omega_n[N]$ on the moduli space ${\cal M}_{0,n}$. In particular, the planar scattering form $\Omega_{\phi^3}^{(n{-}3)}[\alpha]$ is obtained by pushing forward the Parke-Taylor form $\omega^{\rm WS}_{n}[\alpha]$.

Recall that given any form $\omega_n(\sigma)$ on moduli space, its pushforward is given by substituting and summing over all solutions of the scattering equations
\be
\omega(\sigma)\quad\rightarrow\quad
\sum_{{\rm sol.}\;\sigma}\omega_n(\sigma)
\ee
Note that two forms $\omega_n$ and $\omega_n'$ are pushed to the same forward if and only if they are equivalent on the support of the scattering equations. We therefore ``equate'' moduli space forms $\omega_n(\sigma),\omega_n'(\sigma)$ that are equivalent on the support of the scattering equations for which
\be
\omega_n (\sigma) \simeq \omega'_n (\sigma) \quad \implies\quad \sum_{{\rm sol.}\;\sigma}~\omega_n (\sigma)=\sum_{{\rm sol.}\;\sigma}~\omega'_n (\sigma)
\ee

We now wish to classify all forms on moduli space that pushforward to projective scattering forms. Recall from Appendix~\ref{app:bcj} that every projective form can be expanded in a basis of $(n{-}2)!$ planar scattering forms with coefficients given by kinematic numerators for multi-peripheral graphs:
\be
\Omega^{(n{-}3}[N]=\sum_{\pi\in S_{n{-}2}}N(g_\pi|\pi)\Omega_{\phi^3}^{(n{-}3)}[1,\pi(2),\ldots, \pi(n{-}1),n]
\ee
which can obviously be obtained by pushing forward the following form on moduli space:
\be\label{eq:general_omega}
\omega_n[N]=\sum_{\pi\in S_{n{-}2}}N(g_\pi|\pi)\omega_{\phi^3}^{\rm WS}(\pi)
\ee
In this way, we can construct a worldsheet form that gives {\it any} projective form as a linear combination of Parke-Taylor forms with different orderings.

Two important worldsheet forms are the YM and NLSM forms, which are determined by the corresponding CHY half-integrand. More precisely, we claim that
\be\label{YMNLSM}
\Omega^{(n{-}3)}_{\rm YM}=\sum_{{\rm sol.~}\sigma} d\mu_n~{\rm Pf}'\Psi_n   \qquad \Omega^{(n{-}3)}_{\rm NLSM}=\sum_{{\rm sol.~}\sigma} d\mu_n~{\rm det}' A_n
\ee 
where $d\mu_n:=d^n \sigma/{\rm vol~[SL}(2)]$ and ${\rm Pf}' \Psi_n$ and ${\det }' A_n$ are the reduced Pfaffian and determinant (both permutation invariant), respectively, as defined in~\cite{Cachazo:2013hca}.
\be
{\rm Pf}'\Psi_n:=(-1)^{i{+}j}\frac{{\rm Pf}|\Psi_n|^{i,j}_{i,j}}{\sigma_{i,j}}\,\quad {\rm det}' A_n:=\frac{\det |A_n|^{i,j}_{i,j}}{\sigma_{i,j}^2}\,\quad{\rm for~any}~1\leq i<j\leq n\,
\ee 
Here $\Psi_n(\sigma, \epsilon, p)$ is the $2n \times 2n$ matrix built from polarizations and momenta
\be
\Psi_n:=\left(
\begin{array}{cc}
A&-C^T\\
C&B
\end{array}
\right)\,
\ee
where $A_{a,b}, B_{a,b},C_{a,b}$ are $n \times n$ block matrices given by:
\be
\begin{split}
A_{a,b}&:=\begin{cases}
\frac{p_a\cdot p_b}{\sigma_{a,b}}&a\neq b\\0&a=b
\end{cases}\,\qquad
B_{a,b}:=\begin{cases}
\frac{\epsilon_a\cdot\epsilon_b}{\sigma_{a,b}}&a\neq b\\0&a=b
\end{cases}\,\quad\\
C_{a,b}&:=\begin{cases}
\frac{\epsilon_a\cdot p_b}{\sigma_{a,b}}&a\neq b\\-\sum_{c\neq a}C_{a,c}&a=b
\end{cases}\,
\end{split}
\ee
An important property of these worldsheet forms is that, on the support of scattering equations, ${\rm Pf}' \Psi_n$ is manifestly gauge invariant~\cite{Cachazo:2013hca} and $\det' A_n$ has the Adler zero~\cite{Cachazo:2014xea}:
\be\label{condition}
{\rm Pf}' \Psi_n (\epsilon^\mu_i \to \epsilon^\mu_i + \alpha p^\mu_i)={\rm Pf}' \Psi_n (\epsilon^\mu_i)\, \qquad \lim_{p_i^\mu \to 0} \det{}'A_n= 0\,
\ee
The uniqueness of the YM and NLSM forms under the conditions discussed above implies that there is a unique equivalence class of worldsheet forms for each theory, which by \eqref{condition} must be given by ${\rm Pf}' \Psi_n$ and $\det' A_n$, respectively. 
Finally, it is well known that both ${\rm Pf}' \Psi_n$ and $\det' A_n$ can be expanded in terms of Parke-Taylor forms with coefficients given by BCJ numerators~\cite{Cachazo:2013iea}, which reaffirms the result already found in Eq.~\eqref{eq:general_omega}. For example, the $n{=}4$ forms are  
\ba
\det{}' A_4 d\mu_4=\frac{s^2 d\mu_4}{\sigma^2_{12}\sigma_{34}^2} \;\;&\rightarrow&\;\; s\,t\,\left(\omega^{\rm WS}_{\phi^3} (1234)+ \omega^{\rm WS}_{\phi^3}(1324)\right) = s\,t\,\omega^{\rm WS}_{\phi^3}(1423)\,\nl
{\rm Pf}'\Psi_4 d\mu_4 \;\;&\rightarrow&\;\; N_s~\omega^{\rm WS}_{\phi^3} (1234) - N_u~\omega^{\rm WS}_{\phi^3} (1324)\,\nonumber
\ea

\subsection{Extended Positive Geometry for Gluons and Pions?}

It is clear that the gluon and pion scattering forms are fundamental objects, with a canonical purpose in life directly in kinematic space as well as on the worldsheet. What we are still missing is the complete connection of these scattering forms with positive geometries. The obstacle is the most obvious one: while the forms are dictated by god-given properties of gauge-invariance/Adler zero and projective invariance, they are not {\it canonical forms} which must have not only logarithmic singularities but also unit leading residues. 
This may be taken as an invitation to {\it e.g.} further ``geometrize'' the polarization vectors---something we have already seen as a critical part of the amplituhedron story in four dimensions---or there may be other ways to more naturally tie the ``prefactors'' in both YM and the NLSM to the underlying (associahedron) geometry universally associated with the poles of (planar) cubic graphs. 

\section{Summary and Outlook}
\label{sec: outlook}
Let us quickly recap the main ideas we have discussed in this paper. 
\begin{itemize}
\item Scattering amplitudes are better thought of as ``scattering forms''---differential forms on kinematic space. 
\item The kinematic associahedron is the analog of the amplituhedron for bi-adjoint $\phi^3$ theory at tree level, and the tree amplitude is the canonical form of this associahedron.
\item The associahedron geometry makes manifest properties of bi-adjoint $\phi^3$ amplitudes such as factorization and ``soft'' limit. It also provides new representations of the amplitudes from triangulations of the geometry, with the Feynman diagram expansion being one particular triangulation.
\item The tree-level open string moduli space is an associahedron, and scattering equations provide a diffeomorphism between the worldsheet and kinematic associahedra. Furthermore, the pushforward of the Parke-Taylor form---the canonical form of the worldsheet associahedron---gives the tree scattering form for the bi-adjoint scalar theory.
\item ``Color is Kinematics'': the differential forms for cubic graphs satisfy Jacobi relations identical to color factors, thus a color-dressed amplitude is dual to a scattering form and partial amplitudes are obtained as pullbacks of the form to appropriate subspaces. 
\item It is natural to study scattering forms in the big kinematic space, and for a form to be projectively well defined, kinematic numerators must satisfy the same Jacobi identities as color factors. 
\item Two primary examples are the scattering forms for Yang-Mills and the Non-Linear Sigma Model. These forms are uniquely fixed by standard power-counting, gauge invariance/Adler zero conditions, and projectivity. 
\end{itemize}

There are many obvious unanswered questions and open avenues of investigation suggested by our results.  For instance: Is there a complete geometrization of scattering forms for YM and the NLSM that brings the polarization vectors into the geometry? This question is of course also relevant to the search for geometries connected to gravity amplitudes. While we do not have any natural scattering forms due to the absence of color, the amplitudes can be obtained using the double-copy construction a la BCJ~\cite{Bern:2008qj,Bern:2010ue}. More precisely, for a double copy of the form $L\otimes R$ between theories $L$ and $R$, the amplitude for the product theory can be obtained directly from either $\Omega^{(n{-}3)}_L$ for the $L$ theory or $\Omega^{(n{-}3)}_R$ for the $R$ theory by replacing the wedge products $W(g)$ with appropriate kinematic numerators:
\be
{\bf M}^{L \otimes R}_n=\sum_{\text{cubic }g}~\frac{N_L(g)~N_R(g)}{\prod_{I\in g} s_I}=\Omega_{L}^{(n{-}3)}|_{W(g)\to N_R(g)}=\Omega_{R}^{(n{-}3)}|_{W(g) \to N_L(g)}\,,
\ee
For example, we obtain gravity from the product YM $\otimes$ YM, Born-Infeld theory from the product YM $\otimes$ NLSM and the so-called special Galileon theory from NLSM $\otimes$ NLSM~\cite{Cachazo:2014xea}. Along this line, it is very tempting to connect our worldsheet picture for the open string to the ambitwistor string~\cite{Mason:2013sva,Adamo:2013tsa,Ohmori:2015sha, Casali:2015vta}, and related worldsheet methods using scattering equations~\cite{Casali:2016atr, Siegel:2015axg} which are exclusively for the closed string. Furthermore, could we understand the double-copy construction, and possible geometries for gravity amplitudes in a way similar to the Kawai-Lewellen-Tye relations connecting open- and closed-string amplitudes~\cite{Kawai:1985xq} (See \cite{Huang:2016bdd, Mizera:2017rqa} for related ideas)?

We have seen the YM scattering form as pushforward of the Pfaffian form on the worldsheet, which is unique gauge invariant under the assumptions provided. What is even more remarkable is that the full-fledged {\it open string amplitude} can be obtained by directly integrating the Pfaffian on the worldsheet associahedron (with Koba-Nielsen factor~\cite{Koba:1969kh} as a natural regulator for logarithmic divergences)! While this can be shown by string theory calculation ({\it c.f.} \cite{Mafra:2011nw}), there must be deep conceptual reasons why the gauge-invariant object for gluon scattering in YM theory completely dictates the infinite series of higher-dimensional corrections from superstrings! It would be highly desirable to understand this fact better and to see in general how integrals and pushforwards of worldsheet forms are related to each other~\cite{strings}.

Let us end with a few suggestions for immediate avenues of progress which are more continuously connected to the themes introduced in this paper. 

\paragraph{General $d\log$ Projective Forms: Permutohedra and Beyond}
Recall that a scattering form Eq.~\eqref{dloggen} is called {\it $d\log$ scattering form} if it is projective and every kinematic numerator is either $0$ or $\pm 1$. A classification of all such forms is then equivalent to solving the Jacobi relations $\eqref{Jacobi}$ provided $N(g|\alpha_g)\in\{0,\pm 1\}$.
\be\label{dloggen}
\Omega^{(n{-}3)}_{\,d\log} (S)=\sum_{{\rm cubic~}g}~N(g)\;\Omega^{(n{-}3)}(g)(S)
\ee

While we do not have a complete classification, we can discuss some general properties. To every $d\log$ scattering form, we assign a connected graph $\Upsilon$ consisting of a vertex for every cubic graph $g$ whose numerator $N(g)$ is non-zero, with a line between any two vertices related by mutation. Furthermore, projectivity is satisfied precisely if every vertex is adjacent to exactly $(n{-}3)$ lines ({\it i.e.} the graph is ``simple''), and a sign flip occurs between any two vertices related by mutation. In particular, walking along any closed path in the graph $\Upsilon$ should return us back to the same sign. Note that this does not imply that the path must be of even length, since the sign at the initial vertex depends on its propagators which may have been reordered by the sequence of mutations.

The simplest example of of $d\log$ scattering forms is of course the planar scattering form $\Omega^{(n{-}3)}_{\phi^3}[\alpha] $ whose connected graph $\Upsilon$ is given by the skeleton of the $\alpha$-ordered associahedron, also known as the Tamari lattice~\cite{tamari_lattice}. In fact, for every $n$, the Tamari lattice provides the smallest number of vertices possible. However, the Tamari lattice is only the beginning of a large class of examples. For $n{=}4$, there is only one possible topology for the graph ({\it i.e.} a line segment). For $n{=}5$, we have seen possible topologies are pentagon, hexagon, octagon and nonagon. 


For all $n$, a large class of possible connected graphs are given by the skeleton of the ``Cayley polytopes'' discussed in~\cite{Gao:2017dek} whose $d\log$ scattering forms were obtained by pushing forward Cayley functions (expressed as a form on moduli space) via the scattering equations. The Cayley polytopes are polytopes constructed directly in kinematic space, of which the kinematic associahedron is one example. Furthermore, much of our associahedron discussion generalizes word-for-word to the Cayley polytopes, a summary of which is provided below.

\begin{itemize} 
\item 
The Cayley polytope (whose skeleton is the connected graph $\Upsilon$) is constructed directly in kinematic space $\K_n$ by intersecting a $(n{-}3)$-dimensional subspace with the positive region defined by setting $s_I\geq 0$ for every propagator appearing in the cubic graphs.
\item The pullback of the $d\log$ scattering form to the subspace gives the canonical form of the Cayley polytope.
\item The scattering form can be obtained as the pushforward of a form on moduli space ${\cal M}_{0,n}$. 
\end{itemize}
\begin{figure}[!htb]
\centering
\begin{overpic}[width=8cm]
{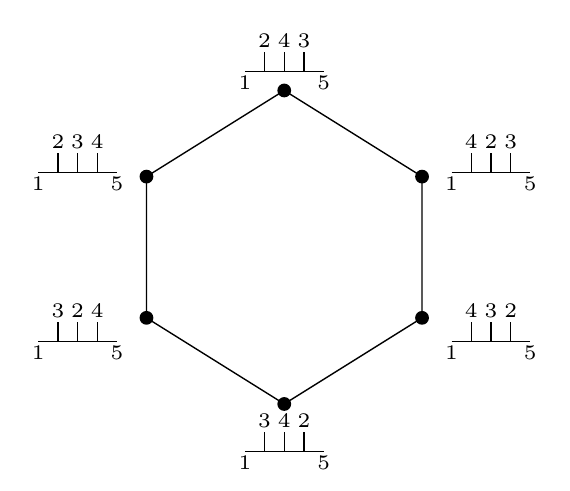}
\end{overpic}
\begin{overpic}[width=8cm]
{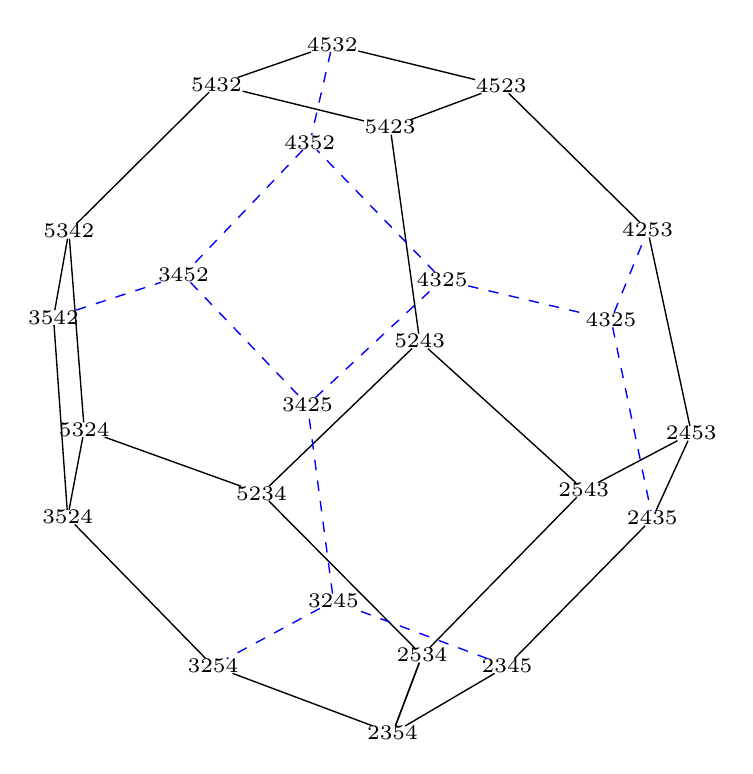}
\end{overpic}
\caption{\label{perm56} Permutohedra for $n{=}5$ (top) and $n{=}6$ (bottom)}
\end{figure}

Here we present the construction for one example: {\it permutohedron} ${\cal P}_n$~\cite{permutohedron}, which is the Cayley polytope with largest number of vertices for any $n$, where each of the $(n{-}2)!$ vertices corresponding to a multi-peripheral cubic graph with respect to $1$ and $n$ as shown in Figure~\ref{mtpphgrphpi}. 

We begin by defining the top-dimensional ``positive region'' where all possible poles of the multi-peripheral graphs are positive:
\be
s_{1a_1\cdots a_m}\qquad\text{for $m=1,\ldots, n{-}3$ and $2\leq a_1<\cdots<a_m\leq n{-}1$}
\ee
where every cut corresponds one of the $(2^{n{-}2}{-}2)$ facets of the permutohedron. Furthermore, the subspace is given by the following $(n{-}2)(n{-}3)/2$ conditions:
\be\label{subspaceperm}
s_{ij}\qquad\text{ is a negative constant for}\quad 2\leq i<j\leq n{-}1\,,
\ee
which are the analog of non-adjacent constants for the associahedron case. One can prove that the intersection of the positive region with the subspace gives the permutohedron by showing geometric factorization on all possible boundaries. Note that ${\cal P}_{n{=}4}$ is a line segment; and ${\cal P}_{n{=}5}$ is a hexagon while ${\cal P}_{n{=}6}$ is a truncated octahedron, as shown in Figure~\eqref{perm56}.  

Similar to that for associahedron, the (projective) scattering form for ${\cal P}_n$ is given by
\be\label{permutohedron}
\Omega_{{\cal P}_n}^{(n{-}3)}=\sum_{\pi \in S_{n{-}2}} {\rm sgn}(\pi)\bigwedge_{a=2}^{n{-}2}d\log s_{1\pi(2)\cdots \pi(a)}
\ee
where ${\rm sgn}(\pi)$ is the signum of permutation $\pi$. Furthermore, the pullback of the scattering form to the subspace (denoted $Q_n$) gives the canonical form of the permutohedron:
\be
\Omega_{{\cal P}_n}^{(n{-}3)}|_{Q_n}=\left(\sum_{\pi\in S_{n{-}2}} 
\frac{1}{
\prod_{a=1}^{n{-}2}s_{1\pi(2)\cdots \pi(a)}
}
\right)d^{n{-}3}s
\ee
Finally, the scattering form can be obtained as a pushforward of the following form on moduli space:
\be
\omega_{{\cal P}_n}:=\sum_{\pi\in S_{n{-}2}}{\rm sgn}(\pi)\omega_n^{\text{WS}}(\pi)
\ee
as suggested by Eq.~\eqref{eq:general_omega}.

The discussion provided above can be generalized to all Cayley polytpes studied in~\cite{Gao:2017dek}.
which belong in the much larger class of {\it generalized permutohedra} studied by Postnikov~\cite{Postnikov, Postnikov2}. In on-going discussions with Postnikov we have learned that our construction for these polytopes are equivalent to his under a natural change of variables. It is likely that a corresponding $d\log$ scattering form exists for generalized permutohedra. Moreover, for recent studies of worldsheet forms that are relevant to our construction, see~\cite{Early:2017lku, Cachazo:2017vkf}.


\paragraph{Massive Scalar Amplitudes, Non-logarithmic Forms}
While we have ostensibly focused on amplitudes for massless particles, for the bi-adjoint $\phi^3$ theory in particular it is clear that there is no obstruction to dealing with the scattering of massive particles. One interesting point about doing this is the following: we know that we can generate {\it e.g.} $\phi^4$ couplings from a cubic theory once massive particles are integrated out. Now, suppose we started with a $\phi^4$ theory; there are already small subtleties on how to geometrize the scattering form in this case related to the fact that the form simply does not have logarithmic singularities, and have singularities at infinity.  The addiction to ``forms with logarithmic singularities'' is perhaps the central obstacle to seeing connections to positive geometries. But if we generate the quartic coupling by integrating out massive scalars in a cubic theory, the full theory {\it does} have logarithmic singularities, and so we can ``sneak up'' on the hard problem of dealing with non-logarithmic singularities by regulating them as logarithmic ones which are then sent to infinity. Furthermore, the scattering forms for the NLSM have non-trivial residues on all the poles as well as poles at infinity; it would again be fascinating to find a purely geometrical characterization of these residues. 

\paragraph{Loops}
Furthermore, we can immediately start to explore scattering forms and possible positive geometries associated with the loop integrand for {\it e.g.} the bi-adjoint scalar theory. We can attempt to mimic the steps needed to ``upgrade'' the amplitude to a differential form at loop level. An early and obvious source of annoyance is what to do about bubble topologies, since naively including them would give double and higher poles, thus ruining the logarithmic singularities of the form. It is perhaps reasonable to then sum over all diagrams excluding these bubbles. At four points and one loop, this leaves us with a sum over five $d\log$ forms. Can these forms be made to be projective, and is there a positive geometry in the extended kinematic space of loop and Mandelstam variables attached to the loop scattering forms?  

Going beyond the bi-adjoint scalar case, one can consider scattering forms for loop integrands in gauge theories and more general theories with color. In particular, it should be straightforward to write down forms for one-loop maximally supersymmetric Yang-Mills amplitudes in general dimensions, since there is no contribution from bubbles. Similarly we expect these forms for loop integrands to have a worldsheet origin that may be related to scattering equations and ambitwistor strings at loop level~\cite{Adamo:2013tsa, Casali:2014hfa, Geyer:2015bja, He:2015yua, Geyer:2015jch, Cachazo:2015aol, Geyer:2016wjx, He:2016mzd,He:2017spx, Gomez:2016cqb}. 

\paragraph{Scattering Forms, Amplituhedron and Twistor Strings in Four Dimensions}

We have now seen {\it two} notions of scattering forms. In the story of the amplituhedron the forms play the role of combining different helicity amplitudes (as does the super-amplitude) into a single object, while in this paper the differential forms are tied to the geometrization of color. How are these pictures related to each other? There must be a connection, not only for moral reasons, but for more pragmatic and technical ones. We know that the scattering equations and the CHY formula for gluon amplitudes transition smoothly in four-dimensional spacetime to the Roiban-Spradlin-Volovich (RSV) equations and the twistor-string formulas for ${\cal N}=4$ SYM scattering amplitudes~\cite{Roiban:2004yf}. The latter is deeply connected to the geometry of the positive Grassmannian and the amplituhedron, while we have exposed the connection of the former to the worldsheet associahedron. Making progress on these particular questions will undoubtedly need some conceptually new ideas. 

On the other hand, first steps have been taken in identifying scattering forms for (tree-level) super-amplitudes in ${\cal N}=4$ SYM and the ``amplituhedron'' in ordinary, four-dimensional momentum space; these are $(2n{-}4)$-forms $\Omega_n^{(2n{-}4)}$ encoding all helicity amplitudes in the space of $\{\lambda_a, \tilde\lambda_a~|~a=1,2, \cdots,n\}$ subject to momentum conservation, and the $(2n{-}4)$-dimensional ``amplituhedron'' lives in a ``positive region'' in the space with correct ``winding numbers''~\cite{KITP, strings}. In close analogy with our associahedron story, there is strong evidence that the four-dimensional scattering equations (RSV) provide a diffeomorphism from $G_{>0}(2,n)$ (the twistor-string worldsheet) to the ``amplituhedron'' in momentum space; its canonical form, or the pullback of $\Omega_n^{(2n{-}4)}$ to the subspace where it lives, is then given by the pushforward of the cyclic form of $G_{>0}(2,n)$~\cite{KITP, strings}. We leave the study of these exciting questions for future investigations. 

\section*{Acknowledgments}

We would like to thank Freddy Cachazo, Nick Early, Yvonne Geyer, Peter Goddard, Steven Karp, Thomas Lam, Alex Postnikov, Oliver Schlotterer, Hugh Thomas, Lauren Williams, Ellis Yuan and Yong Zhang for stimulating discussions. SH would also like to thank Nima Arkani-Hamed and the Institute for Advanced Study for their hospitality. GY would also like to thank the Department of Physics at Tsinghua University for four great years of undergraduate education. We would also like to thank the anonymous referee for helpful comments. The work of NAH is supported by the DOE under grant DOE DESC0009988. YB is supported by the Department of Physics, Princeton University and a NSERC postgraduate scholarship. SH's research is supported in part by the Thousand Young Talents program and the Key Research Program of Frontier Sciences of CAS.
\appendix

\section{A Quick Review of Positive Geometries and Canonical Forms}
\label{app:pos}

In this section, we provide a quick review of {\it positive geometries} and {\it canonical forms}, which were introduced in~\cite{Arkani-Hamed:2017tmz} by two of the authors of the present paper and Thomas Lam.

\subsection{Definitions}
\label{app:pos_def}

Loosely speaking, a positive geometry $\A$ is a {\it real, oriented, closed geometry} with boundaries of all codimensions. In particular, each boundary of a positive geometry is again a positive geometry. For instance, polytopes are positive geometries with linear boundaries. More generally, a positive geometry can have curved boundaries defined by polynomials of higher order. A more rigorous definition of positive geometry was introduced in~\cite{Arkani-Hamed:2017tmz} as a semi-algebraic variety with some topological assumptions.

The crucial point is that every positive geometry has a {\it unique} differential form $\Omega(\A)$ defined on its ambient space called its canonical form, satisfying the following properties:
\begin{enumerate}
    \item
    It is meromorphic, with simple poles precisely along the boundaries of the geometry.
    \item
    For any hyper-surface $H$ containing a boundary $\B$ of $\A$, the residue along $H$ is given by
    \be\label{eq:canon_res}
    \Res_H\Omega(\A)=\Omega(\B)
    \ee
    \item
    If $\A$ is a point, then $\Omega(\A)=\pm 1$ depending on the orientation.

\end{enumerate}

Assuming that the ambient space does not admit any non-zero holomorphic top forms, the canonical form is unique for each positive geometry. For this reason, the positive geometry is usually embedded in (real) projective space $\Proj^N(\R)$ rather than (real) Euclidean space $\R^N$ where holomorphic top forms exist in abundance. But since $\R^N$ can be embedded in $\Proj^N(\R)$ via $x\rightarrow (1,x)$, it is convenient to visualize projective space as Euclidean space with a hyperplane at infinity.

One trivial property of canonical forms is that for any pair of positive geometries $\A$ and $\mathcal{B}$, we have
\be\label{eq:canon_prod}
\Omega(\A\times \mathcal{B})
=\Omega(\A)\wedge\Omega(\mathcal{B})
\ee
In addition, canonical forms have two important properties which we now discuss: {\it triangulation} and {\it pushforward}.

\subsection{Triangulations}
\label{app:pos_triang}

Given a {\it subdivision} of a positive geometry $\A$ by finitely many pieces $\A_a$, the canonical form satisfies
\be\label{eq:triangulation}
\Omega(\A)=\sum_a \Omega(\A_a)
\ee
We often refer to a subdivision as a {\it triangulation} even if the pieces $\A_a$ are not simplices. Since the right hand side is independent of the choice of triangulation, we say that:
\be
\text{The canonical form is {\it triangulation independent}.}
\ee
The intuition behind Eq.~\eqref{eq:triangulation} is that the {\it spurious poles} appearing on the right hand side cancel while the {\it physical poles} are identical on both sides. This is not as obvious as it may seem. Naively it is tempting to think that spurious poles cancel in pairs along the boundary between any two adjacent pieces of the triangulation, but this is generically false as multiple pieces may be needed to cancel a spurious pole. See Section 3 of~\cite{Arkani-Hamed:2017tmz} for a careful derivation.

\subsection{Pushforwards}
\label{app:pos_push}
Consider a map $\phi:\A\rightarrow \B$ between positive geometries of the same dimension. Given a form $\omega$ on the ambient space of $\A$, we can {\it push} it to a form $\eta$ on the ambient space of $\B$ via the map $\phi$:
\be\label{eq:push_def}
\omega(a)\;\;\xrightarrow[]{\phi}\;\;\eta(b):= \sum_{\text{roots $a$}}\omega(a)
\ee
where for any $b\in\B$ we sum over all {\it complex} roots $a$ of $b=\phi(a)$, where $\phi$ is analytically continued. This is called a {\it pushforward}, also denoted by
\be
\phi_*(\omega) := \eta
\ee

Suppose, furthermore, that $\phi$ is a {\it diffeomorphism} between the interior of the two positive geometries, but possibly {\it finitely-many-to-one} when analytically continued. We claim that the map pushes the canonical form of the domain to the canonical form of the image:
\be\label{eq:push_claim}
\Omega(\A)\;\;\xrightarrow[]{\phi}\;\;\Omega(\B)
\ee
We therefore say that:
\be\label{eq:push}
\text{The pushforward preserves canonical forms.}
\ee
This claim has been proven in certain examples where the boundary structures of $\A$ and $\B$ are well understood~\cite{Arkani-Hamed:2017tmz}. However, it remains an outstanding challenge to prove it in the most general case. Some ideas for doing so is discussed in Section 4 of~\cite{Arkani-Hamed:2017tmz}, which involves a ``blowup'' procedure.

For computational purposes, the pushforward can be expressed in a more useful way. Let $a_i$ denote coordinates on $\mathcal{A}$ and $b_i$ coordinates on $\mathcal{B}$ for $i=1,\ldots, D$. Also let
\ba
\omega := f(a)d^Da\;\;\;\;\eta := g(b)d^Db
\ea
denote the top forms. Then
\be
g(b)=\int d^D a f(a)\;\delta^D(b-\phi(a))
\ee
where the integral sign is simply an instruction to sum over all roots on the support of the delta function. This is the {\it delta function expression} of the pushforward. It is important that the $b_i$ variables appear with unit Jacobian in the delta functions.

Before ending this section, we generalize the pushforward to the case where $\dim\B\geq \dim \A$. Consider a set of scalar equations $\Phi_i(a,b)=0$ with $a\in\A$ and $b\in\B$. Here $\Phi$ acts as an {\it implicit function} between the positive geometries, rather than a direct map. We assume that there are $\dim\A$ independent equations, and that for any $b\in\B$, there are {\it finitely} many complex roots $a\in\A$. Now, given a form $\omega$ on the ambient space of $\A$, we can push it to a form $\eta$ on the ambient space of $\B$ via $\Phi$:
\be
\omega(a)\;\;\xrightarrow[]{\Phi}\;\;\eta(b):= \sum_{\text{roots $a$}}\omega(a)
\ee
As before, we can denote the pushforward as:
\be
\Phi_*(\omega) := \eta
\ee

If $\dim\A=\dim\B$ and $\Phi_i(a,b)=b_i-\phi_i(a)$, then we recover~\eqref{eq:push}.

\subsection{Projective Polytopes and Dual Polytopes}
\label{app:pos_poly}
We discuss the properties of {\it convex polytopes as positive geometries}. While polytopes are most easily visualized in Euclidean space $\mathbb{R}^m$, for the present discussion it is more convenient to embed the polytope in projective space $\mathbb{P}^m(\mathbb{R})$ via $x\rightarrow (1,x)$. Let $Y=(1,x)$ denote a point in projective space with components $Y^A$ indexed by $A=0,\ldots, m$. Furthermore, let $W$ denote points in the {\it dual space} with components $W_A$, and we define the contraction $Y\cdot W:= Y^AW_A$ where the repeated index $A$ is implicitly summed.

Now consider a convex polytope $\mathcal{A}$ with vertices $Z_i=(1, Z_i')$. Then the interior of $\A$ is given by all positive linear combinations of the vertices in projective space:
\be
\mathcal{A}=\left\{\sum_i C_iZ_i \;\;|\;\; C_i>0\right\}
\ee
Note of course that the coefficients generically form a redundant representation of the interior. Furthermore, the polytope can be cut out by linear equations of the form $Y\cdot W_j\geq 0$ for some collection of dual vectors $W_j$. The facets of the polytope are therefore given by $Y\cdot W_j=0$.

Furthermore, we construct the {\it dual polytope} $\A^*$ as the convex polytope in the {\it dual projective space} whose vertices are given by the dual vectors $W_j$. It follows that the interior of $\A^*$ is the set of all positive linear combinations of the dual vectors:
\be
\mathcal{A}^*=\left\{\sum_j C_jW_j \;\;|\;\; C_j>0\right\}
\ee
It can be shown that $\A^*$ is precisely the set of all points $W$ cut out by the inequalities $W\cdot Z_i\geq 0$, implying that the facets of the dual polytope are given by $W\cdot Z_i=0$. This leads us to an important fact about the duality of polytopes:
\be
\text{The facets of $\A$ are dual to the vertices of $\A^*$, and vice versa.}
\ee
More generally, we have:
\begin{enumerate}
\item
\text{The codim-$d$ boundaries of $\A$ correspond to the $(d{-}1)$-boundaries of the dual $\A^*$.}
\item
\text{Any two boundaries of $\A$ differing by one dimension are adjacent precisely if}\\
\text{their duals are adjacent.}
\end{enumerate}
It follows that the dual of every simple polytope is simplicial, and vice versa. Recall that a polytope of dimension $m$ is called {\it simple} if every vertex is adjacent to exactly $m$ facets (or equivalently $m$ edges); and a polytope is called {\it simplicial} if every facet is a simplex. We leave the derivation as an exercise for the reader.

Having established the dual polytope $\A^*$, we find a direct connection to the canonical form of the original polytope $\A$---the canonical form is determined by the volume of the dual. For any $Y$ on the interior of $\A$, we define a $Y$-dependent measure on the dual space:
\be\label{eq:dVol}
d\text{Vol} := \frac{\left<Wd^mW\right>}{(Y\cdot W)^{m{+}1}}
\ee
where the angle brackets denote the determinant $\left<Wd^mW\right>:=\det\left(W,dW,\ldots, dW\right)$. The ($Y$-dependent) volume of the dual $\A^*$ is therefore
\be
\text{Vol}(\A^*):=\int_{W\in\A^*}d\text{Vol}
\ee
Then, as shown in Section 7 of~\cite{Arkani-Hamed:2017tmz}, the canonical form of the polytope $\mathcal{A}$ is determined by the volume of the dual polytope $\mathcal{A}^*$:
\be
\Omega(\A)={\rm Vol}(\A^*)\;\left<Yd^mY\right>/m!
\ee
which we also write as
\be
\aOmega(\mathcal{A})=\text{Vol}(\A^*)
\ee
where $\aOmega(\mathcal{A})$ is called the {\it canonical rational function} and is the coefficient of the universal factor $\left<Yd^mY\right>/m!$ appearing in the canonical form $\Omega(\mathcal{A})$. For convenience, the canonical rational function defined here is normalized differently from the one defined in the original reference~\cite{Arkani-Hamed:2017tmz}. In particular, the volume of a dual simplex $\Delta^*$ with vertices $W_1,\ldots, W_{m{+}1}$ is given by
\be\label{eq:simplex_vol}
\text{Vol}(\Delta^*)=\frac{\left<W_1\cdots W_{m{+}1}\right>}{\prod_{j=1}^{m{+}1}(Y\cdot W_j)}
\ee
which can be computed by integrating Eq.~\eqref{eq:dVol} over all $Y=C_1W_1+\cdots+C_mW_m+W_{m{+}1}$ parameterized by $C_1,\ldots, C_m>0$. In order for the integral to converge, it suffices to put $Y$ inside the simplex by requiring $Y\cdot W_j>0$ for every $j$. The canonical form is therefore
\be\label{eq:canon_simplex}
\Omega(\Delta)=
\frac{\left<W_1\cdots W_{m{+}1}\right>}{m!\prod_{j=1}^{m{+}1}(Y\cdot W_j)}\left<Yd^mY\right>=\bigwedge_{j=1}^m d\log\left(\frac{Y\cdot W_j}{Y\cdot W_{m{+}1}}\right)
\ee
where the equivalence of the last two expressions can be seen by applying a ${\rm GL}(m{+}1)$ transformation to fix the $W_j$'s to the identity matrix. The last expression is antisymmetric in the $W_j$'s for all $j$ even though the appearance of $W_{m{+}1}$ appears to break this symmetry. Alternatively, the canonical form can be expressed in terms of the vertices $Z_1,\ldots, Z_{m{+}1}$ of the simplex as follows:
\be\label{eq:canon_simplex_Z}
\Omega(\Delta)=\frac{\left<Z_1\cdots Z_{m{+}1}\right>^m}{m!\prod_{i=1}^{m{=}1}\left<YZ_1\cdots \hat{Z}_i\cdots Z_{m{+}1}\right>}\left<Yd^mY\right>
\ee
where the hat denotes omission. This formula can be derived by substituting $(W_j)_A=\epsilon_{AA_1\cdots A_m}Z_{j{+}1}^{A_1}Z_{j{+}2}^{A_2}\cdots Z_{j{+}m}^{A_m}$ into Eq.~\eqref{eq:canon_simplex}, whereby the facet $Y\cdot W_j=0$ is assumed to be adjacent to the vertices $Z_{j{+}1},\ldots,Z_{j{+}m}$. Note that the index on the vertices are labeled modulo $(m{+}1)$. More generally, the volume of a dual {\it polytope} $\A^*$ can be obtained by triangulation into simplices and summing over the volume of each simplex.

We summarize the key points as follows:
\begin{enumerate}
\item
Every convex polytope $\A$ has a {\it dual polytope} $\A^*$.
\item
The canonical form of the polytope $\Omega(\A)$ is determined by the volume of the dual polytope ${\rm Vol}(\A^*)$.
\end{enumerate}

\subsection{Simple Polytopes}
\label{app:pos_simple}


We now compute the canonical form of simple polytopes for which a simple formula exists. Let $\A$ denote a convex simple polytope of dimension $m$. We claim that the canonical form can be expressed as a sum over all vertices:
\begin{equation}\label{eq:canon_simple}
\Omega(\A)=\sum_{{\rm vertex} Z} \sign(Z) \bigwedge_{a=1}^m d\log \left(Y\cdot W_a\right)
\end{equation}
where for every vertex $Z$ the dual vectors $W_a$ correspond to the $m$ adjacent facets. Furthermore, the $\sign(Z)\in\{\pm 1\}$ denotes the orientation of the facets $W_1,\ldots,W_{n{-}3}$, which of course is antisymmetric. It is important that the polytope be simple, for otherwise the expression would be ill-defined.

We derive Eq.~\eqref{eq:canon_simple} by induction on dimension $m$. For $m=0$, $\A$ is an isolated point and the canonical form is simply $\pm 1$ depending on its orientation. Now suppose $m>0$, and our claim has been proven for all dimensions less than $m$. It suffices to argue that~\eqref{eq:canon_simple} has the correct first order poles and residues, since the canonical form is uniquely defined by such properties. Clearly, it has poles on the facets of the polytope, as required. Furthermore, for any facet $F$ given by $Y\cdot W=0$, the residue of Eq.~\eqref{eq:canon_simple} along $Y\cdot W=0$ is
\be
\sum_{Z'} \sign(Z')\bigwedge_{a=1}^{m{-}1}d\log(Y\cdot W_a)
\ee
where we sum over all vertices $Z'$ adjacent to the facet $F$. But by the induction hypothesis this is the required canonical form $\Omega(F)$, thus completing the derivation.

\subsection{Recursion Relations}
\label{app:pos_rec}
We argue that the canonical form of a convex polytope $\A$ of dimension $m$ can be obtained from the canonical forms of its facets. This provides a recursion relation for the canonical forms of polytopes. Combined with the factorization properties of the kinematic associahedron as discussed in Section~\ref{sec:fac}, this provides recursion relations for the amplitude as shown in Section~\ref{sec:rec}.

However, there is an obvious difficulty. The canonical form of a facet is only defined on the hyperplane containing that facet, while the canonical form of $\A$ is defined on the whole space. We resolve this issue by pulling back the facet canonical form via a projection map.
For each facet $F$ of $\A$, let $W_F\cdot Y=0$ denote the hyperplane containing $F$ for some dual vector $W_F$. We pick a reference point $Z_*$ on the interior of $\A$, and we establish a deformation $Y\rightarrow \hat{Y}$ given by
\be\label{eq:rec_shift}
\hat{Y}=Y-\frac{(Y\cdot W_F)}{(W_F\cdot Z^*)}Z_*
\ee
which can be visualized by drawing the straight line crossing $Y$ and $Z_*$, and recognizing $\hat{Y}$ as the intersection between the line with the hyperplane $Y\cdot W=0$. Hence, the deformation projects points onto the hyperplane along the direction of the reference point. We can therefore pullback the canonical form $\Omega(F)$ of the facet, thus giving us a form on the whole space which we denote by $\hat{\Omega}(F)$.

We now argue that the canonical form $\Omega(\A)$ can be obtained from the $\hat{\Omega}(F)$ forms by employing a little ``trick'', which instructs us to factor out the universal factor $\left<*Yd^{m{-}1}Y\right>$ from $\hat{\Omega}(F)$ and replace it by a different form:
\be\label{eq:num_replace}
\left<*Yd^{m{-}1}Y\right>\rightarrow \frac{(Z_*\cdot W)}{(Y\cdot W)}\left<Yd^mY\right>/m
\ee
We denote this ``replacement procedure'' as an operator $D_W$ giving $\hat{\Omega}(F)\rightarrow D_W\hat{\Omega}(F)$. While this is not a differential operator, it increases the rank of the form by one.

The $D_W$ operator may seem unfamiliar, but it has a simple geometric interpretation. For any facet $F$, the $D_W\hat{\Omega}(F)$ is nothing more than the canonical form of the polytope given by the convex hull of $F$ with $Z_*$ which we denote by $\A(Z_*,F)$:
\be\label{eq:DW_canon}
D_W\hat{\Omega}(F)=\Omega(\A(Z_*,F))
\ee
We show this in the case where $F$ is a simplex with vertices $Z_1,\ldots, Z_{m}$ whose canonical form (See Eq.~\eqref{eq:canon_simplex_Z}) is given by
\be\label{eq:canon_facet}
\Omega(F) = \frac{\left<XZ_1\cdots Z_m\right>^{m{-}1}\left<XYd^{m{-}1}Y\right>}{(m{-}1)!\prod_{a=1}^m\left<YXZ_1\cdots \hat{Z}_a\cdots Z_m\right>}
\ee
where the hat denotes omission and $X$ is an arbitrary vector for which $W\cdot X\neq 0$. Also, the point $Y$ is restricted to the hyperplane where the facet lives. It can be shown that $\Omega(F)$ is independent of $X$. The pull back via the deformation $Y\rightarrow \hat{Y}$ is therefore
\be
\hat{\Omega}(F) = \frac{\left<Z_*Z_1\cdots Z_m\right>^{m{-}1}\left<Z_*Yd^{m{-}1}Y\right>}{(m{-}1)!\prod_{a=1}^m\left<YZ_*Z_1\cdots \hat{Z}_a\cdots Z_m\right>}\ee
which is most easily obtained by setting $X=Z_*$ in Eq.~\eqref{eq:canon_facet} and realizing that the deformation term in $\hat{Y}$ is absorbed in the brackets by the $Z_*$ so that $\hat{Y}$ can be replaced by $Y$ wherever it appears. Applying the replacement operator $D_W$ then gives
\be\label{eq:DW_simplex}
D_W\hat{\Omega}(F) = \frac{\left<Z_*Z_1\cdots Z_m\right>^m\left<Yd^mY\right>}{m!\left<YZ_1\cdots Z_m\right>\prod_{a=1}^m\left<YZ_*Z_1\cdots \hat{Z}_a\cdots Z_m\right>}
\ee
where we have substituted $W_A=\epsilon_{AA_1\cdots A_m}Z_1^{A_1}\cdots Z_m^{A_m}$ because the hyperplane $Y\cdot W=0$ is spanned by the vertices $Z_1,\ldots, Z_m$. But in light of Eq.~\eqref{eq:canon_simplex_Z}, we find that Eq.~\eqref{eq:DW_simplex} is precisely the canonical form of $\A(Z_*,F)$ as claimed. For the more general case where $F$ is a generic polytope, we derive Eq.~\eqref{eq:DW_canon} by triangulating $F$ in terms of simplices and applying the preceding argument to each simplex.

Finally, we propose that the canonical form $\A$ is given by
\be\label{eq:rec}
\Omega(\A)=\sum_{\text{facet }F} D_W\hat{\Omega}(F)
\ee
which follows directly from the fact that $\A$ is triangulated by the polytopes $\A(Z_*,F)$.

\section{Vertex Coordinates of the Kinematic Associahedron}

We now provide a recursive algorithm for deriving the vertices of the associahedron $\A_n$. Consider a vertex $Z_0$ corresponding to a triangulation of the $n$-gon. Our goal is to work out all planar components $X_{i,j}$ of $Z_0$ in terms of the non-adjacent constants $c_{kl}$. Our strategy is to compute the components one planar basis ({\it i.e.} a basis of planar variables given by the diagonals of any triangulation) at a time by starting with the $Z_0$ basis where all components vanish, and applying a sequence of mutations. Since every planar basis can be reached by such a sequence, this establishes a recursive procedure for computing all planar components. It suffices then to discuss how the components are related by mutation. Consider a mutation $Z\rightarrow Z'$ like the one shown in Figure~\ref{fig:diag_flip} (top) where $X_{i,k}$ mutates to $X_{j,l}$. From Eq.~\eqref{eq:XXXXc} we find
\be
X_{j,l}=X_{j,k}+X_{i,l}-X_{i,k}+\sum_{\substack{i\leq a<j\\k\leq b<l}}c_{ab}
\ee
which computes $X_{j,l}$ from the basis of $Z$, thus completing the algorithm.

Here we present the vertex coordinates for the kinematic associahedron $\A_{n{=}5}$ from Figure~\ref{fig:kin_assoc} (top right) in the basis $Y=(1,X_{13},X_{14})$:
\ba
Z_A&=&(1,0,0)\\
Z_B&=&(1,0,c_{14}+c_{24})\\
Z_C&=&(1,c_{13}+c_{14},c_{14}+c_{24})\\
Z_D&=&(1,c_{13}+c_{14},c_{14})\\
Z_E&=&(1,c_{13},0)\\
\ea

\section{BCJ Relations and Dual-basis Expansion from Projectivity}
\label{app:bcj}

We argue that the requirement of {\it projectivity} has two important consequences for scattering forms.
\begin{itemize}
    \item 
    The partial amplitudes satisfy BCJ relations.
    \item
    Every projective scattering form can be written as a linear combination of planar scattering forms $\Omega_{\phi^3}^{(n{-}3)}[\alpha]$ like Eq.~\eqref{eq:dualbasis}.
\end{itemize}

We derive the first claim by directly applying a local ${\rm GL}(1)$ transformation $s_I\rightarrow \Lambda s_I$ to the DDM form Eq.~\eqref{eq:DDM_form}. Invariance under the transformation implies
\be\label{eq:vanishes}
\sum_{\pi\in S_{n{-}2}}{\rm sgn}(\pi)M_n(\pi)\sum_{i=2}^{n{-}2}({-}1)^iz(\pi,i)\bigwedge_{j=2;\;j\neq i}^{n{-}2}dz(\pi,j)=0
\ee
where ${\rm sgn}(\pi)$ is the signum of the permutation and
\ba
M_n(\pi)&:=&M_n(1,\pi(2),\ldots, \pi(n{-}1),n)\\
z(\pi,i)&:=&s_{1,\pi(i)}+s_{\pi(2),\pi(i)}+\cdots+s_{\pi(i{-}1),\pi(i)}
\ea
Furthermore, we pull back to a subspace where
\be
\left\{ds_{ij}=0\;\text{for}\;1\leq i<j{-}1<n{-}1\right\}\cap\{ds_{n{-}2,n{-}1}=0\}
\ee
We find that the only permutations $\pi$ that contribute are $\pi_i$ for $k=2,\ldots,n{-}1$. For $2\leq k\leq n{-}2$ we have
\ba
\pi_k=(1,2,\ldots, k{-}1,n{-}1,k,k{+}1,\ldots, n{-}2,n)
\ea
and for $k=n{-}1$ we have $\pi_{n{-}1}=(1,2,\ldots, n)$. Moreover, for $2\leq k\leq n{-}2$ only the $i=k$ term in Eq.~\eqref{eq:vanishes} contributes, giving
\be\label{eq:part1}
M_n(\pi_k)z(\pi_k,k)\left\{(-1)^{n{-}1}\bigwedge_{j=2;\;j\neq k}^{n{-}2}dz(\pi_k,j)\right\}
\ee
where we applied ${\rm sgn}(\pi_k)=(-1)^{n{-}k{-}1}$. For $k=n{-}1$, however, all values of $i$ contribute in Eq.~\eqref{eq:vanishes}, giving
\be\label{eq:part2}
M_n(\pi_{n{-}1})\left[-\sum_{i{=}2}^{n{-}2}z(\pi_{n{-}1},i)\right]\left\{(-1)^{i{-}1}\bigwedge_{j=2;\;j\neq i}^{n{-}2}dz(\pi_{n{-}1},j)\right\}
\ee
We leave it as an exercise for the reader to show that the expressions in curly braces appearing in Eq.~\eqref{eq:part1} and Eq.~\eqref{eq:part2} are identical on the pullback. Furthermore, the square bracket expression in Eq.~\eqref{eq:part2} is equivalently
\be
[\cdots] = z(\pi_{n{-}1},n{-}1)
\ee
which follows from the kinematic identity $\sum_{1\leq i<j\leq n}s_{ij}=0$. Finally, combining the contributions for all $k$ gives
\be
\sum_{k=2}^{n{-}1}M_n(\pi_k)z(\pi_k,k)=0
\ee
Or equivalently
\be
\sum_{k=2}^{n{-}1}(s_{1,n{-}1}{+}s_{2,n{-}1}{+}\cdots{+}s_{k{-}1,n{-}1})M_n(1,\ldots, k{-}1,n{-}1,k,\ldots, n{-}2,n)=0
\ee
which we recognize as one of the fundamental BCJ relations. By pulling back to other subspaces, we can derive all BCJ relations. It follows that partial amplitudes of projective scattering forms satisfy BCJ relations.

We now move on to derive the second claim. Recall that the Jacobi identities impose linear relations for the kinematic numerators, leaving a basis of $(n{-}2)!$ independent elements. In particular, every numerator can be expanded in a basis of numerators corresponding to all multi-peripheral graphs with respect to $1$ and $n$ (See Figure~\ref{mtpphgrphpi}):
\be
N(\pi):=N(g_\pi|1,\pi(2),\ldots, \pi(n{-}1),n)\qquad \text{for $\pi\in S_{n{-}2}$}
\ee
Thus, every numerator has an expansion of the form
\be
N(g|\alpha_g) = \sum_{\pi\in S_{n{-}2}}M(g|\alpha_g;\pi)N(\pi)
\ee
for some coefficients $M(g|\alpha;\pi)\in \{0,\pm 1\}$. By the color-kinematics duality, the color factors must then obey the same expansion:
\be
C(g|\alpha_g) = \sum_{\pi\in S_{n{-}2}}M(g|\alpha_g;\pi)C(\pi)
\ee
where
\be
C(\pi):=C(g_\pi|1,\pi(2),\ldots, \pi(n{-}1),n)\qquad \text{for $\pi\in S_{n{-}2}$}
\ee
Substituting this into the (color-dressed) bi-adjoint scattering form Eq.~\eqref{eq:scalar_form} and extracting the coefficient of $C(\pi)$ ({\it i.e.} the only term contributing to the ordering $\pi$ in the standard trace decomposition) gives the following expansion for the planar scattering form:
\be
\Omega_{\phi^3}^{(n{-}3)}[\pi]=\sum_{\text{cubic }g}M(g|\alpha_g;\pi)\Omega^{(n{-}3)}(g|\alpha_g)
\ee
It follows that for an arbitrary (projective) scattering form, we have
\ba
\Omega^{(n{-}3)}[N]&=&\sum_{\text{cubic }g}N(g|\alpha_g)\Omega^{(n{-}3)}(g|\alpha_g)\\
&=&\sum_{\text{cubic }g}
\;
\sum_{\pi\in S_{n{-}2}}N(\pi)M(g|\alpha_g;\pi)\Omega^{(n{-}3)}(g|\alpha_g)\\
&=&\sum_{\pi\in S_{n{-}2}}N(\pi)\Omega_{\phi^3}^{(n{-}3)}[\pi]
\ea
which is a linear combination of planar scattering forms, as promised.

\bibliographystyle{utphys}
\bibliography{refs}

\end{document}